%% file: main.tex
\newcommand{\teff}{T$_{\mathrm{eff}}$}

\newcommand{\logg}{log~$g$}
\newcommand{\feh}{[Fe/H]}

\newcommand{\afe}{[$\alpha$/Fe]}

\newcommand{\acronym}[1]{{\small{#1}}}

\newcommand{\degree}{$^{\circ}$}
\newcommand{\hii}{\hbox{{\rm H}\kern 0.1em{\sc ii}{\rm }}}

\documentclass[twocolumn]{aastex701}

\graphicspath{{./}{figures/}}
\usepackage{multirow}

\begin{document}

\title{The Nineteenth Data Release of the Sloan Digital Sky Survey }
\shorttitle{SDSS-V DR19}


\correspondingauthor{Keith Hawkins} 
\email{keithhawkins@utexas.edu} 
 
\author{SDSS Collaboration} 
\affiliation{Sloan Digital Sky Survey}
\email{spokesperson@sdss.org}
 
 \author[0000-0002-5864-1332]{Gautham Adamane Pallathadka}
\affiliation{William H.\ Miller III Department of Physics \& Astronomy, Johns Hopkins University, 3400 N Charles St, Baltimore, MD 21218, USA}
\email{gadaman1@jhu.edu}

\author[0000-0001-8341-3940]{Mojgan Aghakhanloo}
\affiliation{Department of Astronomy, University of Virginia, Charlottesville, VA 22904-4325, USA}
\email{mvy4at@virginia.edu}

\author[0000-0003-1908-8463]{James Aird}
\affiliation{Institute for Astronomy, University of Edinburgh, Royal Observatory, Edinburgh EH9 3HJ, UK}
\email{james.aird@ed.ac.uk}

\author[0009-0000-0733-2479]{Andrés Almeida}
\affiliation{Department of Astronomy, University of Virginia, Charlottesville, VA 22904-4325, USA}
\email{tac6na@virginia.edu}

\author[0009-0000-3962-103X]{Singh Amrita}
\affiliation{Departamento de Astronom\'ia, Universidad de Chile, Camino del Observatorio 1515, Las Condes, Santiago, Chile}
\email{amrita@das.uchile.cl}

\author[0000-0003-4524-9363]{Friedrich Anders}
\affiliation{Institut de Ci\`encies del Cosmos (ICCUB), Universitat de Barcelona (UB), Mart\'i i Franqu\`es 1, E-08028 Barcelona, Spain}
\affiliation{Departament de Física Qu\`antica i Astrof\'isica (FQA), Universitat de Barcelona (UB), Mart\'i i Franqu\`es 1, E-08028 Barcelona, Spain}
\affiliation{Institut d'Estudis Espacials de Catalunya (IEEC), Esteve Terradas, 1, Edifici RDIT, Campus PMT-UPC, 08860 Castelldefels, Spain}
\email{fanders@fqa.ub.edu}

\author[0000-0002-6404-9562]{Scott F. Anderson}
\affiliation{Department of Astronomy, University of Washington, Box 351580, Seattle, WA 98195, USA}
\email{sfander@uw.edu}

\author[0000-0002-6270-8624]{Stefan Arseneau}
\affiliation{Department of Astronomy \& Institute for Astrophysical Research, Boston University, 725 Commonwealth Ave., Boston, MA 02215, USA}
\email{arseneau@bu.edu}

\author[0000-0001-5179-980X]{Consuelo Gonz\'alez~\'Avila}
\affiliation{Las Campanas Observatory, Ra\'{u}l Bitr\'{a}n 1200, La Serena, Chile}
\email{consuelo@carnegiescience.edu}

% ...

 \collaboration{all}{SDSS-V Collaboration}



\begin{abstract}
Mapping the local and distant Universe is key to our understanding of it. For decades, the Sloan Digital Sky Survey (SDSS) has made a concerted effort to map millions of celestial objects to constrain the physical processes that govern our Universe. The most recent and fifth generation of SDSS (SDSS-V) is organized into three scientific ``mappers". Milky Way Mapper (MWM) that aims to chart the various components of the Milky Way and constrain its formation and assembly, Black Hole Mapper (BHM), which focuses on understanding supermassive black holes in distant galaxies across the Universe, and Local Volume Mapper (LVM), which uses integral field spectroscopy to map the ionized interstellar medium in the local group. This paper describes and outlines the scope and content for the nineteenth data release (DR19) of SDSS and the most substantial to date in SDSS-V. DR19 is the first to contain data from all three mappers. Additionally, we also describe nine value added catalogs (VACs) that enhance the science that can be conducted with the SDSS-V data. Finally, we discuss how to access SDSS DR19 and provide illustrative examples and tutorials.

\end{abstract}



\section{An Introduction to the Ecosystem of the Sloan Digital Sky Survey} 
\label{sec:Introduction}

Over the last 27 years, the various generations of the Sloan Digital Sky Survey (SDSS) have collected a vast amount of photometric and spectroscopic data for millions of individual objects. From celestial objects in the Solar System to galaxies and quasars in the distant Universe, SDSS has made it a mission to map the cosmos. The first generation of SDSS \citep[SDSS-I,][]{York_2000_sdss} performed a photometric (specifically in the $ugriz$ bandpasses) and low resolution (R = $\lambda/\Delta\lambda \sim 1800$), optical (3800 - 9200~\AA) spectroscopic survey of stars, galaxies, and quasars. This survey was conducted at the 2.5m telescope located at Apache Point Observatory (APO). Following the success of SDSS-I, SDSS-II extended and completed the photometric goals of the first generation survey \citep[e.g.,][]{Frieman_2008_sdss2supernovae} while adding low-resolution, targeted optical spectroscopy for hundreds of thousands of stars in the Milky Way through the Sloan Extension for Galactic Understanding and Exploration \citep[SEGUE,][]{Yanny_09_SEGUE} survey. 

The diverse and highly informative spectra generated from SDSS-II helped pave the way for a new, third generation of SDSS \citep[SDSS-III,][]{Eisenstein_11_sdss3overview}, which was based solely on spectroscopy. The spectroscopic instruments of SDSS-III were updated and expanded to several fiber-fed, multi-object spectrographs (MOS). These MOS instruments used ``plug plates" (i.e., round metal plates that were drilled with holes to accept fibers) to place each fiber in its correctly targeted place on the sky. This enabled a detailed exploration for large samples of stars in the Milky Way and extragalactic sources. The original, low-resolution, optical spectrographs were updated and custom built \citep[e.g.,][]{Smee_2013} to conduct the Baryonic Oscillation Spectroscopic Survey of SDSS-III \citep[BOSS,][]{Dawson_2013_boss, Dawson_2016_eboss}. Following the successful use of the original spectrographs to observe not just galaxies but also stars (through SEGUE-1), these new ``BOSS" spectrographs were also used to conduct the SEGUE-2 survey \citep[e.g.][]{Rockosi_2022_segue2}, which primarily explored the old stellar populations in the Milky Way to constrain its formation and assembly. SDSS-III included not only a rebuild of the original SDSS optical spectrographs but also new instrumentation \citep{Wilson_2019_apogeespectrographs}. A moderate resolution (R$\sim$22500) near-infrared ($H$ band) spectrograph was built and installed to conduct the first generation of the Apache Point Galactic Evolution Experiment \citep[APOGEE-1]{Majewski_2017_apogeeoverview}, which aimed to further explore red giant stars in the distant Milky Way. SDSS-III also added a higher resolution (R$\sim$11,000) optical instrument to carry out the Multi-object APO Radial Velocity Exoplanets Large-area Survey \citep[MARVELS][]{Ge_2008_marvels}.

SDSS-IV \citep{Blanton_2017_sdss4} built on the excitement of the new instrumentation of the previous generation and added an additional APOGEE spectrograph to the Las Campanas Observatory (LCO) located in the southern hemisphere for the APOGEE-2 survey \citep{Zasowski_2017_apogee2targeting}. This enabled SDSS-IV to conduct one of the first dual-hemisphere near-infrared spectroscopic surveys. While SDSS-IV retired the spectrograph used for MARVELS, it added an optical integral field unit (IFU) to conduct a survey of the properties and assembly of nearby, low-redshift galaxies in the Mapping Nearby Galaxies at APO (MaNGA) survey \citep{Bundy_2015_manga}. SDSS-IV was completed in 2020 and subsequently reported its results in the seventeenth data release (DR17) of SDSS \citep{Abdurrouf_2022}. 

The current generation of SDSS \citep[SDSS-V,][Kollmeier et al, in press]{Kollmeier_2017_sdss5} began in 2020 and has focused on executing a panoptic, dual-hemisphere spectroscopic survey of stars, galaxies, quasars and a host of celestial objects in between. The broad goals of SDSS-V  are to: (i) understand the formation and evolution of the Milky Way and its stars, (ii) probe black hole growth across cosmic time, and (iii) constrain the the physical processes that govern the interstellar medium (e.g., stellar feedback, chemical enrichment, etc.). These goals are being achieved through 3 science-oriented `mappers’: Milky Way Mapper (MWM, Johnson et al., in prep., see section~\ref{subsec:MWM}), Black Hole Mapper (BHM, Anderson et al., in prep., see section~\ref{subsec:BHM}), and Local Volume Mapper \citep[LVM,][see also section~\ref{subsec:LVM}]{Drory2024}. 

Executing SDSS-V, particularly given its large number of targets and regularity of observing specific time-resolved objects required innovative leaps in instrumentation. While previous iterations of SDSS (I-IV) relied on fiber-fed, multi-object spectrographs (MOS) that used plug plates, to increase the speed at which observations can be taken, SDSS-V has developed a Focal Plane System (FPS) that instead positions fibers that feed various spectrographs using robotic positioners \citep{Pogge_2020_fps}. SDSS-V's use of robotic fiber positioners removed the need to drill plug plates. Although SDSS-V initially intended to use robotic fiber positioners and the FPS from its beginning, the onset of the COVID-19 pandemic at the start of the survey made this transition impossible. As a result, SDSS-V began its operations with the continued use of the plug plate system through the plate program and then gradually moved to the FPS infrastructure (see section~\ref{subsec:plateprogram} for more details) . 

The FPS system was installed and commissioned on both the du Pont telescope in the southern hemisphere in late 2022, which feeds the APOGEE-S and a BOSS spectrograph, and on the 2.5m SDSS telescope in the northern hemisphere at the Apache Point Observatory in 2022, which feeds the APOGEE-N and a northern BOSS spectrograph. In addition to this innovative step in infrastructure, a new telescope and wide-field IFU instrumentation system, known as the Local Volume Mapper Instrument \citep[LVM-I,][Konidaris et al, in prep.]{Herbst_2020_lvmi}, was installed in the southern hemisphere to carry out the LVM program. This program, new in SDSS-V, aims to explore gas, star-forming regions, and the interstellar medium in the Milky Way and nearby local group systems. The LVM instrument was commissioned in mid-2023 and the main science survey began late in the same year. In total, SDSS-V is operating with a low-resolution, optical spectrograph (BOSS) in the North and South (BOSS-N, BOSS-S), a moderate-resolution near-infrared spectrograph in the North and South (APOGEE-N, APOGEE-S), and a wide-field IFU system in the South (LVM-I).

Together, these instruments (i.e., APOGEE-S, APOGEE-N, BOSS-N, BOSS-S, and LVM-I) have been used to observe more than 10$^5$ celestial objects to achieve the ambitious objectives of SDSS-V. In the first SDSS-V data release, SDSS DR18 \citep{almeida2023}, detailed targeting information was presented for the BHM and MWM programs along with a modest number of new SDSS spectra ($\sim$25000), specifically for the BHM program.  

This paper will describe the nineteenth (and most recent) data release (hereafter DR19) of SDSS, which represents the first substantial public release of new data from the SDSS-V ecosystem. This release contains new data from all three SDSS-V mappers. To facilitate the discussion and contextualization of DR19, we begin by describing the science objectives of SDSS-V and its individual mapper programs in Section~\ref{sec:Scienceobjs}. The targeting that underlies DR19 is described in Section~\ref{sec:target}. With the targeting in hand, we describe in Section~\ref{sec:pipeline} the reduction and analysis pipelines used to turn the observed data into science-ready products. In section~\ref{sec:scope}, we outline the scope of DR19. In Section~\ref{sec:data.access}, we describe how the data and various Value Added Catalogs (VACs, outlined in Section~\ref{sec:VAC}) can be publicly accessed. To illustrate various ways in which the data can be used and accessed, in Section~\ref{sec:demos}, we describe a set of Python-based tutorials for each mapper. Finally, we summarize DR19 in Section~\ref{sec:Summary}. DR19 is also documented at \url{https://www.sdss.org/dr19/}.

\section{The Science Objectives of SDSS-V and its Mappers} \label{sec:Scienceobjs}
\subsection{Milky Way Mapper} \label{subsec:MWM}

The Milky Way Mapper (MWM) is observing stars across the Hertzsprung-Russell diagram (HRD) to decode Galactic history, understand stellar systems architecture, and  probe stellar physics. To achieve each of these in a combined MWM survey, we target stars in thirteen overarching programs. These programs include:

\begin{itemize}
    \item {\bf Galactic Genesis:} This program builds off the success of APOGEE-1 and APOGEE-2 in chemical cartography. SDSS-V aims to push chemical cartographic maps of the Milky Way to much larger distances than previous surveys. Whereas APOGEE-1 and -2 observed $<10^6$ unique stars, Galactic Genesis is forecasted to observe up to 3 million stars with APOGEE, with a significantly higher fraction being luminous red giants. This is accomplished by targeting redder, more evolved (and thus brighter) giant stars and removing dwarfs by using Gaia DR3. With this number of stars, Galactic Genesis will achieve 100 stars/pc$^2$ in disk out to 15 kpc, to constrain the chemodynamic history of the Galaxy.
    
    \item {\bf Magellanic Genesis:} Our nearby satellites, the Magellanic clouds for example, offer a unique laboratory to explore other galaxies and how they are affected by accretion processes. To this end, SDSS-V aims to observe individual RGB and AGB stars of stars in the Magellanic clouds with both BOSS and APOGEE. While no data for Magellanic Genesis is being released in DR19, we expect early results from this program will come with the first release of LCO data in the next data release.
    
    \item {\bf White Dwarfs:} White dwarfs (WDs) provide an invaluable record of intermediate mass star formation over the history of the Galaxy that is difficult to probe by other means. Subclasses of white dwarfs probe planetary debris, stellar magnetism, and complex binary interactions. SDSS-V will obtain spectra for thousands of WDs with the BOSS spectrographs. 
    
    \item {\bf Solar Neighborhood Census:} To understand the star formation history of the lowest mass objects, neither red giants nor WDs are suitable. By observing an unbiased, magnitude-limited sample of stars within $<$ 100 pc, the solar neighborhood census addresses questions of the metallicity distribution function and kinematics of the least luminous stars and provides a bridge to the more luminous stellar populations observed throughout the Galaxy. Most spectra will come from the BOSS instrument, but for the brighter targets APOGEE spectra will also exist. 
    
    \item {\bf Halo:} While 98\% of stars reside in the disk and bulge, the 2\% of stars in the halo reveal information on the Galaxy's merger history and early nucleosynthesis. In addition to the bright red giants observed by Galactic Genesis, the halo program expands MWM's view by observing fainter, more distant red giants and high-proper dwarfs with BOSS and also prioritizing metal-poor (\feh\ $< -1$) candidates from a variety of techniques.
    
    \item {\bf Young Stellar Objects (YSOs):} MWM combines the deep view of the Galaxy's history with a snapshot of current star formation for stars of all masses. The YSO program focuses on stars whose youth (age $<$ 100~Myr) can be identified by colors, clustering, or variability. These stars are observed with both APOGEE and BOSS.
    
    \item {\bf OB Stars:} SDSS-V aims to observe a large (10$^4$) population of O- and B-type stars, which spend little time getting to the main sequence and have maximum ages of $\sim$150 million years making them very young. These objects are crucial because they can help trace the young Milky Way. They also trace the sites of recent massive star formation their high binary fractions make them excellent laboratories to study the properties of binary systems.  This population of stars are being observed both with APOGEE and BOSS.
    
    \item {\bf Dust:} Star-forming regions, spiral arms, and cool gas throughout the thin disk of the Galaxy harbor dust. The dense sampling of Galactic Genesis provides an opportunity to learn about the dust distribution, the kinematics of the interstellar medium, and spatially-resolved reddening curves.

    \item {\bf Galactic eROSITA Sources:} Multi-wavelength studies of stars, especially compact objects, can be invaluable in understanding them. While SDSS-V operates in the optical and near-infrared, extending to other wavelengths is extremely valuable. As such, SDSS-V has a partnership with the extended ROentgen Survey with an Imaging Telescope Array \citep[eROSITA,][]{Predehl2021} which is on board the Spectrum Roentgen Gamma satellite \citep[SRG,][]{Sunyaev2021}. The eROSITA mission has produced a large, uniform survey of the X-ray sky. eROSITA sources in MWM include YSOs and compact binary systems, which have been observed with BOSS and APOGEE.
    
    \item {\bf Binary Systems:} MWM adopts a multi-pronged approach to probe stellar system architecture particularly for binary stars. To further explore the diversity of binary systems in the Milky Way, MWM has targeted these systems for radial velocity monitoring. For radial velocity variability, SDSS-V observations range from few-epoch or short-base line observations to long baselines with $>$ 10 visits. 
    
    \item {\bf Compact Binaries:} The systematic study of accreting compact white dwarf binaries (aCWDBs) is important for two main reasons: (i) it will deliver the necessary input for models of close binary evolution, and (ii) it will eventually allow to synthesize the Galactic Ridge X-ray Emission. For both cases, one needs large (N$>$ 10$^3$) and largely unbiased samples. SDSS-V aims to observe spectra to fulfill this need by using the followup spectroscopy to filter out candidate objects identified with X-ray emission.

    \item {\bf Planet Hosts:} The chemical abundance patterns of planet hosting stars is critical to better understand the formation pathways of exoplanets and their frequency across the Milky Way. Achieving a large, homogeneously derived set of stellar abundances of planet hosting stars is the key to achieving this. SDSS-V aims to observe hundreds to thousands of planets hosting starts with APOGEE to build a homogeneous set of stellar abundances of planet hosting stars. Additionally, with the change in TESS cadences to much faster times for the full-frame images and decrease in number of special slots, we refocused on observing planet hosts. Spectra for these targets will be released starting in the next data release.
    
    \item {\bf Asteroseismic Red Giants:} Asteroseismology, the study of stellar oscillations, constrains the density, surface gravity, and evolutionary states of red giant stars. Building on efforts from SDSS-III and SDSS-IV, this program combines the asteroseismic measurements from space missions like Kepler and TESS with the spectroscopy measured by MWM in order to estimate accurate and precise masses, radii, and ages for these stars. This effort allows for studies of stellar interiors and evolution, serves as a calibration sample for the broader MWM study, and serves as a training set for machine learning and label-transfer activities, including and in particular the estimation of the ages for large numbers of stars and stellar populations.
\end{itemize}

In addition to the main programs in MWM, there are also open-fiber programs. These are an ad hoc set of programs that fill any unused fibers that are left over after the initial allocation of core science programs. These programs are proposed internally within the collaboration. Open fiber targets for MWM were discussed in \cite{almeida2023} and in Kollmeier et al., in press.




\subsection{Black Hole Mapper}
\label{subsec:BHM}
Quasars and Active Galactic Nuclei (AGN) are among the main interests of the Black Hole Mapper (BHM) in SDSS-V. These objects are highly luminous and powered by accretion onto supermassive black holes (SMBHs). Quasars/AGN trace the growth and evolution of SMBHs spanning both cosmic distances and times; they are also excellent laboratories for exploring the physics of accretion and related processes and the short timescales that govern the vicinity near the SMBH. BHM aims to better understand the demographics, evolution, and physics of accreting SMBHs, through several core programs that span two broad astrophysically motivated approaches. The first approach is to explore the physics of SMBHs through time-resolved spectral monitoring studies, which will encompass $\sim$10$^{4.5}$ previously known quasars with multiple additional optical spectral epochs added in SDSS-V. These data will temporally probe spatial scales and physics otherwise challenging to examine. The second approach focuses on combined X-ray and optical studies that add spectroscopic characterizations to $\sim$10$^{5.5}$ actively accreting SMBHs selected primarily by their X-ray emission. This multiwavelength approach will enable SDSS-V to probe both quasar/AGN demographics and the physics of accreting SMBHs. BHM will primary rely on BOSS optical spectroscopy from both APO and LCO. The time-domain aspect of the BHM program will have a greater emphasis in the Northern sky while the spectra of eROSITA X-ray source candidate optical counterparts will be emphasized at LCO in the Southern sky.

The BHM time-resolved spectral monitoring approach features two main subprograms which include the reverberation mapping (RM), and the All-Quasar Multi-Epoch Spectroscopy (AQMES) programs. Combined, RM and AQMES jointly enable studies requiring repeat spectral sampling that span a wide range of cadence and timescales, from days to decades. 

The RM subprogram ultimately aims to target $>$1000 confirmed quasars across at least 4 dedicated fields, each with at least 100-150 spectral epochs expected in total during the duration of SDSS-V. A primary goal of the RM program is to measure the masses of SMBHs -- their most fundamental parameter. Spectroscopic lags between the continuum and broad-line region (BLR) are key to measuring the mass of a SMBH. SDSS-V builds on earlier pioneering SDSS RM MOS investigations \citep{Shen2015,Shen_etal_2024}, advancing RM measures to a significantly larger sample encompassing a broad range of quasar parameters (e.g., luminosity, redshift, etc.). The RM program in SDSS-V will cover a relatively small area of sky ($\sim$25 deg$^2$) but with intense, high-cadence, time-domain repeat optical spectroscopy in BHM. 

The AQMES program within BHM is a time-domain program which includes two components: a wide-area ($\sim$2000 deg$^2$ of sky) but low-cadence tier (AQMES-wide or AQW) and a modest-area ($\sim$200 deg$^2$) and modest-cadence tier (AQMES-medium or AQM). Both of these will also add new spectral epochs in SDSS-V to $i<19.1$ quasars/AGN with at least one previous epoch of SDSS I to IV optical spectra. These targets have been selected mainly from the DR16 quasar catalog of \citet{Lyke2020}. AQMES-wide is targeting BOSS spectra for $\sim$20,000 such quasars, typically adding 1-2 epochs during SDSS-V. When combined with the extant earlier-epoch SDSS spectra this will yield $\sim$1-10 yr timescales, which enable SDSS-V to probe the BLR dynamics of the most massive quasar SMBHs, statistics of changing look quasars, broad absorption line (BAL) disappearance and emergence, etc. AQMES-medium targets $\sim$2000 quasars, with $\sim$10 epochs of BOSS optical spectra added in SDSS-V, probing also down to $\sim$1-month to $\sim$1-year timescales, adding sensitivity to unfolding BLR structural and dynamical changes.

Complementary to the time-resolved spectral monitoring approach discussed above, there is another primary BHM program that will focus on a multiwavelength exploration of quasars. The SPectroscopic IDentfication of ERosita Sources (SPIDERS) program aims to characterize a large sample of hundreds of thousands of X-ray sources through an extensive program of SDSS-V BOSS optical spectroscopic follow-up, across $\sim$10$^4$ degrees of sky. The SPIDERS project itself has two main science subthemes: the follow-up of X-ray point-like sources, which are mainly quasars/AGN and the spectra of X-ray/optical galaxy cluster candidates. The former is aimed at the study of quasar SMBH demographics, evolution, and physics (though also includes X-ray emitting stars) while the latter will also include clustering- and cosmology-related studies. The SPIDERS program relies on the wide and deep X-ray surveys performed with the SRG/eROSITA X-ray telescope \citep[see][]{Predehl2021, Merloni2012, Merloni2024}. In particular, BHM emphasizes BOSS optical spectra in the sky hemisphere (approximately Galactic longitude $180 < l < 360$ deg) of eROSITA. 

In addition to SPIDERS, there was a joint SDSS/eROSITA endeavor that initially used early eROSITA performance validation observations from a smaller and deeper pilot field of the SRG/eROSITA X-ray telescope, known as the ‘eROSITA Final Equatorial Depth Survey’ or eFEDS field \citep{Brunner2022} with the optical counterparts identified  in \citet{Salvato2022}. The SDSS/eFEDS survey served as an early pathfinder for the SPIDERS component of the BHM program, and was the centerpiece of the spectra released in the previous SDSS-V data release, DR18 \citep{almeida2023}. The eFEDS X-ray footprint covers $\sim$140 deg$^2$ centered near RA, Dec = (135, +1)~deg, and includes the ‘GAMA09’ Field \citep{Driver2009}. SDSS-V used early plate observations at APO to collect BOSS optical spectroscopy for counterparts to point-like and extended X-ray sources \citep{Salvato2022, Klein2022}. The SDSS-V/eFEDS pilot sample reached an X-ray flux limit approximately twice as deep as that being pursued by the main SPIDERS survey in BHM, though the optical magnitude limit is comparable. The early eFEDS spectroscopic sample, included 24,000 spectra for $\sim$12,000 X-ray sources, most with secure spectroscopic redshifts and classifications.


Expanding on SDSS/eFEDS, the key component of SPIDERS going forward is its hemisphere survey, which seeks to obtain highly complete ($>85\%$) optical spectroscopic information (redshifts, spectral classifications, etc.) for a large sample ($>$250,000) of point-like X-ray selected targets over $\sim$10$^4$ deg$^2$ of high Galactic latitude sky. The SPIDERS sample will reach X-ray flux limits of a few $\times 10^{-14}$ erg/s/cm$^2$ and optical magnitudes of $i\sim 21.3$. In addition, over that same sky area, SPIDERS in BHM aims to obtain redshifts for $\sim 10^4$ clusters of galaxies, selected via their X-ray and optical properties. Most of the SPIDERS hemisphere sky is in the South, with optical BOSS spectroscopy done mainly from LCO (using FPS); hence, most of the SDSS-V data for SPIDERS will be part of future data releases. However, a smaller portion (a couple thousand deg$^2$ in area) of the SPIDERS optical spectroscopic follow-up is being carried out at APO. 

\subsection{Local Volume Mapper}
\label{subsec:LVM}
The Local Volume Mapper (LVM; \citealt{Drory2024}) is a groundbreaking project aimed at deepening our understanding of the ionized interstellar medium and stellar feedback processes, with a focus on connecting physics from sub-parsec to kiloparsec scales across the Milky Way, Magellanic Clouds, and Local Volume Galaxies. Its central scientific goal is to measure stellar feedback at the energy injection scale, in order to understand what regulates the efficiency of star formation, and map how energy and heavy elements from stars become dispersed throughout the interstellar medium. Using optical integral field unit (IFU) spectroscopy, the LVM measures the physical conditions of the ionized gas, such as gas density, temperature, and ionization, resolving the internal structure of individual ionized nebulae. This approach enables the LVM to link individual feedback sources (massive stars and supernova remnants) with large-scale structures in the Milky Way and nearby galaxies, ultimately creating a comprehensive, high-resolution spectral atlas of ionized gas.

The construction of the new LVM-Instrument (LVM-I) as part of SDSS-V included the development and commissioning of completely new telescope and spectrograph systems \citep{Konidaris2024, Herbst2024, 2024SPIE13094E..03B}.  The LVM observing footprint is divided or tiled, so a given observation taken during the LVM survey with this IFU system is called a tile and has a unique tile ID. Each tile has a $\sim$0.5$^\circ$ field of view that is observed with 1801 science fibers each with 35.3\arcsec diameter, obtaining spectra with $R\sim 4000$ across the full optical wavelength range (3600-9800\AA). The tile pointings are selected to create a mosaic over the Milky Way and Magellanic Clouds, with one row of fibers overlap between tiles for cross-calibration. This observing strategy ensures full spectroscopic coverage over the regions of interest. Reduction of all data (see section~\ref{subsec:LVMDRP}) is carried out using a specially developed Data Reduction Pipeline (Meija et al. in prep.; \citealt{Drory2024}), which results in spectrophotometrically calibrated row-stacked spectra (RSS). A data analysis pipeline is also being developed and in an advanced state \citep{SanchezDAP}, but these outputs are not included in DR19. The current DR19 release includes the reduced RSS data from a single LVM tile, as a preview of the data format that is achieved by this new facility (see Section \ref{sec:lvmpreview}).

\section{MOS targeting} \label{sec:target}

\subsection{MOS target product 
}\label{sec:mos_target_product}
Similar to DR18, we release our targeting information as the MOS target product, hosted as part of the SDSS Catalog Archive Server (CAS) and accessible via CASJobs, which is an online workbench for large scientific catalogs. The MOS product includes the results of our catalog cross-matching as well as additional tables used for target selection. The cross-matching procedure is described in detail in \citet{almeida2023} and we will provide only a brief summary here. The input catalogs, drawn mainly from the literature, but also includes some curated target lists, are ingested sequentially starting with the TIC v8 \citep{Stassun_2019_TIC_v8} table. Catalogs with higher spatial resolution and depth are generally processed first. For each new catalog, we use known target identifiers to relate the new targets to ones already ingested. Targets that cannot be matched in this manner are matched to existing targets using a spherical 2D cone search with a radius that depends on the spatial resolution of the catalog. Finally, entries in the new catalog that cannot be matched to any existing targets are considered new physical objects, added to {\tt mos\_catalog} and assigned a unique {\tt catalogid}. Table~\ref{tab:catalogs} lists the catalogs used for cross-matching. DR19 contains targeting information that is associated with cross-match versions v0.1 and v0.5. 
Among these are a number of legacy tables from previous iterations of the SDSS project (e.g., the MaNGA DAP and DRPall catalogs). For each cross-matched catalog table, {\tt mos\_X}, there is a corresponding {\tt mos\_catalog\_to\_X} table. The column {\tt catalogid} of the table {\tt mos\_catalog} matches the column {\tt catalogid} of the table {\tt mos\_catalog\_to\_X}. The column {\tt target\_id} of the  table {\tt mos\_catalog\_to\_X} matches the primary key column (i.e. the original catalog identifier) of the table {\tt mos\_X}. (See Figure~2 of \citealt{almeida2023}). Different cross-match runs will result in new {\tt catalogid} entries for the same object; this issue is addressed by the introduction of a new {\tt sdss\_id} identifier (see Section~\ref{sec:sdssid}) that relates the {\tt catalogids} from different cross-match runs for a single source.

The MOS product includes other tables that, while not cross-matched, are used for target selection. These tables are listed in Table~\ref{tab:catalogs_noxmatch} and their targets can be joined one-to-one to an existing {\tt catalogid} via the primary keys of cross-matched tables.

Targets to be observed by each mapper are algorithmically selected from the cross-matched catalogs and grouped into ``cartons'' with specific cadence and sky brightness requirements. The targets in this list of cartons are then assigned to specific robotic fiber positioners via the {\tt robostrategy} software (see Section~\ref{sec.fps}). Table~\ref{tab:tables_targetdb} lists the tables used to store target selection and fiber assignment information. For brevity, we only include the cartons which are new in DR19 and describe them in Appendix~\ref{sec:appendix}. The prior cartons are described in Sections 5, 6, 7, and 8 of \cite{almeida2023}.

Finally, a set of tables are used to store information about the telescope operations, including observed fields, cadencing, and completion status. These tables are listed in Table~\ref{tab:tables_opsdb}.





\subsubsection{MOS Targeting Generations}
\label{sec:mos_targeting_generation}

As the SDSS-V survey has progressed, we have modified, reorganized, added, and removed various target cartons.
In order to track which set of cartons (and their versions) were in use at any one time, \citet{almeida2023}
introduced the concept of a ``targeting generation". A targeting generation is simply a collection of (versioned) cartons together with meta-data telling \texttt{robostrategy} how to use each carton for survey planning. SDSS DR19 includes one targeting generation associated with SDSS-V plate operations (``v0.plates", see also section~\ref{subsec:plateprogram}), one associated with FPS commissioning activities (``v0.5.epsilon-7-core-0''),  and three associated with early FPS science operations (``v0.5.2", ``v0.5.3", and ``v0.5.5", see also section~\ref{sec.fps}). Targeting generation ``v0.5.3" was previously released as part of SDSS DR18 \citep{almeida2023}, the rest are new in DR19. Targeting generations are described by three database tables: \texttt{ mos\_targeting\_generation}, which lists all targeting generations; \texttt{mos\_targeting\_generation\_to\_carton}, which associates each targeting generation to a set of cartons; and \texttt{mos\_targeting\_generation\_to\_version}, which links targeting generations to specific \texttt{robostrategy} runs.

\subsubsection{eROSITA X-ray information used for target selection}

The SDSS DR19 MOS target product includes catalogs of candidate targets derived from the combination of eROSITA X-ray data and supporting optical/IR surveys (see table \ref{tab:catalogs_noxmatch}). These eROSITA `superset' catalogs have been used in support of target selection (SDSS-V plate and early FPS phases) by the BHM/SPIDERS and MWM/eROSITA science programs. The X-ray information in these tables is derived from very early reductions of the eFEDS \citep[][]{Brunner2022}, and the first eROSITA All Sky Survey \citep[eRASS1, Western Galactic hemisphere][]{Merloni2024}. The X-ray catalogs were cross-matched to several optical/IR catalogs via algorithms optimized to select candidate AGN \citep{Salvato2022}, clusters of galaxies \citep{Rykoff2014, IderChitham2020}, coronally emitting stars \citep{Freund_2022, Freund_2024} and X-ray emitting compact objects \citep{Schwope_2024}. We strongly advise that the eROSITA-based catalogs released as part of SDSS DR19 are used only to reconstruct the selection function of eROSITA targets within the SDSS-V survey. For other use cases, we direct the reader to i) the eROSITA early data release (EDR), DR1 public data release \citep{Brunner2022, Merloni2024}, and Salvato et al., 2025 (submitted), or to ii) section \ref{sec:sdss5-vac.0004} of this work, where we present data products derived from the combination of eROSITA and SDSS-V spectroscopic data.

\begin{deluxetable*}{lp{9.0cm}l}
\label{tab:catalogs}
\tablecaption{Cross-matched input catalogs for the catalog database.  These catalogs were also used for target selection. The catalogs prefixed with `(+)' are new in DR19. For a table {\tt mos\_X} in the below list, there is a corresponding table {\tt mos\_catalog\_to\_X} which is not listed here for brevity (see text).}
\tablehead{\colhead{Table Name\tablenotemark{1}} &
\colhead{Catalog Description} & \colhead{Primary Key\tablenotemark{1}} }
\startdata
{\tt mos\_allstar\_dr17\_synspec\_rev1} & (+) SDSS DR17 APOGEE Spectroscopy \citep{Abdurrouf_2022} & {\tt apstar\_id} \cr
{\tt mos\_allwise} & AllWISE source catalog \citep{Cutri_2013_allwise} & {\tt cntr} \cr 
{\tt mos\_bhm\_csc} & (+) Counterparts to Chandra Source Catalog 2.0 \citep{Evans_2024} & {\tt pk} \cr
{\tt mos\_bhm\_efeds\_veto } & Vetoed targets in eFEDS field  & {\tt pk} \cr
{\tt mos\_bhm\_rm\_v0\_2 } & Southern Photometric Quasar Catalog, subset \citep{Yang_2023} & {\tt pk} \cr
{\tt mos\_bhm\_rm\_v0} & (+) Southern Photometric Quasar Catalog, subset \citep{Yang_2023} & {\tt pk} \cr
{\tt mos\_catalog} & List of cross-matched targets & {\tt catalogid} \cr
{\tt mos\_catwise2020} & CatWISE2020 source catalog \citep{Marocco2021} & {\tt source\_id} \cr  
{\tt mos\_gaia\_dr2\_source} & (+) Gaia DR2  \citep{Gaia_DR2_2018} & {\tt source\_id} \cr
{\tt mos\_glimpse} & GLIMPSE I, II, 3D source catalog \citep{Churchwell_09_glimpses} & {\tt pk} \cr  
{\tt mos\_guvcat} & GALEX UV Unique Source Catalog \citep{Bianchi_2017_GALEXuvcat} & {\tt objid} \cr  
{\tt mos\_legacy\_survey\_dr8} & DESI Legacy Imaging Surveys DR8 \citep{Dey_2019_DESIsurveys} & {\tt ls\_id} \cr  
{\tt mos\_mangatarget} & (+) SDSS-IV MaNGA targets \citep{Bundy_2015} & {\tt mangaid} \cr
{\tt mos\_marvels\_dr11\_star} & (+) MARVELS DR11  \citep{Alam_2015} & {\tt starname} \cr
{\tt mos\_marvels\_dr12\_star} & (+) MARVELS DR12  \citep{Alam_2015} & {\tt pk} \cr
{\tt mos\_mastar\_goodstars} & (+) MaNGA Stellar Library (MaStar)  \citep{Abdurrouf_2022} & {\tt mangaid} \cr
{\tt mos\_panstarrs1} & Pan-STARRS1 DR2 \citep{Flewelling_2020_PS1,Magnier_2020_PS1data} & {\tt catid\_objid} \cr  
{\tt mos\_sdss\_dr13\_photoobj\_primary} & SDSS DR13 Imaging Catalog, primary objects \citep{Albareti2017_sdss_dr13} & {\tt objid} \cr  
{\tt mos\_sdss\_dr16\_specobj} & SDSS DR16 Optical Spectroscopy \citep{Ahumada2020_sdss_dr16} & {\tt specobjid} \cr  
{\tt mos\_sdss\_dr17\_specobj} & (+) SDSS DR17 Optical Spectroscopy \citep{Abdurrouf_2022} & {\tt specobjid} \cr
{\tt mos\_skies\_v1} & (+) Candidate Sky locations v1 (this work) & {\tt pix\_32768} \cr
{\tt mos\_skies\_v2} & Candidate sky locations v2 \citep[][section 5.4]{almeida2023} & {\tt pix\_32768} \cr  %
{\tt mos\_skymapper\_dr2} & SkyMapper DR2 source catalog \citep{Onken_2019_skymapperDR2} & {\tt object\_id} \cr  
{\tt mos\_supercosmos} & SuperCOSMOS Sky Survey source catalog \citep{Hambly_2001_supercosmos} & {\tt objid} \cr  
{\tt mos\_tic\_v8} & TESS Input Catalog v8 \citep{Stassun_2019_TIC_v8} & {\tt id} \cr 
{\tt mos\_twomass\_psc} & (+) Two Micron All Sky Survey Point Sources \citep{Skrutskie_2006} & {\tt pts\_key} \cr
{\tt mos\_tycho2} & Tycho-2 source catalog \citep{Hog_2000_tycho2} & {\tt designation} \cr  
{\tt mos\_unwise} & (+) unWISE source catalog \citep{Schlafly_2019} & {\tt unwise\_objid} \cr
{\tt mos\_uvotssc1} & Swift UVOT Serendipitous Source Catalog \citep{Yershov_2014_xmmswift} & {\tt id} \cr  
{\tt mos\_xmm\_om\_suss\_4\_1} & XMM OM Serendipitous UV Source Survey v4.1 \citep{Page_2012_xmmUVsource} & {\tt pk} \cr 
\enddata
\tablenotetext{1}{The Table Name and Primary Key are denoted as \texttt{mos\_X} and \texttt{id\_X}, respectively, in Figure~2 of \cite{almeida2023}.}
\end{deluxetable*}

\begin{deluxetable*}{lp{8.5cm}p{2.0cm}}
\label{tab:catalogs_noxmatch}
\tablecaption{ External catalog tables which are used for target selection but that are not involved in the cross-match process.}
\tablehead{\colhead{Table Name}&
\colhead{Catalog Description} & \colhead{Primary Key}}
\startdata
{\tt mos\_best\_brightest} & Curated list of bright stars & {\tt cntr} \cr
{\tt mos\_bhm\_csc\_v2} & Counterparts to Chandra Source Catalog 2.0 \citep{Evans_2024} & {\tt pk} \cr
{\tt mos\_bhm\_rm\_tweaks} & Results of visual inspection of RM targets & {\tt pkey} \cr
{\tt mos\_bhm\_spiders\_agn\_superset} & eROSITA/eFEDS AGN candidates \citep{almeida2023} & {\tt pk}\cr
{\tt mos\_bhm\_spiders\_clusters\_superset} & eROSITA/eFEDS cluster candidates \citep{almeida2023} & {\tt pk}\cr
{\tt mos\_cataclysmic\_variables} & Cross-match of VSX \citep{Watson2006} with Gaia DR2 & {\tt ref\_id} \cr
{\tt mos\_ebosstarget\_v5} & Targets from the SDSS/eBOSS survey \citep{Dawson_2016_eboss} & {\tt pk} \cr
{\tt mos\_erosita\_superset\_agn} & eROSITA/eRASS1 AGN candidates \citep{almeida2023} & {\tt pkey}\cr
{\tt mos\_erosita\_superset\_clusters} & eROSITA/eRASS1 cluster candidates \citep{almeida2023} & {\tt pkey}\cr
{\tt mos\_erosita\_superset\_compactobjects} & eROSITA/eRASS1 compact objects \citep{almeida2023} & {\tt pkey}\cr
{\tt mos\_erosita\_superset\_stars} & eROSITA/eRASS1 stellar candidates \citep{almeida2023} & {\tt pkey}\cr
{\tt mos\_gaia\_assas\_sn\_cepheids} & Gaia-ASAS-SN Classical Cepheid Sample \citep{Inno2021} & {\tt source\_id} \cr
{\tt mos\_gaia\_dr2\_ruwe} & Gaia DR2 Renormalised Unit Weight Error \citep{Gaia_DR2_2018} & {\tt source\_id} \cr
{\tt mos\_gaia\_dr2\_wd} & Gaia DR2 White Dwarf Candidates \citep{GentileFusillo2019} & {\tt source\_id} \cr
{\tt mos\_gaia\_unwise\_agn} & Gaia/unWISE AGN candidates from \citep{Shu2019} & {\tt source\_id} \cr
{\tt mos\_gaiadr2\_tmass\_best\_neighbour} & Gaia DR2-2MASS best neighbors \citep{Gaia_DR2_2018} & {\tt source\_id} \cr
{\tt mos\_geometric\_distances\_gaia\_dr2} & Geometric distances for Gaia DR2 \citep{Gaia_DR2_2018} & {\tt source\_id} \cr
{\tt mos\_mangadapall} & MaNGA Data Analysis Pipeline \citep{Westfall2019} & {\tt pk} \cr
{\tt mos\_mangadrpall} & MaNGA Data Reduction Pipeline \citep{Law2016} & {\tt (mangaid, plate)} \cr
{\tt mos\_mastar\_goodvisits} & MaNGA Stellar Library \citep{Yan2019} & {\tt mangaid} \cr
{\tt mos\_mipsgal} & The MIPSGAL Spitzer survey \citep{Carey2009} & {\tt mipsgal} \cr
{\tt mos\_mwm\_tess\_ob} & TESS CVZ sources \citep{Ricker2022} & {\tt gaia\_dr2\_id} \cr
{\tt mos\_sagitta} & Photometrically identified pre-main sequence stars \citep{McBride2021} & {\tt source\_id} \cr
{\tt mos\_sdss\_apogeeallstarmerge\_r13} & APOGEE All Star R13 \citep{Abdurrouf_2022} & {\tt apogee\_id} \cr
{\tt mos\_sdss\_dr16\_qso} & SDSS DR16 QSO catalog \citep{Lyke2020} & {\tt pk} \cr
{\tt mos\_sdssv\_boss\_conflist} & Early reductions of SDSS-V BOSS data, used for targeting only & {\tt pkey} \cr
{\tt mos\_sdssv\_boss\_spall} & Early reductions of SDSS-V BOSS data, used for targeting only & {\tt pkey} \cr
{\tt mos\_sdssv\_plateholes} & Catalog of SDSS-V plate holes, used for targeting only & {\tt pkey} \cr
{\tt mos\_sdssv\_plateholes\_meta} & Catalog of SDSS-V plates, used for targeting only & {\tt yanny\_uid} \cr
{\tt mos\_skymapper\_gaia} & SkyMapper DR1.1 \citep{daCosta2023} & {\tt skymapper\_ object\_id} \cr
{\tt mos\_tess\_toi} & Curated list of observed TESS targets & {\tt pk} \cr
{\tt mos\_tess\_toi\_v05} & Curated list of observed TESS targets & {\tt pkey} \cr
{\tt mos\_yso\_clustering} & Clustered young stars in Gaia DR2  \citep{Kounkel2019} & {\tt source\_id} \cr
{\tt mos\_zari18pms} & Photometrically identified pre-main sequence stars \citep{Zari2018} & {\tt source} \cr
\enddata
\end{deluxetable*}

\begin{deluxetable*}{ll}
\label{tab:tables_opsdb}
\tablecaption{Tables for telescope operations}
\tablehead{\colhead{Table Name}&
\colhead{Table Description}}
\startdata
{\tt mos\_opsdb\_apo\_camera} & BOSS red, BOSS blue, and APOGEE  \cr
{\tt mos\_opsdb\_apo\_camera\_frame} & An exposure of one camera, with Signal-to-Noise (SNR) estimate \cr
{\tt mos\_opsdb\_apo\_completion\_status} & [not started $|$ started $|$ done]  \cr
{\tt mos\_opsdb\_apo\_configuration} & A realized configuration of the FPS robots, associated with a design  \cr
{\tt mos\_opsdb\_apo\_design\_to\_status} & Design completion tracking, with MJD \cr
{\tt mos\_opsdb\_apo\_exposure} & Spectrograph exposures, associated with APOGEE or BOSS red and BOSS blue \cr
{\tt mos\_opsdb\_apo\_exposure\_flavor} & [science $|$ flat $|$ arc $|$ dark $|$ etc] \cr
\enddata
\end{deluxetable*}

\begin{deluxetable*}{lp{10cm}}
\label{tab:tables_targetdb}
\tablecaption{ Tables for targeting, catalogid, and sdss\_id. The table {\tt mos\_catalog} is listed in the below table and also in Table~\ref{tab:catalogs}.}
\tablehead{\colhead{Table Name}&
\colhead{Table Description}}
\startdata
{\tt mos\_assignment} & Specific robot to carton-to-target mapping per design   \cr
{\tt mos\_cadence} & Time between observations and on-sky requirements  \cr
{\tt mos\_cadence\_epoch} & Cadence parameters for each epoch   \cr
{\tt mos\_carton} & Carton name, associated program, and date created \cr
{\tt mos\_carton\_to\_target} & Associations between all targets and cartons they belong to \cr
{\tt mos\_catalog} & List of cross matched targets \cr
{\tt mos\_catalogdb\_version} & Version for catalog cross matches  \cr
\texttt{mos\_category} & [sky $|$ standard $|$ science $|$ open fiber $|$ etc]  \cr
{\tt mos\_design} & Association of up to 500 assigned robots with a field exposure sequence \cr
{\tt mos\_design\_mode} & Observing modes e.g. bright, dark\_rm, dark\_faint, etc  \cr
{\tt mos\_design\_mode\_check\_results} & Results of {\tt mugatu} validation (Medan et al. in prep.) \cr
{\tt mos\_design\_to\_field} & Mapping of designs to field exposure sequence \cr
{\tt mos\_field} & Location and position angle on sky  \cr
{\tt mos\_hole} & Position in FPS wok (a wok is a fixture holding all the robots)  \cr
{\tt mos\_instrument} & BOSS or APOGEE \cr
{\tt mos\_magnitude} & Optical and infrared magnitudes for a target \cr
{\tt mos\_mapper} & MWM or BHM  \cr
{\tt mos\_observatory} & LCO or APO  \cr
{\tt mos\_obsmode} & On sky observing constraints, e.g. sky brightness and airmass  \cr
{\tt mos\_positioner\_status} & Positioner status labels  \cr
{\tt mos\_revised\_magnitude} & Updated magnitude, supersedes magnitude table if exists  \cr
{\tt mos\_sdss\_id\_flat} & A pivoted version of {\tt mos\_sdss\_id\_stacked}, each row has a unique (catalogid, sdss\_id) \cr
{\tt mos\_sdss\_id\_stacked} & The combined crossmatches, unique objects are labeled with the sdss\_id  \cr
{\tt mos\_sdss\_id\_to\_catalog} & A table which allows one to find catalog information based on sdss\_id  \cr
{\tt mos\_target} & Targets selected by at least one carton \cr
{\tt mos\_targetdb\_version} & Versioning for carton and robostrategy runs \cr
{\tt mos\_targeting\_generation} & MOS Targeting Generation, see section \ref{sec:mos_targeting_generation} \cr
{\tt mos\_targeting\_generation\_to\_carton} & Links targeting generations to cartons \cr
{\tt mos\_targeting\_generation\_to\_version} & Links targeting generations to robostrategy runs \cr
\enddata
\end{deluxetable*}

\subsection{SDSS-V plate program }
\label{subsec:plateprogram}
SDSS-V science observations with plates began at APO on the night of October 23rd, 2020 and continued through June 27th, 2021. Although SDSS-V was not originally planned to involve plate observations, schedule impacts due to the onset of the COVID-19 pandemic necessitated an initial plate observation phase. This period enabled SDSS-V to use on-sky time to begin collecting data for its core science programs, and in the course of designing plates for those observations, to update and extend SDSS targeting processes to support SDSS-V's expanded footprint, target classes, and observing modes.  A very small fraction of the observing time ($\lesssim$1\%; a few half-nights) was reserved for dedicated test observations, such as collecting BOSS spectra with fibers that were intentionally offset from bright target stars, to test the feasibility of obtaining science spectra for stars that would saturate the detector if observed directly, and to measure the excess flux in adjacent fibers targeting blank sky positions to refine design limits that ensure high spectrophotometry.  Otherwise, the vast majority of the plate program's on-sky time was used to acquire science data, with plates whose design phase included all necessary preparatory work and targeting updates (e.g., identifying standards and sky fiber positions for BOSS observations in the Galactic Plane; adjusting magnitude limits for BOSS standards to support observations during bright time; defining the priority order of different target classes on each plate).   
 
As part of the SDSS-IV/V transition process, one of the BOSS spectrographs at the APO 2.5m was removed and sent to the Observatories of the Carnegie Institution of Washington (OCIW) for upgrades.  At APO, the data collected during the SDSS-V Plate Program were from the remaining BOSS spectrograph \citep{Smee_2013_bossspectrographs} and the APOGEE spectrograph \citep{Wilson_2019_apogeespectrographs}. To enable routine APOGEE and BOSS co-observing, the observing cartridges or `carts' that hold each plate and fiber assembly on the telescope need to be able to have fiber assemblies attached and positioned correctly to feed fibers into the APOGEE and BOSS spectrographs.  Six carts - 3 which had been previously used for MaNGA-APOGEE co-observing, and 3 formerly APOGEE-only carts  were converted to APOGEE-BOSS carts by attaching BOSS slitheads from formerly eBOSS-only carts - were available at the beginning of the plate program.  By the end of January, 2021, three additional MaNGA-APOGEE carts had been converted for APOGEE-BOSS observations.  The number of carts available limited the number of plates that could be plugged and available for observations over the course of a night. As a result, the number of fields observed per night, and the total integration time per plate, varied over the course of the plate program, particularly during the first several months. On occasion, BOSS- or APOGEE-only plates were also loaded into remaining eBOSS- and APOGEE-only carts to more efficiently utilize dark or bright time.  Given the time required for the upgrades to the southern BOSS spectrograph, and the significant logistical hurdles involved in transporting plates to LCO during the first year of COVID-19 supply chain disruptions, SDSS-V made the strategic decision to {\it NOT} observe with APOGEE-S after SDSS-IV concluded observations in Jan. 2021.

Targeting procedures evolved over the course of the SDSS-V plate program, as the survey transitioned from a workflow involving regular ($\sim$ monthly) `plateruns' toward the algorithmic \& database-centric workflow that the MOS target product uses to support FPS observations.  In the platerun model, scripts, external catalog queries, and curated target lists were used to design and drill a discrete set of plates, each of which was optimized for observation at a specific hour angle to account for changes in the plate scale due to atmospheric refraction (an effect which is corrected for on a by-positioner basis in the FPS era). By necessity, the SDSS-V plate program maintained the platerun model, but moved to ingest targets, flux and telluric standards, and sky fiber positions from the cartons defined in the ``v0.plates'' targeting generation. Information about the cartons that were implemented for (or over the course of) the plate program are provided in Appendix~\ref{sec:appendix}. 

Typical SDSS-V plates were drilled for a total of 800 fibers: 500 for the BOSS spectrograph and 300 for the APOGEE spectrograph.  Similar to the {\tt design\_mode} and {\tt obsmode} parameters that govern the properties and observability of FPS designs, the plate program's observing plan was designed around a discrete set of plate types that emphasized specific types of targets, locations on the sky, and constraints on observing conditions. Each plate was designed either for bright time or dark time observations, and the number of plates of each type that were designed aimed to ensure that the plate program would efficiently utilize the range of LSTs and sky brightnesses available throughout the SDSS-V plate program.  The 8 plate types that were regularly designed were the following: 
\begin{itemize}
    \item{\textbf{AQM and AQW:} Plates optimized for BOSS observations of extragalactic targets in the BHM AQMES-medium and AQMES-wide footprints, aiming for multiple $\sim$1 hour observations per month and observation season. Toward the end of the plate program, AQMES-bonus fields that were not originally scoped for the FPS program were designed to make use of available LSTs. }
    \item{\textbf{eFEDS:} Plates optimized to complete the SDSS-IV/V \textit{eROSITA} Final Equatorial Depth Survey by obtaining single-epoch $\sim$2 hour BOSS spectra during dark time over a wide area (140 deg$^2$) \textit{eROSITA} performance verification field near RA, Dec=9H, +2 deg. Dominated by extra-galactic X-ray emitting sources (AGN, clusters of galaxies), but with a significant number of Galactic stars, and compact objects \citep[see for details][]{Salvato2022}.  The eFEDS plate design process is described in more detail in Sec. 7.3 of \citet{almeida2023}, with analysis of the completeness and purity of the resulting redshift catalog presented along with initial science results in \citet{Aydar2025}. }
    \item{\textbf{GG:} Plates targeting MWM Galactic Genesis fields, optimized for short ($\sim$30m) bright-time APOGEE observations of red giants in the Galactic Plane.}
    \item{\textbf{RM:} Plates targeting 3 BHM RM fields (i.e., SDSS-RM, COSMOS and XMM-LSS), optimized for multi-epoch (several observations per lunation, over the full observing season) deep (1-2 hour exposures, scheduled in dark time) BOSS observations of extragalactic targets.  Multiple plates were designed for each field to enable observations over a variety of hour angles}
    \item{\textbf{RV:} Plates optimized for multiple APOGEE observations of MWM stellar targets, aiming for multiple $\sim$1 hour observations per lunation and observation season.}
    \item{\textbf{TESS:} Plates targeting few-epoch bright time APOGEE observations of stars in the TESS Northern Contiguous Viewing Zone}
    \item{\textbf{YSO:} Plates targeting few-epoch bright time APOGEE and BOSS spectra of young stellar objects in regions of active or recent Galactic star formation. }
\end{itemize}

SDSS-V plates were designed and drilled using scripts and workflows shared by the SDSS-IV team,  which similarly built off workflows developed and refined during each of the prior iterations of SDSS.  On each of the plate types listed above, the cartons from which targets were drawn and their priority order were customized to suit the plate's primary science goals. Other plate design parameters, such as the number of fibers reserved for flux/telluric standards and sky fibers, were similarly varied to suit each plate's intended science goals and on-sky target density.  Several other lower level plate design parameters (e.g., the assumed ambient plate temperature; the effective wavelength each fiber's position was optimized for; the epoch assumed in correcting each target's fiber position for proper motion) and algorithms used in assigning fibers to sources (e.g., sorting APOGEE fibers to minimize brightness differences between adjacent fibers, and their associated crosstalk) were also adjusted over the course of the plate program to improve the efficiency and data quality obtained from the plate program. The design scripts used in each plate run, as well as input files such as the list of target types that governed priorities for fiber assignments, are archived in a  `five\_plates' github repo \url{https://github.com/sdss/five_plates}, so that the selection effects associated with each plate's targeting process can be documented and reproduced. 

In total, 265 plates were observed between Oct. 23, 2020 and June 27, 2021. Each plate was typically observed on 3-4 distinct nights, for a total of 965 unique plate-epochs. The resulting Plate Program data includes $\sim$246k APOGEE and $\sim$199k BOSS spectra of $\sim$56k and $\sim$89k unique targets, respectively. 

\subsection{FPS program}  \label{sec.fps}

The FPS program began in February 2022 at APO and August 2022 at LCO. These programs are planned using the {\tt robostrategy} software (Blanton et al. in prep.; Medan et al. in prep.), which plans the set of observations to perform at each location on sky throughout the survey, and selects which targets should be assigned fibers. The {\tt roboscheduler} software (\citealt{donor24a}) schedules individual observations in real time each night. We report the results of the planning process and the resulting observations in the {\tt robostrategy} file outputs, which are placed in the database.

We define a set of 12,358 fields across the sky. For each field, we have a planned cadence, which includes a number of epochs, a number of individual designs per epoch, and a preferred timing of the epochs. In good conditions, we expect each design to result in a single observation, which consists of one BOSS exposure and two APOGEE exposures, with BOSS and APOGEE taking concurrently. Within each design there is an assignment of either the BOSS or APOGEE fiber in each robot positioner to a specific target.

Depending on its desired observing conditions, each design has a ``design mode'' assigned defining the set of calibration target requirements and the required observing conditions, most notably the sky brightness requirements, which are primarily driven by the lunation. The planned observations are stored in the public database in a manner that parallels the full database. The relevant tables are listed in Table~\ref{tab:tables_targetdb}. The {\tt mos\_field} table describes the fields and the planned cadences, and each field is associated with one or more entries in the {\tt mos\_design} table, through the {\tt mos\_design\_to\_field} table. Since the FPS program has gone through several iterations of planning, each field may have had several cadences, and there is one entry in {\tt mos\_field} for each {\tt robostrategy} plan. For each design, there are a set of entries in the {\tt mos\_assignment} table that describe which {\tt carton\_to\_target} entries were used to assign each fiber. Ancillary information about the design mode and cadence descriptions are also stored in the database. A more complete description of the FPS targeting database, including a visual schema, is provided in S\'anchez-Gallego et al. (in prep.). 

During the night, the {\tt roboscheduler} software decides which designs to observe. The {\tt jaeger} software performs the fiber assignment and reports the final configuration of the robots for each observation. When the observations are performed, they are assessed by on-mountain quick-look reductions to decide if the design is completed. Each morning, the data are archived at the Center for High Performance Computing (CHPC) at the University of Utah and final processing is performed.

The results of the observations are stored in the public database, again in a manner that parallels the internal database. The {\tt mos\_design\_to\_status} table reports the status of each design. The {\tt mos\_configuration} table provides meta-data associated with each configuration of the fiber positioners (usually corresponding to a single observation). The {\tt mos\_exposure} table provides meta-data associated with each exposure.

The observations may not fully reflect the planned observations in several ways. The preferred epoch timing is not always achieved. Sometimes, more than one observation per design is needed to achieve the required signal-to-noise ratio (SNR). In some cases, not all designs of an epoch are observed, because the overall SNR for the primary targets in the epoch has already been achieved. Finally, in some cases the robot positioners or their fibers are damaged and the objects assigned to them are not observed, so that the configurations do not agree with the planned assignments.

The data in the DR19 reductions may also not perfectly reflect all of the exposures and configurations reported in the database. There may be designs observed in exposure whose reductions were excluded due to reduction failures or other data quality issues.

\subsection{Semaphore: SDSS-V Targeting Flags}\label{sec:semaphore}
\input{sections/MOS_Targeting/semaphore}

\subsection{sdss\_id }\label{sec:sdssid}
\input{sections/MOS_Targeting/sdss_id}

\section{Pipelines} \label{sec:pipeline}
\input{sections/pipelines/Overview}

\subsection{BOSS Data Reduction Pipeline}
\input{sections/pipelines/BOSS_DRP}

\subsection{APOGEE Data Reduction Pipeline }
\label{apogee_drp}

There have been many upgrades to the APOGEE data reduction pipeline (DRP) for SDSS-V.  In the last several years, nearly the entire pipeline has been translated from IDL to Python and ``refactored'' (improving the internal structure without changing the external behavior).  A substantial amount of documentation was also added including a ReadTheDocs page\footnote{\url{https://apogee-drp.readthedocs.io/}}.

In the past, arclamp exposure and wavelength solutions were created roughly every two weeks.  In SDSS-V, the reduction pipeline was upgraded to process every arclamp taken on an average wavelength solution (averaging over arclamp exposures taken from $\pm$ 3 days) on a daily basis.  This required a substantial upgrade to the wavelength software to make it very robust and a culling of the arclamp linelists. In addition, the wavelength solutions were upgraded to make use of the new Fabry-Perot Interferometer (FPI) information.  Full-frame (all 300 fibers) FPI exposures are taken during the afternoon and morning calibration sequences, while every science exposure has two fibers dedicated to FPI light which can be used to track small shifts throughout the night.  The average daily wavelength solution is used to full-frame FPI exposures averaged over the 300 fibers to obtain very precise absolute wavelengths for each unique FPI line (there is no absolute wavelength for the FPI lines and they shift slightly over time).  These calibrated FPI wavelengths are then used to refine the wavelength solution for each fiber by using the full-frame FPI information.

New quality assurance checks were written to evaluate the quality of every exposure taken.  Only data passing these checks were processed through the rest of the pipeline. This helps with the complete automation of the entire data reduction process. A new {\tt apogee\_drp} database was created on the Utah SDSS-V servers.  This tracked every aspect of the DRP processing including the status of the processing steps and the results.  This makes it easier for SDSS-V operations staff and collaboration members with Utah accounts to access reduction results directly.

During the plate era, every plate visit required a separate dome flat exposure.  This was used to: (1) trace out the positions of the spectra on the detector and create an empirical PSF, and (2) measure the fiber-to-fiber throughput (mostly used for sky subtraction).  During the FPS era, the gang connector is left connected in the FPS throughout the entire night and dome flats are only taken at the beginning and end of the night (mainly for fiber-to-fiber throughput measurements). As a result, another procedure was needed to generate the empirical PSF for each exposure (as the traces move slightly throughout the night, especially at APO).  Therefore, a ``model PSF'' was created which can be used to generate an empirical PSF for any exposure by measuring the trace positions on the data itself. The quicklook softare on the mountain was rewritten from IDL to Python and was vastly simplified and sped up. A detailed description of the DRP software upgrades will be given in the APOGEE data reduction software paper (Nidever et al., in prep.).

\subsection{Milky Way Mapper Data Analysis Pipeline: \textsc{Astra} Stellar Parameters }
\label{astra}

\textsc{Astra} is a software package that manages and orchestrates the analysis of Milky Way Mapper spectra (Casey et al., in prep.). \textsc{Astra} ingests the reduced data products from the APOGEE and BOSS data reduction pipelines, sends the data through data analysis pipelines, and produces science-ready data products that store stellar parameters, chemical abundances, line measurements, and other quantities. Because MWM has both BOSS and APOGEE spectra over a wide range of \teff, \textsc{Astra} is used to bring together analysis methods into a single software framework.

In Table~\ref{tab:astra_pipe}, we list the pipelines included in \textsc{Astra} and the spectra that they processed. Note that each pipeline was refactored at least slightly to work within the \textsc{Astra} framework.

\begin{deluxetable*}{llll}
\label{tab:astra_pipe}
\tablecaption{List of Data Analysis Pipelines included in the \textsc{Astra} Framework}
\tablehead{\colhead{Pipeline} & \colhead{Reference} & \colhead{Relevant Spectra} & \colhead{Recommended Use}}
\startdata
APOGEENET & \citet{BOSSNET2024AJ....167..173S}  & all APOGEE spectra &  YSOs and stars \\
&  & & with \teff $> 20,000$K\\
\hline
ASPCAP & \citet{ASPCAP.2016AJ....151..144G} & all APOGEE spectra & Stellar parameters for stars with \teff$<20,000$K;\\
& & & elemental abundances \teff$< 7000 $K \\
\hline
The Payne & \citet{ThePayne2019ApJ...879...69T} & all APOGEE spectra & \\
\hline
AstroNN & \citet{AstroNN2019MNRAS.483.3255L} & all APOGEE spectra & \\
\hline
BOSSNet & \citet{BOSSNET2024AJ....167..173S} & BOSS spectra &  Stellar parameters for OBAFGK stars\\
& & from MWM cartons &   \\
\hline
MdwarfType & Casey et al., (in prep.) &  BOSS spectra & Spectral Type from K5.0 to M9.5 \\
& & from MWM cartons & Metallicity class from usdM to dM\\
\hline
SLAM & Qiu et al., (in prep.) & BOSS spectra of & Stellar parameters for M dwarfs \\
 & & M dwarfs & with [Fe/H] $> -0.6$ \\
\hline
SnowWhite & Casey et al., (in prep.) & BOSS Spectra of  & Classification; Stellar parameters for DA\\
 & &  White Dwarfs & \\ 
\hline 
corv & \citet{corv} & BOSS Spectra of & Radial Velocities for DA WDs \\
 & &  White Dwarfs & \\ 
\hline
LineForest & Saad et al. (2024)  & BOSS spectra & Line Indices for H, Li,  \\
& & from MWM cartons & Ca, and other lines particularly for YSOs \\
\hline
\enddata
\end{deluxetable*}

\subsubsection{APOGEE Spectra in DR19}

In DR19, \textsc{Astra} has been used to analyze both SDSS-V APOGEE spectra and legacy SDSS-IV APOGEE spectra from DR17. While they were processed through the same pipelines in \textsc{Astra}, the DR17 spectra were not reprocessed through the DR19 data reduction pipeline. If a star was observed both in SDSS-IV and SDSS-V, the spectra were not co-added across surveys and only the parameters from the SDSS-V data are included in the data release. 

APOGEE spectra in DR19 are run through several pipelines in the \textsc{Astra} framework. We discuss the APOGEE Stellar Parameters and Abundances Pipeline \citep[hereafter ASPCAP,][]{ASPCAP.2016AJ....151..144G} as implemented for DR19 in the most detail below, given its prominence in previous data releases. The other pipelines are summarized first.
\begin{itemize}
\item APOGEENet: The original APOGEENet \citep{Olney2020AJ....159..182O} convolutional neural network was updated by \citet{Sprague2022AJ....163..152S} with an expanded training set, adding hot stars to the sample of cooler dwarfs, red giants, and YSOs. \citet{BOSSNET2024AJ....167..173S} improved labels for OBA stars, improved surface gravities for YSOs, and expanded labels to cover late M-early L brown dwarfs. APOGEENet is the only APOGEE pipeline that incorporates YSOs into its training or synthetic spectra. As YSOs have lower gravities, higher rotation velocities, and higher probability of emission lines and veiling, pipelines such as ASPCAP that are designed for main-sequence and evolved stars can give inaccurate parameters \citep[e.g.,][]{INSYNC2014ApJ...794..125C}.
\item AstroNN: AstroNN \citep{AstroNN2019MNRAS.483.3255L} uses a deep-learning artificial neural network to determine stellar parameters (\teff, \logg, and metallicity) and 18 individual element abundances with associated uncertainty from a rest-frame resampled APOGEE spectrum. For DR19, AstroNN has been refactored to use PyTorch rather than Tensorflow and has been retrained on DR17 ASPCAP results \citep{Abdurrouf_2022}, updating the labels from APOGEE DR14 \citep{Holtzman2018AJ....156..125H}. A more complete description can be found in Casey et al. (in prep.). AstroNN is complemented by AstroNNDist.

\item ThePayne : ThePayne, \citep{ThePayne2019ApJ...879...69T} trained a neural net on synthetic spectra in the APOGEE wavelength region. We use the identical model to predict stellar parameters and abundances for 25 elements between 3000 K $<$ \teff $<$ 8000 K, $-1.5 <$ [Fe/H] $< +0.5 $ and $-1.0 <$ [X/Fe] $< +1.2$ from MWM APOGEE spectra. 

\item ASPCAP:  \textsc{Astra} includes a new version of ASPCAP \citep[][; Casey et al., in prep.]{ASPCAP.2016AJ....151..144G}. 

\end{itemize}

\subsubsection{Changes in ASPCAP from DR17}
ASPCAP has been used to analyze APOGEE data from the the start of the APOGEE-1 survey and is the source of stellar parameters and abundances for many papers. Therefore, we summarize here the main changes to ASPCAP since DR17 to indicate why the DR19 results are not identical to the DR17 results. A more complete description is in Casey et al., (in prep.). 
\begin{enumerate}

\item The logic for the initial guess of the stellar parameters has changed. In SDSS-V DR 19 we use the APOGEENet (Version~2) stellar parameters as the initial guess for FERRE in the coarse stage, instead of the Doppler code Cannon-based initial guess \citep{Doppler2021zndo...4906681N}. 

\item The continuum normalization treatment for the elemental abundances stage is different. In DR17, the continuum was fit during the stellar parameter stage and then re-fit for each element abundance determination. For some elements there are only a few pixels that are not masked, which means FERRE is simultaneously fitting a flexible continuum and an element abundance from just a few pixels. Therefore, in DR19 we fit the continuum during the stellar parameter stage and keep that continuum fixed when determining the elemental abundances.

\item  Several flags have been added and others have been discontinued. In particular,  In DR17 there was a `STAR\_BAD' flag constructed from the `ASPCAPFLAG' values. In DR19 information about the pipeline is stored in `result\_flags' (which is the closest analogue of `ASPCAPFLAG'), and `flag\_bad' is constructed from the `result\_flags' values (where `flag\_bad' is the closest analogue of `STAR\_BAD' in DR17). A complete list of ASPCAP flags is presented in Casey et al. in. prep. 

\item As with each Data Release, changes were made in the ASPCAP post-processing steps that adjust the raw ASPCAP results to match better with more direct measurements of \teff\ and \logg\ and with literature values of metallicity. These are described in Casey et al., (in prep.) and the comparison between DR19 raw values and these calibrators are described in detail in  \citet{Meszaros25_astra_aspcap}.

\item We experienced unresolved issues with FERRE timeouts, which meant a random subset of good spectra will sometimes have no results. 
We developed tools to identify times that FERRE stalls, kill the process, and re-start FERRE with the remaining spectra (randomly sorted). If the set of spectra to be analyzed by FERRE is small enough then this solution was sufficient. However, often we are running FERRE with very large sets of spectra. When this behavior happened repeatedly (e.g., more than 10 times for one FERRE execution), the remaining spectra in that set were considered problematic and will not have results. This could mean that some spectra would be labelled problematic when they weren’t (i.e., they were just part of a set of spectra that were problematic).

\end{enumerate}

\subsubsection{BOSS Spectra in DR19} 
In addition to  \textsc{Astra} being run on APOGEE spectra, it was also run on stellar spectra observed with BOSS. Since BHM obtained many BOSS spectra for extragalactic objects, \textsc{Astra} was not run on all BOSS spectra. Instead, it was only run on BOSS spectra from cartons that targeted stars. These included YSOs, halo stars, and nearby M dwarfs and white dwarfs, as well as stellar open-fiber programs. However, occasionally a star was observed from a carton that  \textsc{Astra} did not process. For example, visual inspection revealed some likely cataclysmic variable (CV) stars from the ``bhm\_spiders\_agn\_lsdr10" carton. 

 \textsc{Astra} runs several pipelines on MWM BOSS data, which include : 
\begin{itemize}

\item BOSSNet: Described in \citet{BOSSNET2024AJ....167..173S}, this pipeline is similar to APOGEENet, with a BOSS training set.

\item MdwarfType: The MDwarfType pipeline (Casey et al., in prep.; Galligan \& Le\'pine, in prep.) compares spectra against a set of semi-empirical M-dwarf template spectra and estimates the spectral subtype from K5.0 to M8.5, evaluated to a half-subtype, and a “metallicity class” with an integer value from 0.5 to 12.5, from most metal-rich to most metal-poor.

\item SLAM: Described in Qiu et al. (in prep.), this pipeline trained the Stellar LAbel Machine (SLAM) on APOGEENet, ASPCAP and LAMOST parameters for M dwarfs in wide binaries with F, G, K primaries. The SLAM provides \teff, \logg, [Fe/H], and [$\alpha$/Fe] for M dwarfs with BOSS spectra for stars with [Fe/H] $> -0.6$

\item The {\tt corv} pipeline: \citep{corv} is optimized for measuring RVs of Hydrogen rich (DA) white dwarfs by linearly interpolating between the Montreal models of DA–type white dwarfs to forward model white dwarf spectra. 

\item SnowWhite: SnowWhite (Casey et al. in prep.) is a pipeline specifically designed for analyzing BOSS spectra of white dwarfs. It consists of both a classifier that reports the type of white dwarf based on line indices and, for DA white dwarfs, a parameter estimator. Stellar parameters are based on model spectrum fits around hydrogen lines, using Gaia astrometry and photometry to differentiate between potentially multimodal solutions.

\item LineForest: LineForest \citep{LineForest2024AJ....167..125S} is a pipeline that measures the integrated line strengths (equivalent widths) of numerous key transitions in BOSS spectra using a neural network. A neural network approach was adopted because other methods attempted, such as Gaussian profile fitting, automated continuum determination, and automated width estimation, could not be generalized across all sources and only worked well for a subset of stars.

\end{itemize}

\subsubsection{astraMWMLite: Single Summary File of Preferred Parameters} \label{subsec:astraMWMLite}

Each of the pipelines listed in Table~\ref{tab:astra_pipe} produces a summary file of relevant stellar parameters and, in some cases, abundances. Each spectrum may be processed by up to four pipelines. In the case of APOGEE, these are ASPCAP, APOGEENet, AstroNN, and ThePayne. In the case of BOSS, these are BOSSNet and SLAM (for M dwarfs) or SnowWhite (for white dwarfs). There are also at least 10$^4$ stars observed with both APOGEE and BOSS. To provide a summary file that has one entry per star with our preferred parameters, \textsc{Astra} creates the astraMWMLite file. This was a widely requested pipeline and data product to provide an initial “stepping stone” for users who do not want to make decisions about the quality of the results from individual pipelines. While astraMWMLite is our attempt to create a summary file that contains the single preferred parameters for an astrophysical source given its observed spectra, there is no guarantee about scale consistency in stellar parameters from different pipelines. This file is ideal for searching through the whole MWM dataset to find particular objects of scientific interest (e.g., wide binaries, the population of solar neighborhood, or young stars in all their manifestations). 

\begin{figure*}[h]
\centering
\includegraphics[width=0.8\textwidth]{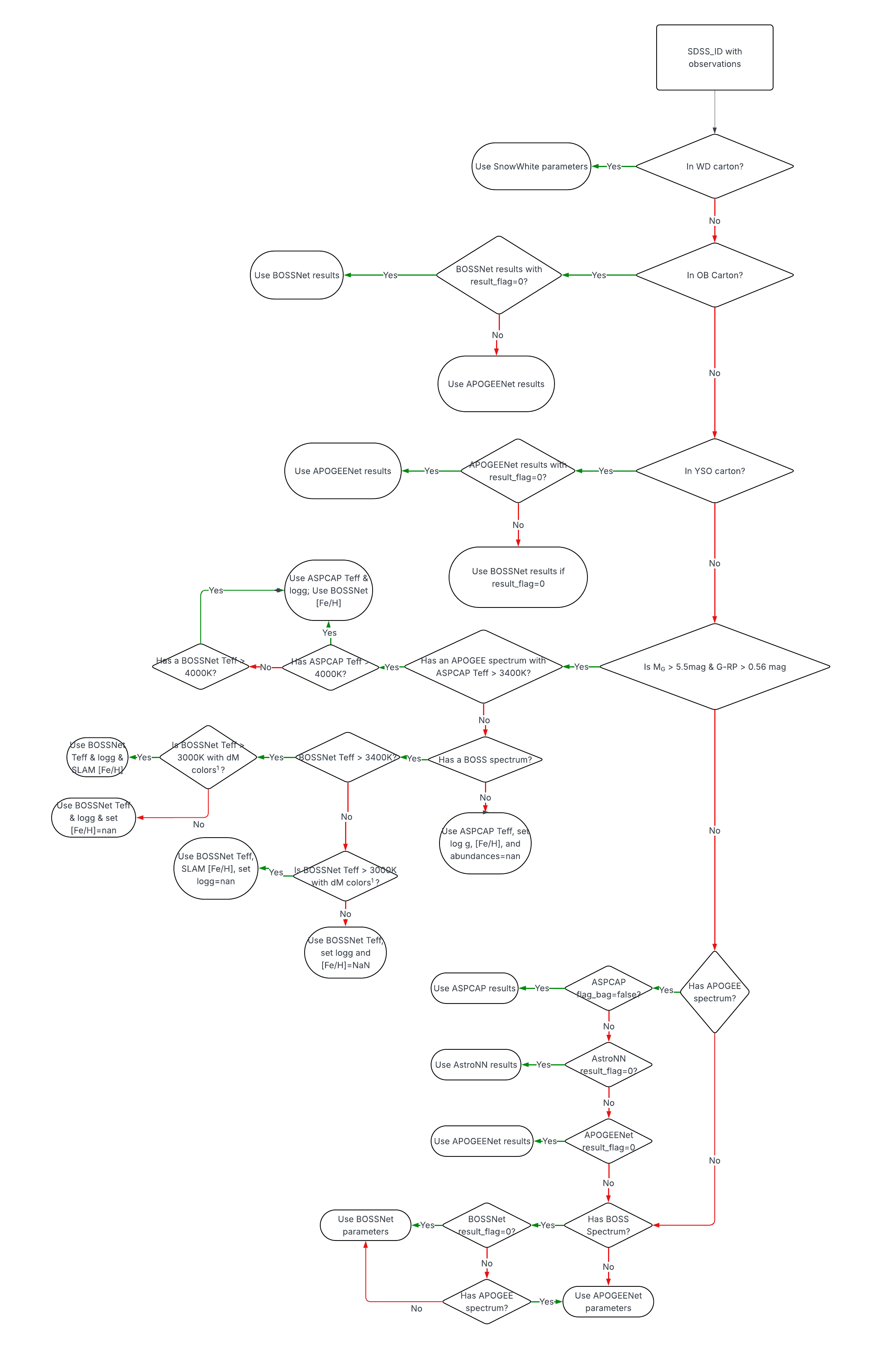}
\caption{This flowchart contains the decision tree used to determine which pipeline and reported parameters \textsc{Astra} are preferred for an individual unique source (i.e., with a unique SDSS\_ID).} 
\label{fig:flow_lite}
\end{figure*}

Figure~\ref{fig:flow_lite} gives the overview of how \textsc{Astra} constructs the astraMWMLite file. For a full description and justification, please see Casey et al., (in prep.).  

\subsection{LVM Data Reduction Pipeline }
\label{subsec:LVMDRP}
This release will introduce LVM data for the first time. In this section, we briefly describe the pipeline to reduce such data. A comprehensive explanation of the LVM DRP will be provided in Mej\'ia-Narv\'aez et al. (in prep.), outlining how it takes a raw exposure taken with a system of 4 telescopes (1 science, 2 sky, and 1 spectrophotometric) to produce flux calibrated and sky subtracted row-stacked spectra (RSS) files.

In a broad sense, a science exposure reduction is carried out in the following sequence of steps:

\begin{itemize}
    \item Preprocessing and detrending
    \item Astrometric solution
    \item Extraction of 1D spectra
    \item Wavelength calibration
    \item Fiber flat-fielding
    \item Flux calibration
    \item Sky subtraction
\end{itemize}

Before any reduction step is performed, the pipeline requires a set of calibration frames, namely:

\begin{itemize}
    \item Bias, pixel mask and pixel flat field
    \item Guider camera frames
    \item Fiber model and fiber centroids and widths (from dome flats)
    \item Wavelength and line spread function (LSF) models (from arcs)
    \item Fiber flat field (from twilight sky flats)
\end{itemize}

The cadence at which calibrations are made is driven by the stability of the instrument. Typically, we take a full set of calibrations every few (1--2) months. This cadence is only disrupted by external events bound to introduce changes in the instrument (e.g., seisms, instrument interventions). We will defer the details about the production of these calibration frames to an upcoming paper. 

The light entering all four telescopes is fed through fibers\footnote{We note that there are  $1801$ science fibers, $119$ sky fibers, and $24$ spectrophotometric fibers.} into three spectrographs labeled $1,2,3$, each with three arms or spectral channels labeled $b,r,z$.  As a result, an LVM raw (unreduced) exposure consists of 9 camera frames, each stored in {\tt FITS} format. Each camera frame has a size of $4080\times4120$ pixels and is divided into 4 quadrants (one for each amplifier), which in turn are subdivided into data, pre-scan and overscan regions. Regarding the fiber arrangement, the image fibers for each camera are grouped in $18$ blocks of $36$ fibers each ($648$ per camera frame).

Assuming all calibrations are in place, the preprocessing starts by subtracting and trimming the overscan regions in each camera frame. The subtraction goes row-per-row, capturing as much as possible any relevant structure. The bias frame is then subtracted from the remaining frame to fully remove any bias structure left behind after the overscan subtraction. The resulting frame is then flat fielded using the pixel flat frame. In order to flag any CCD quasi-static artifacts that cannot be flat fielded (e.g., hot pixels, bad columns, etc.) the pixel mask is added as an extension. At this point, each camera frame is converted to electrons using the gain values in each quadrant.

Part of the detrending consists of the detection and flagging of cosmic rays (CR), based on the Laplacian CR rejection algorithm by \cite{vandokkum01}.\footnote{The typical LVM science exposure is 900s long, so we are bound to detect cosmic rays with our CCDs.} Finally, the detrending step ends with the modeling and subtraction of the stray-light in each camera frame.

Before running the spectral extraction the pipeline accounts for likely sub-pixel thermal shifts of the 2D spectra. This is done by cross-matching a selection of columns against the fiber model. The typical thermal shift per camera frame within a night is $\sim0.05$pix. The extraction implements a fiber fitting algorithm assuming a Gaussian profiles of the 2D spectra. The extracted $9$ camera frames are grouped spectrograph-wise so that we end up with $3$ frames per exposure (one for each spectrograph channel).

The wavelength model also requires accounting for potential thermal shifts. This is done by measuring the (Gaussian) centroids of a selection of strong and relatively isolated sky lines detected in each channel, namely: $b$: $5577$\AA, $r$: $6363$\AA, $7358$\AA, $7392$\AA, and $z$: $8399$\AA, $8988$\AA, $9552$\AA, $9719$\AA. The measured centroids are then compared to the known wavelengths to calculate the correction along the fiber array for each channel. The typical thermal wavelength correction is $\sim0.01$\AA\ within a single night. The LSF is added at this stage as is, with no further corrections. Heliocentric velocity corrections are calculated and added to the header metadata but not applied to the spectra.

At this point in the pipeline each channel frame is fiber flat-fielded and wavelength-rectified, so that all 1D spectra follows the same wavelength dependency (to within $\sim1\,$per cent) and are sampled in a common wavelength grid of $0.5$\AA.

To determine the spectral sensitivity of the instrument, we use the extracted and flat-fielded 1D spectra of 12 individual F-type standard stars taken by the spectrophotometric telescope over the course of one $900$s science exposure ($\sim45$s each after overheads). Our default method of calibration during survey operations is to use these stars, all pre-selected to have Gaia XP spectra available. Given our wide science IFU field of view ($0.5$deg), many fields that we observe will also have Gaia sources. During early science, if an insufficient number of high quality stars were obtained with the spectrophotometric telescope, we have also implemented a technique to use stars within the science field to provide flux calibration. Regardless of the method, the flux calibration for each $b,r,z$ spectral channel is performed independently. The sensitivity curve is calculated as the ratio of the spectrophotometric 1D spectra in instrumental units of electron densities to the Gaia XP spectra. During this step we account for the sky extinction using a fiducial extinction curve. We are currently implementing an approach that carries out stellar template fitting, and achieves higher spectral resolution response functions that are also able to accurately correct for telluric absorption features. At this point, the 3 channel frames are stitched together wavelength-wise, accounting for the errors in the overlapping regions.

The contribution of sky emission to the science spectrum is monitored simultaneous to the science observation via the two sky telescopes. One telescope points at a relatively dark region of the sky that is within 10$^{\circ}$ of the science field, selected from a grid of $768$ locations semi-uniformly sampling the full sky, intended to provide good constraints on the sky continuum level. The second telescope points to a part of the sky that is very dark, also in Milky Way foreground emission, to provide good constraints on the time-variable night sky line emission. The spectra from these two telescopes are analyzed and combined, in order to recover a model of the sky spectrum at the location of the science field (Jones et al. in prep.). The ultimate product of this full DRP is a sky subtracted, flux calibrated RSS with a filename format `lvmSFrame\_$\{expnum\}$.fits', where $\{expnum\}$ is the exposure number for a given observation.

\section{The Scope of SDSS DR19} 
\label{sec:scope}

\begin{figure*}
\centering
\includegraphics[width=1.0\textwidth]{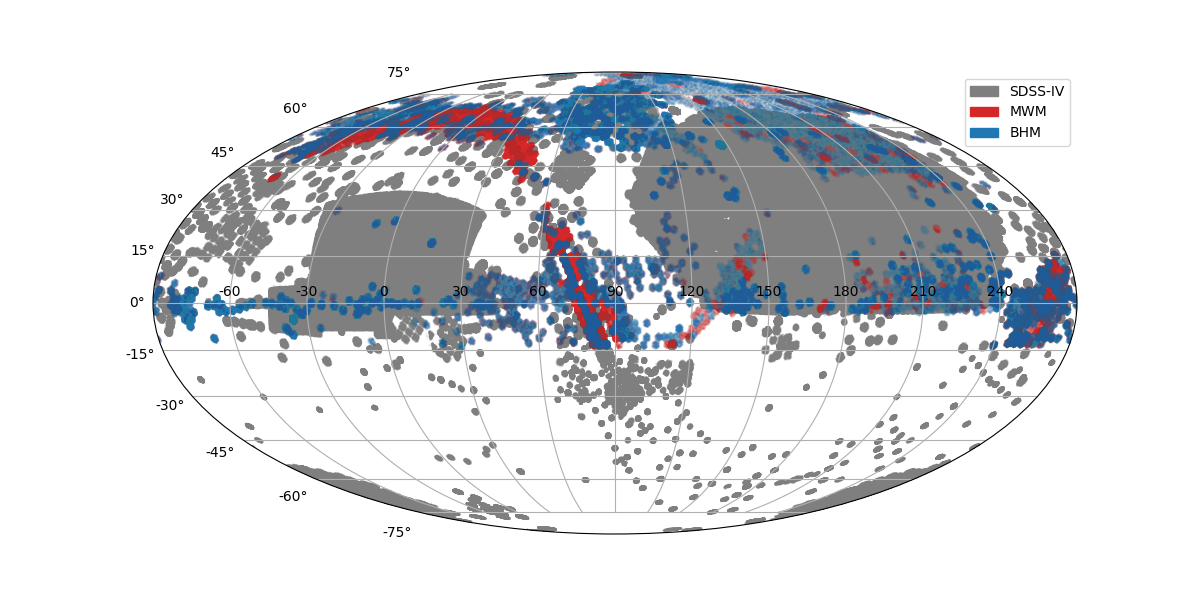}
\caption{The right ascension (RA) and declination (DEC) for targets observed in DR19 MWM (red), DR19 BHM (blue) and legacy observations from SDSS-IV (gray). SDSS-V DR19 contains new spectra from the APO telescope in the Northern hemisphere. }
\label{fig:allsky}
\end{figure*}

With the targeting and data reduction and analysis pipelines in place, in this section we describe the scope of the newest data release from SDSS-V. For context, DR18 \citep{almeida2023}, announced in 2023, was the first public data release of SDSS-V, and as such it laid down the groundwork for future data releases. DR18 introduced the new concept of `cartons': groups of targets that are produced by specific target selection algorithms, to address specific science goals. A complete overview of all the cartons in SDSS-V is available on our website\footnote{\url{https://www.sdss.org/dr19/targeting/flags/}}. Whereas DR18 consisted mostly of targeting data, DR19 now introduces the first set of spectra and data products produced by MWM, and significantly expands the available BHM data set. Targeting data for both MWM and BHM has also been updated (see section~\ref{sec:target}). Finally, we include in DR19 an integral-field data set of the Helix Nebula (NGC~7293), produced by the LVM. In Figure~\ref{fig:allsky}, we plot the sky positions (in RA and DEC) for MWM (red) and BHM (black) observations in DR19. For reference, we also add the sky positions for legacy SDSS data from SDSS-IV. 

All data products in DR19 are available in fits file format from the Science Archive Server (SAS), and many data products are also available in table format on the Catalog Archive Server (CAS). More details on these server systems, and how to use them to access data products, is given in Section~\ref{sec:data.access}. New in DR19 is the Zora/Valis Web Framework (section~\ref{sec:data.dataviz}), which allows users to search for targets and observations, display their spectra, and explore their properties. Also new for DR19 is a set of Python tutorial notebooks, that introduce users to the various data sets in this data release, and show how to access and use the data products. These notebooks are available from the SDSS github repository\footnote{\url{https://github.com/sdss/dr19_tutorials}}, and in SciServer (see section~\ref{sec:data.sciserver}). The tutorials are described in more detail in section~\ref{sec:demos}.

All previous SDSS data releases remain available through the Science Archive Server (see section \ref{sec:data.sas}).

\subsection{Milky Way Mapper data products}
\label{sec:MWMscope}
\begin{figure}
\centering
\includegraphics[width=0.5\textwidth]{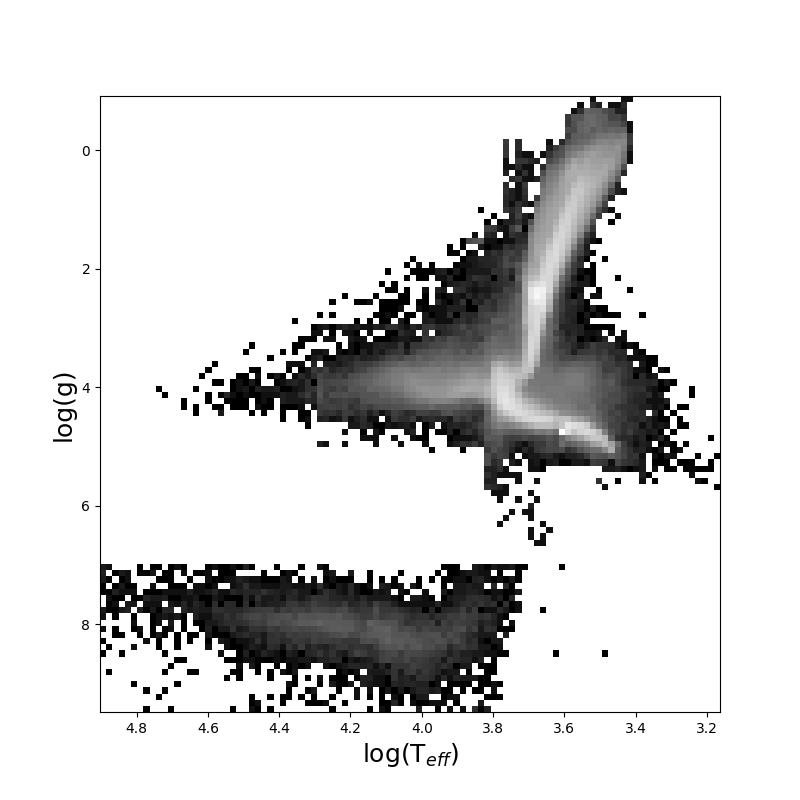}
\caption{The `preferred' \logg\ as a function of \teff\ (Kiel diagram) for MWM spectra run through \textsc{Astra}. Lighter cells represent those which contain significantly more individual objects compared to cells which are darker. `Preferred' \teff\ and \logg\ are drawn from the astraMWMLite file, which uses the decision tree outlined in Figure~\ref{fig:flow_lite} to select the preferred parameters.   }
\label{fig:kiel}
\end{figure}

In DR19, MWM (and BHM) are releasing new spectra obtained at Apache Point Observatory, up to MJD 60130 (5 July 2023). For MWM this corresponds to $\sim$1.2M near-infrared (APOGEE) spectra, and $\sim$800,000 optical (BOSS) spectra, for $\sim$390,00 and $\sim$475,000 unique stars with infrared or optical data, respectively. The main categories of stars are targets for Galactic Genesis, OB stars, YSOs, Solar Neighborhood and white dwarfs. Accumulation of Galactic Genesis slower than expected because of the effects of the COVID-19 pandemic and the required plate program in the first part of SDSS-V (more detail can be found in section~\ref{subsec:plateprogram}). Almost all of these spectra have been run through \textsc{Astra} to extract stellar parameters and atmospheric abundances where possible (see section \ref{astra}). Figure~\ref{fig:kiel} shows the 2 dimensional histogram of `preferred' (see section~\ref{subsec:astraMWMLite} and Figure~\ref{fig:flow_lite}) \logg\ as a function of \teff\ for MWM sources processed through \textsc{Astra} in DR19.

In addition to the spectra and data products in \textsc{Astra}, MWM, together with BHM, are also releasing 9 Value Added Catalogs (VACs). These catalogs are based on DR19 spectra, and contain data products not delivered by the official SDSS data reduction and analysis pipelines (see section \ref{sec:pipeline}), but are compiled by the science teams within the mappers. The VACs are described in detail in section \ref{sec:VAC}. The various ways to access the MWM spectra, data products and VACs are described in section \ref{sec:data.access}. 
\subsubsection{MWM Radial Velocity Stability}
\begin{figure}[h]
\centering
\includegraphics[width=0.5\textwidth]{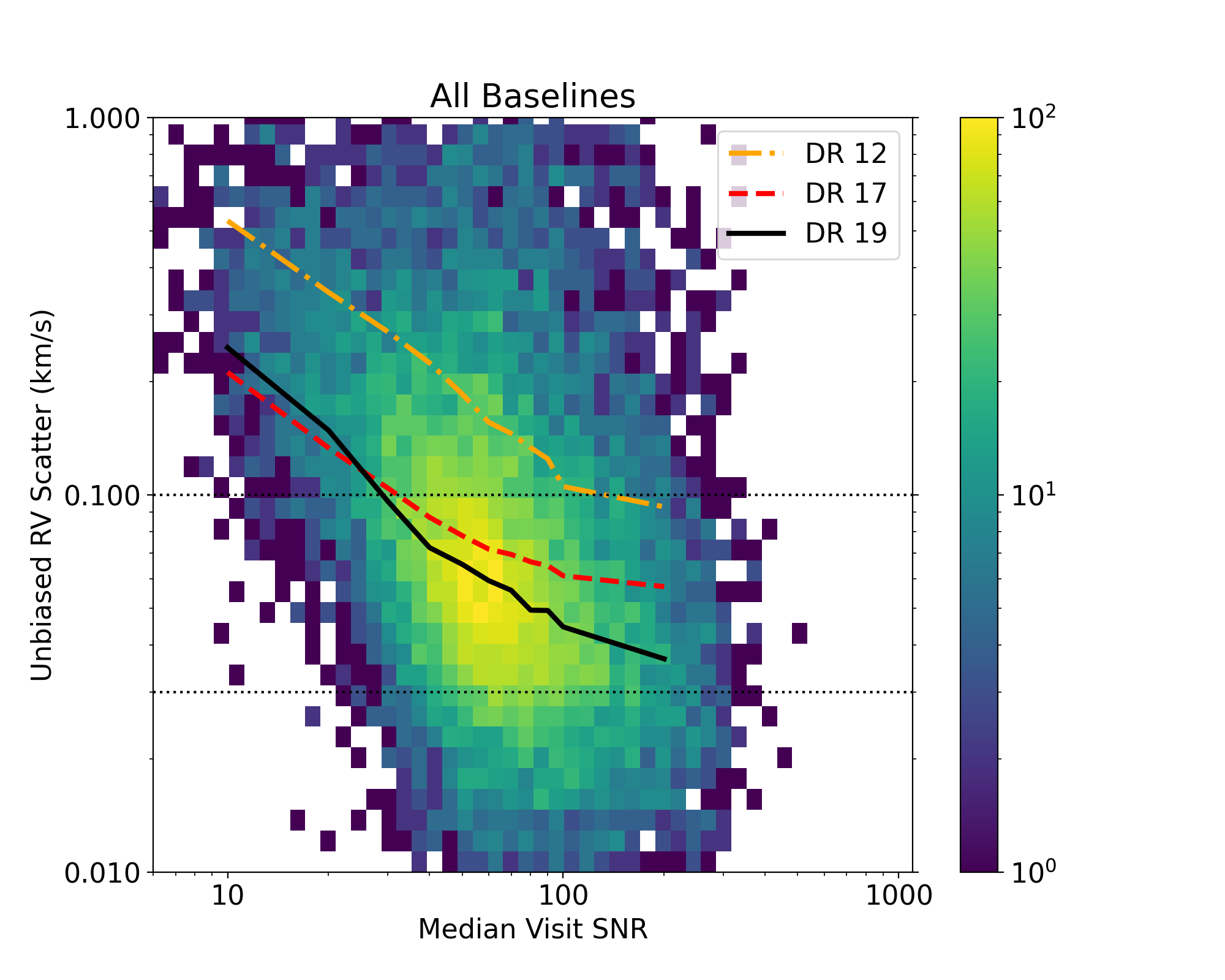}
\caption{The relationship between the median visit SNR versus the unbiased radial velocity scatter for stars with the following cuts: $T_{\rm eff} < 6250$ K, $\logg \: > 1.5$, and unbiased RV scatter $< 1.0$ km/s. The underlying color map shows the number of stars in each logarithmic 2-dimension bin. Each line shows the median unbiased RV scatter in each SNR bin for SDSS Data Releases 12 (orange dot dashed line), 17 (red dashed line), and 19 (black line). The The data is binned by logarithmic x-coordinate (e.g. 10, 20, 30...100, 200 etc.) The horizontal dotted lines are at 100 m/s and 30 m/s respectively.}
\label{fig:apogeerv}
\end{figure}

One major science goal of MWM is to understand the rates of stellar companions (stars, brown dwarfs, and planets) across the HR diagram. In order to characterize what types of companions will be detected, the radial velocity (RV) precision of the APOGEE instrument must be determined. For this analysis, we are focusing on the APOGEE DRP data products.  In order to simplify the analysis, stars were drawn from the APOGEE allStar file that had a minimum of 3 \texttt{ngoodvisits}, taken in the FPS era (after MJD 59550), and  \texttt{starflag} $=0$. The stars were further selected down to have the following stellar characteristics: an $T_{Eff} < 6250$K, $\logg \: > 1.5$, and unbiased \texttt{vscatter} $< 1.0$ km/s. These criteria were to limit ourselves to stars with reasonably narrow lines, small astrophysical RV jitter, and to remove obvious binaries respectively. One complication of this analysis is that the stars in our sample had a wide variety of number of epochs, so comparing the standard deviations (\texttt{vscatter}) from star to star can give artificially low values of velocity scatter for stars with small numbers of epochs. To correct for this problem, we used the $c_4$ parameter from \citet{Holtzman1950} to normalize the \texttt{vscatter} column from the allStar file. The resulting velocity scatter is referred to going forward as the unbiased \texttt{vscatter}. The median visit signal-to-noise (SNR) was calculated for each star. In Figure \ref{fig:apogeerv}, we plot a 2-dimensional histogram of this unbiased \texttt{vscatter} as a function of the median visit SNR. In Figure \ref{fig:apogeerv}, there are three lines showing the median unbiased RV scatter per SNR bin for SDSS DR12 \citep[orange dot dashed line][]{Alam_2015}, DR17 \citep[red dashed line][]{Abdurrouf_2022}, and DR19 (black line). DR 19 shows a significant improvement from DR 12 and DR 17 with a minimum unbiased RV scatter of approximately 41 m/s. This puts us within reach of our final goal of 30 m/s. For an example of future improvements that can be made to reach the 30 m/s goal see section \ref{sec:sdss5-vac.0006}.

\subsection{Black Hole Mapper data products}
\label{sec:BHMscope}
\begin{figure}[h]
\centering
\includegraphics[width=0.45\textwidth]{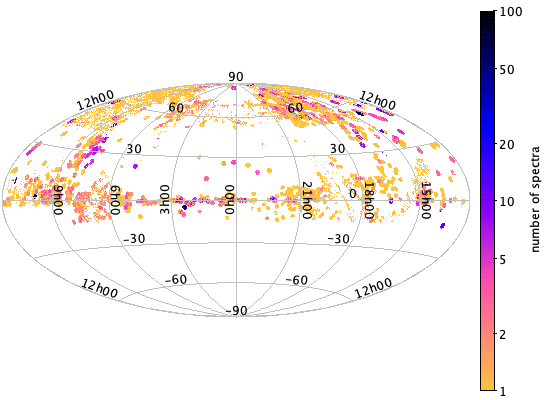}
\caption{Distribution (RA, Dec) on the sky of BHM (and related) optical spectra released in DR19. The color coding indicates the number (logarithmic scaling) of spectra per BHM object/target; many BHM spectra are multi-epoch.}
\label{fig:bhmipl3sky}
\end{figure}

SDSS DR19 is providing $\sim$380,000 BHM-led optical BOSS science spectra taken via both plates (section~\ref{subsec:plateprogram}) and FPS (section~\ref{sec.fps}) from APO (mainly North of declination -15~$\deg$) for $\sim$120,000 distinct objects (more statistics of the spectra released can be found in Table~\ref{tab:bhm_stats}). These sources are mainly quasars/AGN spanning all of BHM's main core science subprograms (see section~\ref{subsec:BHM}), but also include a number of spectral targets of additional complementary science subprograms as well. Some examples include: galaxies in X-ray emitting clusters, X-ray emitting stars, and miscellaneous other targets as briefly described below. Similar to MWM, these spectra extend up to MJD 60130 (5 July 2023). Many of the spectra for BHM in DR19 are also multi-epoch spectra. The DR19 set represents an order of magnitude expansion in BHM related spectra released, compared with DR18 \citep{almeida2023}. The sky distribution of BHM targets/objects with optical spectra released in DR19 is depicted in Figure~\ref{fig:bhmipl3sky} with color coding indicating the number of spectra per target.  

DR19 contains BOSS spectra for three BHM RM fields routinely observed from APO: the COSMOS, SDSS-RM, and XMM-LSS fields (a fourth main RM field, CDFS, is observed from LCO in the South and RM data from that fourth field will also be included in later SDSS-V data releases). Nearly $\sim$110,000 BOSS spectra for $\sim$2400 distinct RM targets are included in DR19, the bulk of which are repeat higher-cadence spectroscopy for the approximately 380 primary RM science targets in each of the COSMOS, SDSS-RM, and XMM-LSS fields. 

In addition to new BHM RM spectra, DR19 contains $\sim$35,000 recent spectra for nearly 10,000 core AQMES quasars (all brighter than $i<19.1$). It also includes additional epochs of spectra of DR16 quasars from the \citet{Lyke2020} catalog that are also closely associated with the AQMES core endeavor, e.g, including new-epoch spectra of quasars fainter than $i>19.1$ within fields observed by SDSS-V. Counting both AQMES core and those other closely related ancillary cases, DR19 provides a spectral dataset that is $\sim4\times$ larger set of AQMES-related quasars and spectra than AQMES core alone: in total, of order 135,000 spectra for 45,000 AQMES-related quasars. 

\begin{figure}[h]
\centering
\includegraphics[width=0.45\textwidth]{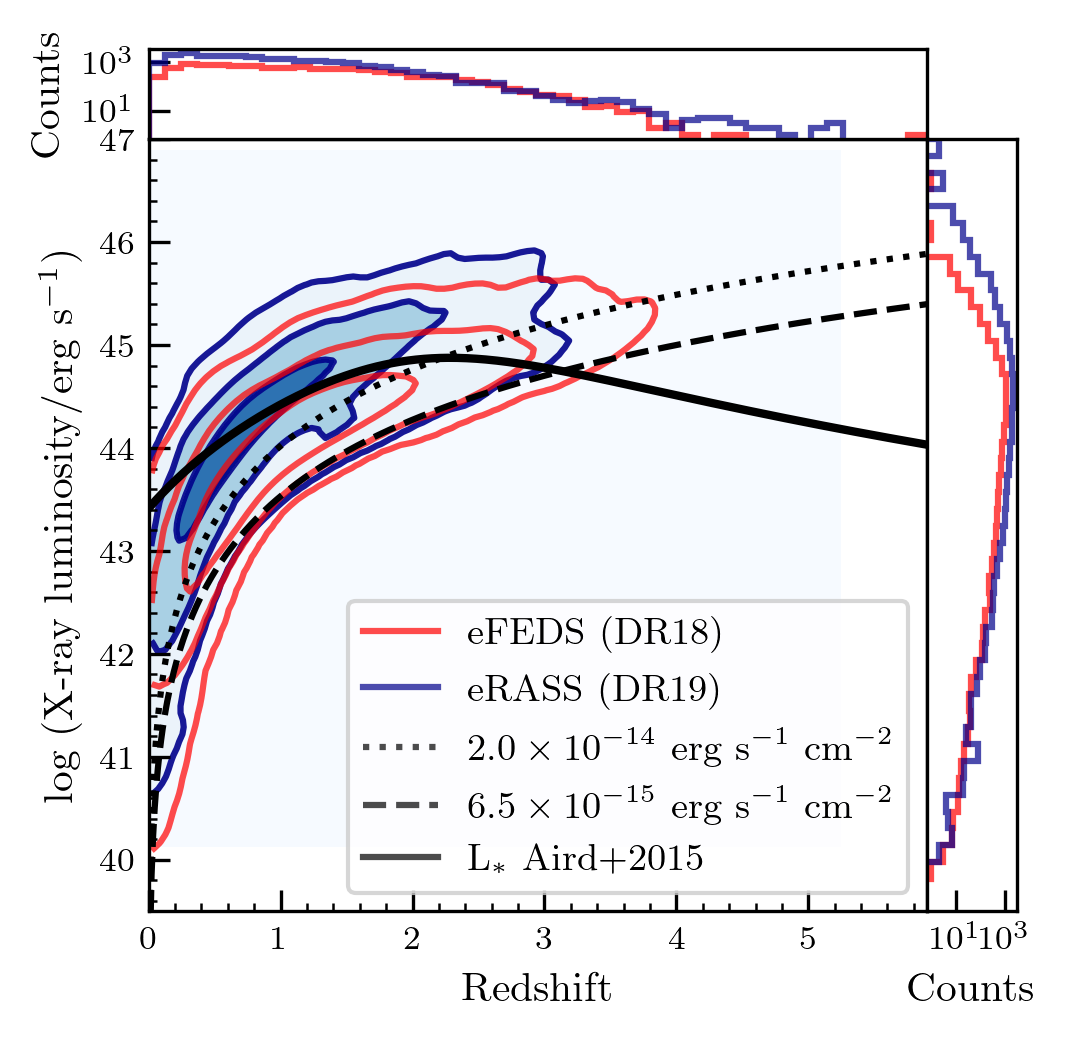}
\caption{Density contours of X-ray sources in the observed X-ray ($0.2-2.3$ keV) luminosity vs. redshift plane. Blue contours represent the sources from eRASS1 released in DR19 as part of the VAC catalog described in section 7.2; red contours represent the sources in the eFEDS sample, released in DR18 \citep{Aydar2025}.
  The dashed (dotted) line is indicative of the X-ray flux limit of the eFEDS (eRASS1) survey, while the filled line indicates the location of the knee of the soft X-ray luminosity function according to the LADE model of \citet{Aird2015}. The top and right panels show the histogram distributions of redshift and X-ray luminosity, respectively, for the DR18 (red) and DR19 (blue) samples.}
\label{fig:SPIDERS_Dr18_Dr19}
\end{figure}

DR19 also contains an additional $\sim$20,000 spectra for $\sim$11,000 X-ray sources from the SPIDERS hemisphere survey at APO. This new dataset is comparable to the earlier pilot eFEDS survey of DR18 and when combined they provide 48,000 total SPIDERS related optical spectra, for $\sim$25,000 distinct X-ray sources through DR19. Figure~\ref{fig:SPIDERS_Dr18_Dr19} shows the distribution of the X-ray sources in the X-ray luminosity-redshift plane, for both the eFEDS sources (DR18, see \citet{Aydar2025}) and the newly released sample (DR19, based on eRASS1). The larger volume probed by the all-sky survey data allows us to explore the region of the parameter space where the most luminous and rare AGN live. More details about the results of SPIDERS in DR19 can be found in its associate VAC, described in section~\ref{sec:sdss5-vac.0004}.

In addition to the spectra released as part of the core BHM science programs: there are of order 19,000 BOSS optical spectra for more than 7000 distinct Chandra Source Catalog (CSC) X-ray sources \citep{Evans_2024}, plus a comparable number of APOGEE spectra relevant to CSC; and, there are a yet further $\sim$66,000 spectra released for $\sim$41,000 distinct objects from "open-fiber" \citep{almeida2023} and also other ancillary spectroscopic targets related to BHM.
For BHM, these latter ancillary targets and spectra are mainly extragalactic objects, for example,
with further ancillary spectra released in DR19 also including many for quasar/AGN candidates from both Gaia unWISE AGN \citep{Shu2019} and from \citet{Yang_2023} photometric samples, and for bright galaxies selected from the DESI Legacy Imaging Surveys \citep{Dey_2019_DESIsurveys}.

The DR19 release (and the earlier DR18) has enabled key BHM science including an exploration of changing look quasars from low and medium cadence SDSS-V time-domain spectra (i.e., akin to AQMES data), and which include both individually-interesting cases and large samples, e.g., as described in, respectively, in \citet{Zeltyn_etal_2022} and \citet{Zeltyn_etal_2024}. There are also high-cadence time domain spectral studies (e.g. from among the RM targets) that include investigations of quasar emission line profile variations in select interesting cases, e.g., \citet{Fries_etal_2023,Fries_etal_2024} and \citet{Stone_etal_2024}, as well as individually-interesting cases of changes in Broad Absorption Line features in BALQSOs, e.g., \citet{Wheatley_etal_2024}. There are also multiple joint X-ray optical studies of the sort enabled by SPIDERS and eFEDS that have featured population, demographics, and astrophysics studies of X-ray selected AGN, including, e.g. \citet{Comparat2023}, \citet{Nandra_etal_2025}, \citet{Waddell_etal_2024}, Rankine et al. (2025), and in providing BOSS spectroscopic verifications relevant to securing photometric redshifts in studies such as those by \citet{Saxena_etal_2024} and \citet{Roster_etal_2024}. 





\begin{deluxetable*}{lrrrr}
\label{tab:bhm_stats}
\tablecaption{Scope of BHM --  Number of DR19 BOSS spectra (per coadding scheme) associated with each BHM (and extrgalactic openfiber) carton 
(considering only the \texttt{FIRSTCARTON} label for each spectrum). The number of unique targets is determined via \texttt{SDSS\_ID}.}
\tablehead{ & & No.  coadded   spectra& & \colhead{Unique}\\
\colhead{Carton Name} & \colhead{daily} & \colhead{epoch} & \colhead{allepoch} & \colhead{targets} }
\startdata
\texttt{bhm\_rm\_known\_spec} &      85975 &      56702 &       1177 &       1457 \\
\texttt{bhm\_aqmes\_med\_faint} &      39845 &      25698 &       1123 &       7818 \\
\texttt{bhm\_aqmes\_wide2\_faint} &      27182 &      21585 &       1220 &      11689 \\
\texttt{bhm\_gua\_bright} &      26773 &      23735 &      14951 &      15307 \\
\texttt{bhm\_gua\_dark} &      22897 &      18259 &      12178 &      12723 \\
\texttt{bhm\_rm\_core} &      21546 &      13059 &        203 &        862 \\
\texttt{bhm\_spiders\_agn\_efeds} &      18055 &      12825 &       9065 &       9200 \\
\texttt{bhm\_aqmes\_med} &      15798 &      10192 &       1259 &       2389 \\
\texttt{bhm\_aqmes\_bonus\_faint} &      13659 &      10846 &       1491 &       7674 \\
\texttt{bhm\_aqmes\_wide2} &      13324 &      10250 &       2584 &       5784 \\
\texttt{bhm\_spiders\_agn\_lsdr8} &      13108 &      10300 &       6891 &       7032 \\
\texttt{bhm\_aqmes\_wide3\_faint} &      12002 &       8998 &        187 &       4094 \\
\texttt{bhm\_csc\_boss\_dark} &      11392 &       7858 &       2891 &       2908 \\
\texttt{bhm\_colr\_galaxies\_lsdr8} &       8812 &       7757 &       6223 &       6328 \\
\texttt{bhm\_csc\_boss} &       6930 &       5023 &       4149 &       4247 \\
\texttt{bhm\_aqmes\_bonus\_dark} &       5783 &       4444 &        107 &       3910 \\
\texttt{bhm\_aqmes\_wide3} &       5664 &       4160 &        193 &       1681 \\
\texttt{openfibertargets\_nov2020\_27} &       4373 &       4161 &       2809 &       3899 \\
\texttt{bhm\_spiders\_agn\_ps1dr2} &       3587 &       2993 &       1988 &       2093 \\
\texttt{bhm\_spiders\_agn\_efeds\_stragglers} &       3234 &       2371 &       1909 &       1974 \\
\texttt{bhm\_spiders\_clusters\_efeds\_ls\_redmapper} &       3178 &       2210 &       1642 &       1666 \\
\texttt{bhm\_rm\_var} &       2049 &       1234 &          2 &         75 \\
\texttt{openfibertargets\_nov2020\_33} &       2018 &       1999 &       1518 &       1875 \\
\texttt{bhm\_spiders\_clusters\_efeds\_sdss\_redmapper} &       1757 &       1186 &        861 &        873 \\
\texttt{manual\_bhm\_spiders\_comm} &       1420 &        355 &        291 &        355 \\
\texttt{bhm\_spiders\_clusters\_efeds\_hsc\_redmapper} &       1370 &        906 &        689 &        696 \\
\texttt{bhm\_spiders\_clusters\_lsdr8} &       1252 &        959 &        684 &        696 \\
\texttt{openfibertargets\_nov2020\_18} &        799 &        777 &        361 &        717 \\
\texttt{bhm\_csc\_boss\_bright} &        749 &        344 &        184 &        198 \\
\texttt{bhm\_aqmes\_bonus\_core} &        741 &        639 &        133 &        348 \\
\texttt{bhm\_spiders\_clusters\_ps1dr2} &        686 &        533 &        391 &        396 \\
\texttt{bhm\_spiders\_agn\_supercosmos} &        681 &        580 &        430 &        436 \\
\texttt{openfibertargets\_nov2020\_11} &        624 &        604 &         58 &        596 \\
\texttt{bhm\_aqmes\_bonus\_bright} &        590 &        233 &         12 &        124 \\
\texttt{bhm\_spiders\_clusters\_efeds\_stragglers} &        431 &        289 &        245 &        249 \\
\texttt{bhm\_spiders\_agn\_gaiadr2} &        168 &        161 &        116 &        119 \\
\texttt{bhm\_rm\_ancillary} &         40 &         25 &          0 &          1 \\
\texttt{bhm\_spiders\_clusters\_efeds\_erosita} &         28 &         16 &         12 &         12 \\
\texttt{bhm\_spiders\_agn\_skymapperdr2} &         28 &         23 &         19 &         20 \\
\texttt{openfibertargets\_nov2020\_26} &         16 &         16 &          8 &         16 \\
\enddata
\end{deluxetable*}

\subsection{Local Volume Mapper data products}
\label{sec:lvmpreview}
To introduce the community to the new LVM data products, we include in DR19 the reduced and calibrated row-stacked spectra (RSS) file associated with a single LVM tile observed on the Helix Nebula.

\begin{figure}[h]
\centering
\includegraphics[width=0.45\textwidth]{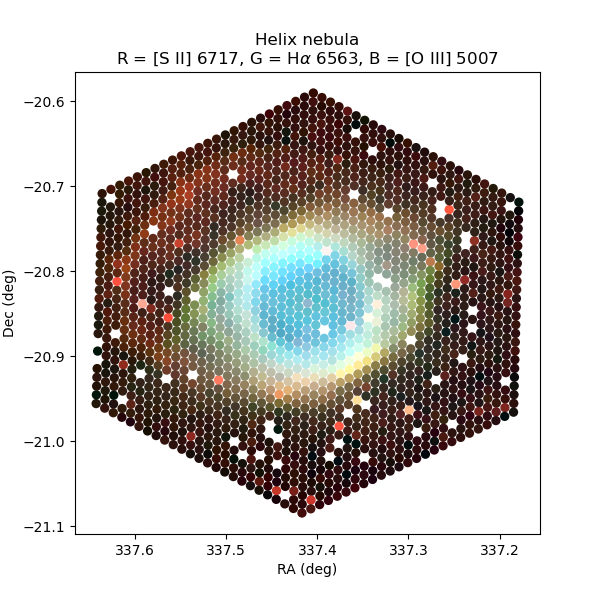}
\caption{Composite red (R), green (G), blue (B) map of the Helix nebula constructed by using spectral windows that capture the three emission lines given in the title. Fibers with known issues have been removed. We include fibers that catch contaminating stars.}
\label{fig:helix}
\end{figure}

The Helix Nebula (NGC~7293) is a well-studied and beautiful planetary nebula (PNe), with a 13~arcmin angular diameter 
that is well matched to the LVM field of view. Located at a distance of 200~pc \citep{Kimeswenger2018}, it is one of the closest and brightest PNe.
The nebula itself is visible as a bright ring of ionized gas, powered by a dying central star, and is an ideal target for optical spectroscopy due to the rich variety of emission lines produced (e.g., hydrogen, helium, oxygen, nitrogen, sulphur). These lines can provide insights into the chemical composition, temperature, density, and kinematics of the nebula \citep{ODell1998, Henry1999}. 
With 35.5\arcsec\ fiber diameters, the LVM provides maps at 0.03~pc resolution across a $\sim$2~pc diameter field of view. In Figure~\ref{fig:helix}, we show a preview of the Helix nebula using data from LVM. Figure~\ref{fig:helix} shows a composite map of the nebula combining information from three emission lines: [S~II] at 6717~\AA (red), [H$\alpha$] at 6563~\AA (green), and [O~III] at 5007~\AA (blue). 

The RSS file structure is a common format for IFU data (also used previously for SDSS-IV MaNGA data releases) where all spectra across the field of view are homogentized in wavelength, and accompanied by a set of metadata that provides associated information, such as their observed location on the sky.
This LVM preview provides a demonstration of the data model that we have developed. 
A tutorial illustrating how to interact with this file format is provided (see Section \ref{sec:lvmtutorial}). The data reduction pipeline is still under active development, but this tile is fully reduced and calibrated with our current latest pipeline version. Future improvements, particularly with regards to sky subtraction and flux calibration, are planned in future releases of the data, and we caution that some lines (e.g. [O\textsc{i}]$\lambda\lambda$6300,6363 and lines with wavelengths above 7000\AA) may still be contaminated by bright sky line emission. 

\section{Accessing SDSS DR19}  
\label{sec:data.access}
\input{sections/data/access}

\subsection{The Science Archive Server}
\label{sec:data.sas} 
\input{sections/data/sas}

\subsection{The Catalog Archive Server}
\label{sec:data.cas}
\input{sections/data/cas}

\subsection{The SDSS Data Model}
\label{sec:data.datamodel}
\input{sections/data/datamodel}

\subsection{The Zora/Valis Web Framework}
\label{sec:data.dataviz}

\input{sections/data/dataviz}

\section{Value Added Catalogs (VACs) } 
\label{sec:VAC}

In this section we will outline the 9 new VACs for DR19, as listed in Table \ref{tab:vacs}, and their associated papers.  The full set of 76 VACs across all data releases is available on the SDSS website at \url{https://www.sdss.org/dr19/data_access/value-added-catalogs/} which can be filtered by SDSS survey or mapper, with SDSS-V VACs currently available for BHM and MWM.

\begin{deluxetable*}{llll}
\label{tab:vacs}
\tablecaption{ Value Added Catalogues New for DR19 }
\tablehead{\colhead{Name} & \colhead{ Data Product } & \colhead{ CAS Tables } & \colhead{ Sec. } }
\startdata
Detailed M Dwarf Elemental Abundances & \href{https://dr19.sdss.org/sas/dr19/env/MWM_MDWARF/elemental_abundances}{MWM\_MDWARF} & {\tt mwm\_mdwarf\_abundances} & \ref{sec:sdss5-vac.0011} \\
Data Level 1 of SDSS-eROSITA & \href{https://dr19.sdss.org/sas/dr19/env/DL1_SDSS_EROSITA}{DL1\_SDSS\_EROSITA} & {\tt DL1\_eROSITA\_eRASS1}
 & \ref{sec:sdss5-vac.0004} \\
Stellar Parameters for BOSS Halo Stars & \href{https://dr19.sdss.org/sas/dr19/env/MWM_MINESWEEPER}{MWM\_MINESWEEPER} & {\tt minesweeper} & \ref{sec:sdss5-vac.0007} \\
Open Cluster Chemical Abundances and Mapping & \href{https://dr19.sdss.org/sas/dr19/env/APOGEE_OCCAM}{APOGEE\_OCCAM} & {\tt occam\_}[\tt{cluster},\tt{member}] & \ref{sec:sdss5-vac.0017} \\
StarHorse Stellar Parameters, Distances, and Extinctions & \href{https://dr19.sdss.org/sas/dr19/env/APOGEE_STARHORSE}{APOGEE\_STARHORSE} & {\tt apogee\_starhorse} & \ref{sec:sdss5-vac.0014} \\
SDSS DA White Dwarf Physical Properties & \href{https://dr19.sdss.org/sas/dr19/env/MWM_WHITEDWARF/da_white_dwarf_properties}{MWM\_WHITEDWARF} & {\tt MWM\_WD\_}[\tt{SDSSV},\tt{eSDSS}]\tt{\_DA\_df} & \ref{sec:sdss5-vac.0008} \\
Spectral Component Separation with MADGICS & \href{https://dr19.sdss.org/sas/dr19/env/APMADGICS}{APMADGICS} & {\tt allVisit\_MADGICS\_}[\tt{th},\tt{dd}] & \ref{sec:sdss5-vac.0006} \\
Starflow Spectroscopic Distances, Masses, and Ages & \href{https://dr19.sdss.org/sas/dr19/env/MWM_STARFLOW}{MWM\_STARFLOW} & {\tt StarFlow\_summary} & \ref{sec:sdss5-vac.0003} \\
Quasar Spectral Properties & \href{https://dr19.sdss.org/sas/dr19/env/BHM_QSOPROP}{BHM\_QSOPROP} & {\tt DR19Q\_prop} & \ref{sec:sdss5-vac.0009} \\
\enddata
\tablecomments{
A complete listing of SDSS VACs is available at \url{https://www.sdss.org/dr19/data_access/value-added-catalogs/}.
}
\end{deluxetable*}

\subsection{Detailed M Dwarf Elemental Abundances}  \label{sec:sdss5-vac.0011}

\citet{Behmard2025} describes a VAC containing detailed elemental abundances for $\sim$17,000 M dwarfs in SDSS-V. This VAC was constructed with an implementation of \emph{The Cannon}, a data-driven method capable of inferring stellar abundances that does not rely on stellar evolution models \citep{Ness2015,Casey2016}. This makes \emph{The Cannon} an excellent choice for characterizing M dwarfs, which have notoriously complex spectra due to the presence of molecular features. In short, \emph{The Cannon} was trained on M dwarfs with FGK dwarf companions that have elemental abundances from \emph{ASPCAP}. We made the assumption that binary companions form from the same birth cloud, and are thus born chemically homogeneous \citep[an assumption that has been shown to be satisfied for most wide binary systems, e.g.,][]{2004A&A...420..683D,2020MNRAS.492.1164H, 2021ApJ...921..118N, 2021NatAs...5.1163S}. This enabled us to tag the M dwarfs with the \emph{ASPCAP} abundances of their FGK companions. These M dwarfs formed a training set for \emph{The Cannon}, which resulted in flux models that were applied to the $\sim$17,000 M dwarfs in SDSS-V with high-quality spectra (see \citealt{Behmard2025} for more details). This allowed us to infer abundances for Fe, Mg, Al, Si, C, N, O, Ca, Ti, Cr, and Ni with median uncertainties of 0.018–0.029 dex.

\subsection{eROSITA and SDSS-V} \label{sec:sdss5-vac.0004}

Multi-wavelength information is crucial for a complete understanding of the Universe.
Considering the X-ray domain, eROSITA \citep{Predehl2021} on board the Spectrum Roentgen Gamma satellite \citep{Sunyaev2021} has produced the largest uniform survey to date due to its wide field telescope designed to provide X-ray spectroscopy and imaging of the entire sky \citep{Merloni2024}. The Spectroscopic Identification of ERosita Sources (SPIDERS) program was designed by selecting objects identified in X-rays with eROSITA to be observed in the optical domain by SDSS-V.

The VAC of eROSITA and SDSS-V is based on the combination of optical (BOSS) spectroscopy from SDSS DR19 (this paper) and X-ray data from the first eROSITA All Sky Survey \citep[DR1;][]{Merloni2024}.
We provide a catalog of spectroscopically observed objects that were targeted in SDSS-V because they are considered to be counterparts to point-like eROSITA X-ray sources (i.e. quasars, Active Galactic Nuclei, stars, and compact objects).
For each eROSITA source that was observed in SDSS-V, we provide the following:
\begin{itemize}
    \item From SDSS: field, MJD, catalog ID, run2D, target index, object type, target flag, fiber position coordinates, redshift (with error and warning), median SNR, class, best fit stellar template, and properties (spectral type, effective temperature, surface gravity, and H$\alpha$ equivalent width from Astra).
    \item From eROSITA: unique X-ray source identifier (\texttt{DETUID}), flux in the full eROSITA band and error, MJD, morphological classification, source detection likelihood, X-ray position estimate, and positional error.
\end{itemize}
For convenience, we also added public photometric data (where available), derived from Gaia DR3 \citep{gaiadr3}, unWISE \citep{unwise}, GALEX \citep{galex}, and 2MASS \citep{2mass}.
We provide separate data tables derived from the \texttt{daily co-add} (individual optical observations of each source) and the \texttt{allepoch} (stack of all the optical observations of the same object) SDSS spectra.
For the \texttt{daily co-add}, we provide information of 15214 unique eROSITA sources, observed in 15494 available optical spectra; for the \texttt{allepoch} file, we provide 28743 optical spectra of 15221 unique X-ray emitting sources.
This VAC contains SDSS spectra obtained via the robotic fibers program; therefore, MJD$>59500$.
For eROSITA targets \citep{Brunner2022} observed within the dedicated SDSS-V/eFEDS plates program \citep{almeida2023}, we refer the reader to the catalog released by \citet{Aydar2025}.

We caution that the X-ray data provided in this VAC is based on the `c946' reductions of the eROSITA All-Sky Survey (eRASS) 1 catalog. This used earlier versions of the X-ray pipeline and calibration model than were used for eRASS1 catalog presented by \citet{Merloni2024}. Hence, there are some differences in the X-ray properties in this VAC w.r.t. the eROSITA DR1 catalog, particularly X-ray flux and uncertainty, as well as detection likelihood. Note also that the X-ray source identifier (\texttt{DETUID}) in this VAC differs from that in eRASS1 DR1 public catalog. Therefore one should cross-match sources based on their position in the sky rather than their IDs.

\subsection{Stellar Parameters for BOSS Halo Stars} \label{sec:sdss5-vac.0007}

The Milky Way Mapper halo survey is expanding the Galactic frontier along two directions: the most distant stars, and the most metal-poor stars. For these stars, it is vital to incorporate all available information about a star (i.e., photometry, spectra, and astrometry) when inferring their fundamental parameters. Furthermore, one of the key science goals of the halo survey is to map out the all-sky dynamics of the halo on the largest scales, which requires a homogeneous sample of stars with spectro-photometric distances.

The MWM \texttt{minesweeper} VAC uses the \texttt{MINESweeper} code to infer stellar parameters for a subset of stars targeted by the halo survey. The \texttt{MINESweeper} code is described and validated in \cite{Cargile2020}. Briefly, it utilizes all available information about a star -- the SDSS-V spectrum, broadband photometry from public surveys, and the \textit{Gaia} parallax --- to find the best-matching stellar parameters. Solutions are constrained to lie on isochrones, limiting the fit to physically-plausible regions of the parameter space. Priors can be placed on various explicit and implicit variables in the stellar model. \texttt{MINESweeper} delivers posterior distributions of the radial velocity $v_\mathrm{r}$, effective temperature $T_\mathrm{eff}$, surface gravity $\log{g}$, metallicity [Fe/H], [$\alpha$/Fe] abundance, and heliocentric distance.

Since \texttt{MINESweeper} is computationally expensive, $\approx 0.5$ core-hours per star, it is only run on a subset of valuable MWM halo stars for which it is most useful.
This includes cartons that target metal-poor stars pre-selected with photometry, and cartons that target distant giants in the outer halo.
The full details of the SDSS-V implementation of \texttt{MINESweeper} will be described in Chandra et al (in prep).

\subsection{Open Cluster Chemical Abundances and Mapping (OCCAM) Catalog} \label{sec:sdss5-vac.0017}

The primary goal of the Open Cluster Chemical Abundance and Mapping (OCCAM) survey \citep[][Otto et al., {\em in prep}]{donor_18,occam_p4,myers_2022} is to create a comprehensive, {\em uniform} infrared-based spectroscopic dataset of key open cluster parameters. This analysis combines the MWM/APOGEE radial velocities and chemical abundances determined with the ASPCAP \citep{ASPCAP.2016AJ....151..144G} pipeline with astrometry membership probabilities using the {\it Gaia} observations \citep{Gaia_DR2_2018} to establish membership probabilities in proper motion, radial velocity, and metallicity for open cluster stars. 
This 4th iteration of the OCCAM catalog provides a larger sample and represents a significant shift in methodology, described as follows.  Previous versions performed the proper motion and position membership determination as part of the overall analysis.  
The new catalog now use the \citet{cg_18} catalog as the starting point of the analysis, which used the {\it Gaia} Data Release 2 observations and 5D astrometry to determine a membership probability for each star.
The \citet{cg_18} membership probability is used as a basis to determine additional membership probabilities from the MWM DR19 radial velocity and [Fe/H] data. As with previous OCCAM VACs, stars that are within $3\sigma$ of the cluster mean in both radial velocity and [Fe/H] space are included in the determination of the bulk cluster parameters. 

This 4th OCCAM VAC includes two datasets, (1) bulk cluster motions \citep[{\it Gaia} Data Release 3,][]{gaiadr3}, chemical abundances, and orbital parameters for 164 open clusters, and (2) the DR19 data used in the membership and bulk parameter determination, including membership probabilities, positional data, {\it Gaia} IDs, SDSS-V IDs, kinematics, and metallicity for individual stars. 
The results and caveats from this VAC will be discussed thoroughly in Otto et al. (2025, {\em in prep}).

\subsection{ {\tt StarHorse}  stellar parameters, distances, and extinctions for APOGEE DR19 Red Giants} \label{sec:sdss5-vac.0014}

This new value-added catalog (VAC) presents the {\tt StarHorse} (SH) run for APOGEE red giants stars in DR19, building on the metholodogy and success of previous releases, particularly DR17 \citep{Queiroz2020,Queiroz2023}.  The SH code is a Bayesian tool that combines spectroscopic, photometric, and astrometric data to infer key stellar parameters such as distances, masses, and extinctions \citep{Queiroz2018,Anders2022}. This VAC contains spectro-photo-astrometric distances, extinctions, temperature, masses and metallicity for 308,860 SDSS-V DR19 APOGEE giant stars, derived using the SH code with calibrated inputs and multi-wavelength data from Gaia, Pan-STARRS1, 2MASS, and AllWISE. 

Parameters are estimated for each unique {\tt sdss\_id} in the data release, provided the SH code successfully converges. If a star has multiple {\tt sdss\_id}s, the ASPCAP results with the highest SNR are used. For each star, SH computes the joint posterior probability distribution function (PDF) over a grid of PARSEC 1.2S stellar models, using input values including ASPCAP-derived effective temperature, surface gravity, metallicity, and alpha-element abundance, as well as Gaia DR3 parallaxes (when available), and the multi-band photometry listed above. ASPCAP effective temperature and surface gravity are calibrated before input to SH. The VAC includes median values of marginalized PDFs for mass, temperature, surface gravity, metallicity, distance, and extinction. The {\tt StarHorse\_INPUTFLAGS} column indicates the input data used, while {\tt StarHorse\_OUTFLAGS} flags possibly uncertain outputs. 

Our DR19 run was limited to red giants ($\log{g} <$ 3.5), which are crucial for the Galactic Genesis program due to their intrinsic brightness and ability to probe distant regions of the Milky Way. To ensure consistency with DR17 and to leverage established asteroseismic calibrations for surface gravity, we applied a calibration to bring the DR19 input parameters ($T_\mathrm{eff}$ and $\log{g}$) onto the DR17 scale. For $T_\mathrm{eff}$, we used the ``raw\_teff'' columns from APOGEE DR19 within the range 4500-7000 K and applied a linear correction: $T_\mathrm{eff}(\mathrm{DR17\text{-}like}) = \mathrm{raw\_teff} \times 0.86463363 + 720.06160957$. Values outside this range are clipped to the nearest boundary. For surface gravities, we applied a calibration within 0.0 $<$ $\log{g}$ $\leq$ 1.8, using: $\log{g}$(DR17-like) $=$ $\log{g}$ $\times$ 0.82897254 $+$ 0.31209658, with clipping at the boundaries. Note that in this case we started from the calibrated $\log{g}$ from DR19, which benefited from asteroseismic calibrated surface gravities. These DR17-like parameters are then used as input for the SH run, and the re-calibrated values are included in the new VAC.

\subsection{Catalogs of SDSS DA White Dwarf Physical Properties} \label{sec:sdss5-vac.0008}

\indent The construction and validation of \textit{The Catalogs of SDSS DA White Dwarf (WD) Physical Properties VAC}, hereafter referred to as the \citet{Crumpler_2025} catalogs, are detailed in depth in \citet{Crumpler_2025}. We summarize the relevant results of that paper and its corresponding VAC here. The \citet{Crumpler_2025} catalogs are the largest catalogs of both spectroscopic and photometric physical parameters of DA white dwarfs available to date, containing 8,545 and 19,257 unique DA WDs observed in SDSS DR19 and previous SDSS data releases, respectively. The code used for all measurements in the \citet{Crumpler_2025} catalogs is publicly available\footnote{\url {https://github.com/nicolecrumpler0230/WDparams}}.

\indent The SDSS-V \citet{Crumpler_2025} catalog is comprised of all SDSS-V DR19 objects identified as DA WDs by the spectral classification algorithm \texttt{SnowWhite} through November 2023. All SDSS-V WD spectra used in this catalog are obtained with the BOSS spectrograph \citep{Smee_2013} using the reduction pipeline v6\_1\_3. The DA WDs in the previous SDSS \citet{Crumpler_2025} catalog, covering WDs observed in SDSS through Data Release 16, were compiled by \citet{Gentile_2021}\footnote{\url {https://cdsarc.cds.unistra.fr/viz-bin/cat/J/MNRAS/508/3877\#/browse}}. \citet{Gentile_2021} visually inspected all spectra, providing spectral classifications of each observation, and \citet{Crumpler_2025} only selected objects classified as DA WDs.

\indent The apparent radial velocities in the \citet{Crumpler_2025} catalogs are measured by fitting the H$\alpha$, H$\beta$, H$\gamma$, and H$\delta$ hydrogen Balmer lines to \citet{Tremblay_2013} model spectra using a $\chi^2$-minimization routine built into the open-source code Compact Object Radial Velocities (\texttt{corv}\footnote{\url {https://github.com/vedantchandra/corv}}). \citet{Crumpler_2025} find that their apparent radial velocities agree to within $7.5$ km/s when compared to previously published SDSS WD catalogs for spectra with SNR $\geq50$.

\indent The spectroscopic surface gravities and temperatures in the \citet{Crumpler_2025} catalogs are measured by fitting the shapes of the first six Hydrogen Balmer series lines (H$\alpha$, H$\beta$, H$\gamma$, H$\delta$, H$\epsilon$, H$\zeta$) using a parametric random forest routine built into the publicly available code \texttt{wdtools\footnote{\url {https://wdtools.readthedocs.io/en/latest/}}} \citep{Chandra_2020_2}. \citet{Crumpler_2025} found that their surface gravity and temperature measurements agree to within $0.060$ dex and $2.4$\%, respectively, when compared to previously published SDSS WD catalogs for spectra with SNR $\geq50$.

\indent The photometric radii and effective temperatures contained in the \citet{Crumpler_2025} catalogs are measured by fitting a combination of cross-matched Gaia DR3 photometry \citep{Gaia_2016}, SDSS Data Release 17 photometry \citep{Abdurrouf_2022}, and \citet{BailerJones_2021} distances to model photometry via $\chi^2$ minimization. The observed photometry is corrected for extinction using the three-dimensional dust map of \citet{Edenhofer_2024} from the \texttt{dustmaps}\footnote{\url {https://dustmaps.readthedocs.io/en/latest/index.html}} Python package \citep{Green_2018} and the extinction curve from \citet{Fitzpatrick_1999} from the \texttt{extinction}\footnote{\url {https://extinction.readthedocs.io/en/latest/}} Python package. The model photometry is created by convolving \citet{Tremblay_2013} model spectra through photometric filter response curves from the \texttt{pyphot}\footnote{\url {https://mfouesneau.github.io/pyphot/}} Python package. \citet{Crumpler_2025} found that their radius and temperature measurements agree to within $0.0005$ $R_\odot$ and $3$\%, respectively, when compared to the \citet{Gentile_2021} catalog for fits to Gaia photometry.

\indent The \citet{Crumpler_2025} catalogs also contain other parameters that may be useful for the DA WD community. These parameters include measured WD masses, a likely binary flag, a stellar population flag, Local Standard of Rest-corrected apparent radial velocities, and asymmetric drift-corrected apparent radial velocities. Additionally, the SDSS-V \citet{Crumpler_2025} catalog has been cross-matched with six other catalogs of DA WD parameters \citep{Koester_2009, Falcon_2010, Anguiano_2017, Kepler_2019, Gentile_2021, Raddi_2022}, and the previous SDSS \citet{Crumpler_2025} catalog has been cross-matched with six other SDSS WD catalogs from previous data releases \citep{Kleinman_2004, Eisenstein_2006, Kleinman_2013, Kepler_2015, Kepler_2016, Kepler_2019}. Already, the \citet{Crumpler_2025} catalogs have been used by \citet{Crumpler_2024} to directly detected the temperature dependence of the mass-radius relation and by Crumpler et al., (in prep.) to search for the spatially-correlated signal ultra-light dark matter would impart on DA WDs.

\subsection{Spectral Component Separation with MADGICS: Marginalize, Don't Subtract!} \label{sec:sdss5-vac.0006}

In the era of big surveys, telescopes, and data, our inferences are becoming increasingly systematics-limited. Many of these systematics are of instrumental origin, for example fiber-to-fiber variation in the LSF introducing a ``fiber-dependence'' in measurements of stellar radial velocities, atmospheric parameters, or abundances \citep[e.g. ][]{2018ApJ_853_198N, Saydjari_2025_AJ}. And, the more ``downstream'' the measurement, the more of these systematics we have to control. For example, accurate and precise interstellar medium absorption line measurements (interstellar Na, diffuse interstellar bands) requires that the instrument, Earth's atmosphere, and star are all well-modeled.

At their core, data reduction pipelines solve a complicated component separation problem. For example these pipelines try to remove the pixel-pixel response function by dividing out flat field images, remove the contribution of the Earth's atmosphere by subtracting an estimate of ``sky'' spectra, and divide by an estimate of the absorption from the Earth's atmosphere. Unfortunately, these pipelines often treat each component separately, and simply subtract a ``best-estimate'' for each component sequentially. However, these ``best'' components can be difficult to estimate, especially when that component is of low SNR. Additionally, simply removing a ``point estimate'' does not provide ``downstream'' tasks any notion of the diversity of effects the removed component could have on the spectrum.

So why not fit these components simultaneously? Or even better, fit them simultaneously with data-driven priors built from vast amounts of survey data? Such a method, if scalable to large datasets, would represent a paradigm shift in spectral data-reduction. In this VAC, we provide the components resulting from a Bayesian component separation pipeline using a method called ``MADGICS,'' Marginalized Analytic Data space Gaussian Inference for Component Separation, developed by \cite{Saydjari_2023_ApJ}. In previous work, we applied this technique to all Gaia DR3 RVS spectra \citep{Saydjari_2023_ApJ}, overcoming massive stellar residual systematics that were dominating Gaia's 8621 \r{A} DIB catalog \citep{GaiaCollaboration_2023_AA}. We have also applied MADGICS to DESI spectra to estimate redshifts for Lyman-alpha emitters \citep{Uzsoy_2025_arXiv}. In this work, we release the results of applying this method to simultaneously model all of the components contributing to an extracted (1-dimensional, wavelength-calibrated) APOGEE spectrum.\footnote{\url {https://github.com/andrew-saydjari/apMADGICS.jl/}}

The core idea of MADGICS is to model each spectrum as a linear combination of components, $x_k$, with a prior on each of these components expressed as a pixel-pixel covariance matrix, $C_k$. By this, we mean that the components and their priors live in the ``data space'' of spectral wavelength bins ($8700 \times8700$ for APOGEE). Then, subject to the constraint that the components sum exactly to the data, we obtain a posterior for each of the components in the model contributing to the spectrum. This allows us to, for example, marginalize over variability in the instrument transfer function, telluric line depths, or stellar types. We can also carefully handle and mitigate effects of fiber-to-fiber LSF variations by applying the LSF models to our priors only when forward modeling the components, meaning when comparing to the data, but not on the inferred components.

In this data release, we provide the mean MADGICS components for several different models (sets of components) applied to all 2.6 million of the uniformly reduced APOGEE spectra (DR 17) at the ``visit'' level.\footnote{The ``visit'' level is a single observational epoch, which only combines contiguous, possibly dither-offset, exposures.} The goal of this VAC release is to provide precise and accurate measurements of DIBs in the APOGEE wavelength range, which will be described in detail in forthcoming work. The other goal of this VAC release is to demonstrate the broadly applicable improvements in spectral reductions obtained using MADGICS, and their impact on inferences of stellar radial velocities, parameters, and abundances. We have already demonstrated that the stellar radial velocities obtained from MADGICS improve the stellar radial velocity precision floor from 100 to 30 m s$^{-1}$ for APOGEE \citep{Saydjari_2025_AJ}. Improvements in the continuum normalization, sky line subtraction, and telluric absorption correction among others will be detailed in the forthcoming pipeline paper. We hope that this VAC release will also serve as a testbed so that we can work with maintainers on how to adapt stellar parameter and abundance inference pipelines to this new regime of spectral data reduction pipelines that deliver component posteriors.

The mean component spectra are provided for each component in a given model. We first model the spectrum as only consisting of the Earth's atmosphere, a star, and noise. The resulting components for this simplest model are: sky continuum, faint sky lines, star continuum, star lines, and residual. We also decompose the spectra with models that include an additional component for one of two diffuse interstellar bands (DIBs) modeled as a Gaussian centered near either 15273 \r{A} or 15672 \r{A}. 

In addition to the ``pure'' component spectra, we also provide an \texttt{apVisit\_v0} component, which is most analogous to the current APOGEE DRP outputs. This spectrum is obtained by first taking a MADGICS decomposition, and then subtracting off the mean component for the sky emission lines and sky continuum and dividing by the stellar continuum. While the resulting spectrum does still suffer from having simply subtracted ``best estimates,'' these estimates are improved by the simultaneous fit and priors on each component. However, one advantage of the \texttt{apVisit\_v0} component, especially for this initial MADGICS VAC release, is that priors on the stellar component only enter implicitly into the resulting spectrum.

\subsection{StarFlow stellar ages and masses for APOGEE DR19} \label{sec:sdss5-vac.0003}

Stellar age estimation is crucial to a complete understanding of the formation and evolution of the Milky Way. The StarFlow VAC builds upon the DistMass catalog from \citet{StoneMartinez_2024}, expanding the use of machine learning for stellar age inference with a more flexible and probabilistic approach. StarFlow provides age estimates for nearly 380,000 evolved stars in SDSS-V DR19, enabling large-scale chemo-dynamical studies across the Galactic disk, bulge, and halo. The catalog was constructed to deliver reliable ages with robust uncertainty estimates by leveraging recent advances in machine learning and asteroseismic calibration.

The StarFlow VAC applies a normalizing flow machine learning model to estimate stellar ages based on APOGEE DR19 stellar parameters and abundances. The model is trained on asteroseismic masses from APOKASC-3 and APO-K2 and captures the full joint distribution of $T_\mathrm{eff}$, $\log g$, [Fe/H], [C/Fe], [N/Fe], and age using a RealNVP architecture. This generative approach allows StarFlow to return full age posteriors for each star and model asymmetric uncertainties that reflect the breakdown of the [C/N]–age relation at old ages.

Uncertainties are incorporated directly during both training and inference via Monte Carlo sampling of the input errors. Validation against asteroseismic and cluster ages shows that typical uncertainties are $\sim$28\%, with 72\% of predicted ages within 1$\sigma$ of asteroseismic values. StarFlow agrees with existing catalogs (\citealp[e.g.][]{Mackereth_19, Leung_2023L}), with known differences at old ages driven by changes in the [C/N] - age relationship at low mass.

The catalog includes maximum likelihood ages, 1$\sigma$ uncertainties, and full posteriors, along with a “training space density” parameter that quantifies how well each star is represented in the training data. A cut at density  $> 3 \times 10^9$ defines the main catalog sample. StarFlow also provides mass estimates trained separately from ages, enabling isochrone-agnostic age inference. 

A full description of the model, validation, and catalog structure is provided in \cite{Stone-Martinez2025}. 

\subsection{Quasar spectral properties for DR19} \label{sec:sdss5-vac.0009}

Optical spectra of extragalactic sources from SDSS have significantly improved our understanding of quasar physics, enabling detailed large-scale statistical studies of SMBH mass, accretion properties, and the inner correlations among observables \citep[e.g.][]{Boroson&Green1992, Schneider_etal_2010, Shen_etal_2011, Dawson_2013_boss, Lyke2020, Wu&Shen2022}. To characterize the physical properties of the quasars in the SDSS DR19 sample, we first compiled a candidate quasar catalog using the {spAll\_v6\_1\_3} file that contains all spectra released in DR19. We restrict to science spectra (excluding sky and standard star spectra) that are either targeted by a BHM firstcarton, or have a pipeline spectral classification of CLASS=`QSO'. We then exclude objects classified as CLASS=`STAR' or `GALAXY' by the BOSS spectroscopic pipeline, unless the object is matched to a source in the DR16 quasar catalog within a $0\farcs2$ radius \citep{Lyke2020}. Our primary interest is broad-line quasars, and objects classified as `GALAXY' have a low yield of broad-line quasars ($\sim 2\%$) -- excluding these objects would only lose a few hundred genuine quasars. We further remove low-S/N spectra with SN\_MEDIAN\_ALL$<1$ (median S/N/pixel across the BOSS spectral range) if the object is targeted by a MWM firstcarton and is not in the DR16 quasar catalog. These low-SNR MWM targeted spectra have a low yield ($\sim 2\%$) of genuine quasars even if their pipeline spectral classification is `QSO'. Finally, repeat spectra for the same object are removed, and only the highest S/N spectrum is retained for visual inspection.

We then visually inspected the initial catalog of $\sim92$k unique quasar candidates to confirm the quasar classifications and correct erroneous pipeline redshifts, especially for low-S/N cases. We only inspected objects that are not already included in the DR16 quasar catalog ($\sim42$k). Among the list of visual inspection candidates, $\sim 60\%$ have SN\_MEDIAN\_ALL$>1$ \& CLASS=`QSO' and are targeted by a BHM firstcarton. This subset of objects ($\sim26$k) has a high yield of genuine quasars ($\sim 95\%$), and we use an automatic approach to reduce the amount of visual inspection required for this subset of high-fidelity quasar candidates. The details of this approach will be described in a forthcoming paper. In short, we fit the spectrum with a quasar template using the reported pipeline redshift after removing the continuum with a spline fit. Poor fits would imply incorrect quasar classification or redshift determination. We rank order the goodness of fit for this subset of high-fidelity candidates, and only visually inspect the first 5\% with the worst fitting quality. We find many of these objects are quasars with absorption lines, and we correct those with catastrophic pipeline redshift failures. Since we did not inspect all of these high-fidelity quasar candidates, it is possible that some misclassified quasars are included in our final VAC. For the other $\sim 40\%$ visual inspection objects ($\sim16$k), we distribute the visual inspection efforts across $\sim 10$ people with quasar expertise, and find an overall yield of $\sim 60\%$ for genuine quasars. Our final input quasar VAC includes 82,363 DR19 objects. 

We measure the quasar spectral properties following previous work \citep{Shen_etal_2011, Shen2019, Wu&Shen2022} using the publicly available code {\tt PyQSOFit} \citep{PyQSOFit} with priors for host galaxy decomposition. In detail, each spectrum is first corrected for the Milky Way extinction using the dust maps \citep{Schlegel1998, Schlafly2011} and then shifted to the rest frame. For quasars at redshift $z < 0.8$, we perform host-galaxy decomposition with priors to remove stellar contamination based on empirical priors \citep{Ren2024}. The continuum is modeled using a combination of a power law, a third-order polynomial, and FeII templates, fitted to line-free regions of the spectrum. Emission and absorption lines are then modeled with multiple Gaussian components, allowing for measurements of peak wavelength, line fluxes, luminosities, FWHM, equivalent widths (EW), and centroid wavelength. Bolometric luminosities are estimated from continuum luminosities at $1350{\rm\AA}$, $3000 {\rm\AA}$, and $5100 {\rm\AA}$ using empirical bolometric corrections in \cite{Richards_etal_2006}. SMBH masses were estimated through single-epoch virial relations \citep{Vestergaard&Peterson2006} based on the broad H$\beta$, MgII, and CIV emission lines and continuum luminosities. We also provide systemic redshift estimates for the quasar sample based on spectral fitting, which account for the average intrinsic velocity shifts of individual emission lines and their luminosity dependence from \cite{Shen_etal_2016b}. A detailed description of the columns included in this DR19 quasar value-added catalog is provided in the online documentation.\footnote{\url{https://github.com/QiaoyaWu/SDSSV_DR19_QSO_VAC}}


\section{Science Demonstrations \& tutorials} \label{sec:demos}
SDSS aims not only to produce large-scale datasets to answer key questions about our Universe but also to ensure that the data are regularly released to the astronomical (and broader) community with high-quality documentation. Together, the data and its documentation enable people around the world to use SDSS to perform scientific studies and outreach projects. New in DR19, we have created a GitHub repo, \url{https://github.com/sdss/dr19_tutorials} to centralize and distribute tutorials using DR19 data. The tutorials in this repository are written as Jupyter notebooks \citep{jupyter}, which allow for detailed, markdown formatted explanations to be included along with the Python code. In addition to GitHub, the tutorials are publicly available at our website at \url{https://www.sdss.org/dr19/tutorials}, and are available for users of SciServer\footnote{\url{https://apps.sciserver.org}}. Users are strongly encouraged to create (free) SciServer accounts; the Jupyter notebook server on SciServer has direct access to all DR19 data as well as the suite SDSS software tools.

\subsection{Survey-Wide Tutorials}

Since the Semaphore product is entirely new for DR19, a detailed tutorial covering these targeting flags has been made available. The \textit{sdss\_id} identifier is also new in DR19, therefore a tutorial on accessing and cross-matching using \textit{sdss\_id} is provided. There is also a tutorial highlighting the catalog cross-matching functionality of SciServer \citep{sciserver_xmatch}. A few tutorials include detailed instructions on using the \texttt{sdss\_access}\footnote{\url{https://sdss-access.readthedocs.io/en/latest/intro.html}} package.

\subsection{Milky Way Mapper : Demos and Tutorials  }

DR19 introduced a variety of new stellar parameter summary files, so tutorials have been made to help users navigate these files. These tutorials give some instruction on how to select a sample from a summary file, then use that sample to make scientifically interesting plots, e.g., \afe vs \feh\ relationships in different parts of the Galaxy.

There are also examples showing how to retrieve and plot both APOGEE and BOSS spectra, as well as one example showing how to find stars observed with both BOSS and APOGEE. For learners and instructors, there is also a beginners notebook available on using MWM data in the classroom.


\subsection{Black Hole Mapper : Demos and Tutorials}

A few tutorials exploring the BOSS pipeline are available, including an exploration of coadded spectra and a demonstration of alternative pipeline model and redshift outputs. An in-depth tutorial demonstrating the VAC associated with QSO properties (see section~\ref{sec:sdss5-vac.0009}) is also provided. Similar as for MWM, we also provide a beginners tutorial for using BHM data in the classroom.

\subsection{Local Volume Mapper : Demos and Tutorials 
}
\label{sec:lvmtutorial}

A tutorial based on the LVM data preview for the Helix Nebula can be found in the ``lvm\_dr19\_helix\_nebula\_tutorial'' directory at: \url{https://github.com/sdss/dr19_tutorials}. It requires the lvmSFrame-00004297.fits file as input, which can be found at: \url{https://dr19.sdss.org/sas/dr19/spectro/lvm/redux/1.1.1/0011XX/11111/60191/}. This introductory tutorial focuses on simple interactions with the RSS data file, the lvmSFrame, which is the final output of the LVM data reduction pipeline. It teaches the user to:

\begin{enumerate}
    \item Access the fits file and display the list of Header Data Units
    \item From the fits file, extract the slitmap table; and from this table, extract the data for fibers without known problems and load it into a data frame
    \item Display a table with the slitmap
    \item Make a map of fibers without known issues with fiber-id labels
    \item Correct the wavelength array for a Doppler shift
    \item Obtain the index within the flux array corresponding to a fiber of interest
    \item For a fiber with a given ID, plot the reduced spectrum, including the flux errors
    \item Zoom into spectral regions around H$\alpha$ and H$\beta$ in the observed and rest frames, note the presence of bad columns
    \item Note that some fibers capture bright stars
    \item Make maps of the emission within spectral windows around emission lines
    \item Make a composite RGB (red/green/blue) map that combines fluxes from three different spectral windows, selected to capture emission lines
    \item Fit a single Gaussian to an emission line to obtain its flux
    \item Make an emission-line ratio map based on Gaussian fits to the lines
\end{enumerate}

%

\section{Summary and Outlook} \label{sec:Summary}
The fifth generation of SDSS (SDSS-V) is pioneering the first panoptic, dual-hemisphere
spectroscopic optical and near-infrared survey of stars, galaxies, quasars, and other celestial objects. In this paper, we have described the nineteenth data release (DR19) of the Sloan Digital Sky Survey. Following targeting information and the release of a modest number of ($\sim$25000) spectra in DR18, SDSS DR19 is the second and most substantial from SDSS-V. To describe SDSS-V DR19, we first outline the scientific aims and objectives of the three core mapper programs (MWM, BHM, and LVM, for more information on these mappers consult section~\ref{sec:Scienceobjs}). After, we describe the targeting procedures for each mapper (section~\ref{sec:target}) and the data reduction and analysis pipelines (section~\ref{sec:pipeline}) used to convert the spectroscopic observations into science-ready products. With the data reduction and analysis pipelines in hand, we produce science-ready spectra and scientific products for a diverse set of celestial objects that are described below. 

In total, SDSS DR19 contains :
\begin{itemize}
    \item optical BOSS spectra and near-infrared APOGEE spectra for 479,081 and 390,676 stars, respectively, across the Hertzsprung-Russell diagram and local group (mostly in the Milky Way and its clusters and satellites), which can be used for a variety of stellar and Galactic astrophysics through the Milky Way Mapper (Section~\ref{sec:MWMscope})
    \item BOSS spectra of 318,123 galaxies and quasars/AGN over a variety of fields and with a wide range of  time baselines (Section~\ref{sec:BHMscope})
    \item A first preview of IFU (tile) data of the Helix Nebula (NGC~7293) from the LVM program (section~\ref{sec:lvmpreview})
    \item Nine Value Added Catalogues (Section~\ref{sec:VAC}) based on DR19 MWM and BHM spectra, offering additional data analysis products prepared by the SDSS-V science teams
    \item A new visualisation app, Zora, to inspect and analyse BHM and MWM spectra and catalogs (Section~\ref{sec:data.dataviz})
    \item Several tutorials (Section~\ref{sec:demos}), available in both GitHub and SciServer, to help users explore the data products in this release
\end{itemize}

While the newest spectra in DR19 comes only from APO telescope in the Northern hemisphere (along with legacy data from both the Northern and Southern hemispheres), DR20 will contain new data with the updated FPS system from both hemispheres. DR20 is also anticipated to include updates from the targeting catalogs and data pipelines and a much larger sample from the new LVM program. 

\section*{Acknowledgments}
We would like to thank the Department of Astronomy at New Mexico State University for
their hospitality during “DocuChili” in June 2024, and the eScience Institute at the University of Washington for hosting us in December 2025 for “DocuOrca“. These events were the main venue for writing the documentation
for DR19 (including this paper), and were organized by Anne-Marie Weijmans and José Sánchez-Gallego. DocuChili and DocuOrca were attended by Scott Anderson, Mike Blanton, Joel Brownstein, Andy Casey, Brian Cherinka, John Donor, Tom Dwelly, Pramod Gupta, Jennifer Johnson, Sean Morrison, Jordan Raddick, José Sánchez-Gallego, Conor Sayres, Anne-Marie Weijmans, and Adam Wheeler, and remotely by Joleen Carlberg, Niall Deacon, Emily Griffith, Keith Hawkins, Juna Kollmeier, Ilija Medan, Riley Thai, and Ani Thakar. 

Funding for the Sloan Digital Sky Survey V has been provided by the Alfred P. Sloan Foundation, the Heising-Simons Foundation, the National Science Foundation, and the Participating Institutions. SDSS acknowledges support and resources from the Center for High-Performance Computing at the University of Utah. SDSS telescopes are located at Apache Point Observatory, funded by the Astrophysical Research Consortium and operated by New Mexico State University, and at Las Campanas Observatory, operated by the Carnegie Institution for Science. The SDSS web site is \url{www.sdss.org}.

SDSS is managed by the Astrophysical Research Consortium for the Participating Institutions of the SDSS Collaboration, including Caltech, The Carnegie Institution for Science, Chilean National Time Allocation Committee (CNTAC) ratified researchers, The Flatiron Institute, the Gotham Participation Group, Harvard University, Heidelberg University, The Johns Hopkins University, L’Ecole polytechnique f\'{e}d\'{e}rale de Lausanne (EPFL), Leibniz-Institut f\"{u}r Astrophysik Potsdam (AIP), Max-Planck-Institut f\"{u}r Astronomie (MPIA Heidelberg), Max-Planck-Institut f\"{u}r Extraterrestrische Physik (MPE), Nanjing University, National Astronomical Observatories of China (NAOC), New Mexico State University, The Ohio State University, Pennsylvania State University, Smithsonian Astrophysical Observatory, Space Telescope Science Institute (STScI), the Stellar Astrophysics Participation Group, Universidad Nacional Aut\'{o}noma de M\'{e}xico, University of Arizona, University of Colorado Boulder, University of Illinois at Urbana-Champaign, University of Toronto, University of Utah, University of Virginia, Yale University, and Yunnan University.

This work is based on data from eROSITA, the primary instrument aboard SRG, a joint Russian-German science mission supported by the Russian Space Agency (Roskosmos), in the interests of the Russian Academy of Sciences represented by its Space Research Institute (IKI), and the Deutsches Zentrum f\"ur Luft- und Raumfahrt (DLR).
The SRG spacecraft was built by Lavochkin Association (NPOL) and its subcontractors, and is operated by NPOL with support from the Max Planck Institute for Extraterrestrial Physics (MPE).
The development and construction of the eROSITA X-ray instrument was led by MPE, with contributions from the Dr. Karl Remeis Observatory Bamberg \& ECAP (FAU Erlangen-N\"urnberg), the University of Hamburg Observatory, the Leibniz Institute for Astrophysics Potsdam (AIP), and the Institute for Astronomy and Astrophysics of the University of T\"ubingen, with the support of DLR and the Max Planck Society.
The Argelander Institute for Astronomy of the University of Bonn and the Ludwig Maximilians Universit\"at Munich also participated in the science preparation for eROSITA.
The eROSITA data shown here were processed using the eSASS software system developed by the German eROSITA consortium.

\begin{contribution}
All authors are listed alphabetically according to the  SDSS-V publication policy. 
\end{contribution}

\appendix
\section{Newly released targeting cartons}
\label{sec:appendix}
In the sections below we describe the target cartons that are newly released in SDSS DR19.
The majority of the new target cartons are associated with the initial year of SDSS-V plate operations at APO (targeting generation ``v0.plates'').
In addition, we describe a small number of new (FPS-era) target cartons which fall outside the ``v0.5.3'' targeting generation that was previously presented by \citet{almeida2023}.


\subsection{MWM Plate Targets}
\label{sec:mwm_cartons_appendix}

The MWM plate cartons are described in detail in the upcoming MWM target selection paper (De Lee et al., in prep) and on the SDSS DR 19 website: \url{https://www.sdss.org/dr19/mwm/programs/cartons/}. In Table \ref{tab:mwm_plate_cartons}, the new plate program cartons that were not previously listed in \citet{almeida2023} are presented along with a short statement about their selection criteria, what instrument was used, and the number of targets that satisfy the carton selection function in the targeting database. The number of targets ultimately observed for each carton will be smaller than this value.

\begin{deluxetable*}{llll}
\label{tab:mwm_plate_cartons}
\tablecaption{New Milky Way Mapper Plate Program Cartons}
\tablehead{\colhead{\multirow{2}{*}{Carton Name}} & \colhead{\multirow{2}{*}{Selection Summary\tablenotemark{1}}} &
\colhead{\multirow{2}{*}{Instrument}} & \colhead{Available} \\ [-0.2cm]
\colhead{} & \colhead{} & \colhead{} & \colhead{Targets\tablenotemark{2}}}

\startdata
	mwm\_cb\_300pc & Compact binaries within 300pc from Gaia+GALEX & BOTH & 89492 \\
    mwm\_cb\_cvcandidates & Compact binaries from literature  & BOTH & 5167 \\
    mwm\_cb\_gaiagalex & Compact binaries from Gaia+GALEX & BOTH & 261570 \\
    mwm\_gg\_core & Luminous Giants across the sky based on G-H color cuts & APOGEE & 5459267 \\
    mwm\_halo\_bb & A sample of metal-poor giants selected based on optical and infrafred photometry & BOSS & 646940 \\
    mwm\_halo\_sm & A sample of metal-poor stars selected based on SkyMapper Ca K photometry & BOSS & 138298 \\
    mwm\_planet\_tess & TESS TOI and 2 minute Cadence Follow-up  & APOGEE & 177494 \\
    mwm\_rv\_long\_bplates & Short Exposure Plate Radial Velocity Monitoring -- Long Temporal Baseline & APOGEE & 6397 \\
    mwm\_rv\_long-bplates & Short Exposure Plate Radial Velocity Monitoring -- Long Temporal Baseline & APOGEE & 4370 \\
    mwm\_rv\_long-fps & Radial Velocity Monitoring -- Long Temporal Baseline & APOGEE & 111233 \\
    mwm\_rv\_long-rm & RM Plate Radial Velocity Monitoring -- Long Temporal Baseline & APOGEE & 1344 \\
    mwm\_rv\_short-bplates & Short Exposure Plate Radial Velocity Monitoring -- Short Temporal Baseline & APOGEE & 50567 \\
    mwm\_rv\_short-fps & Radial Velocity Monitoring -- Short Temporal Baseline & APOGEE & 6283604 \\
    mwm\_rv\_short-rm & RM Plate Radial Velocity Monitoring -- Short Temporal Baseline & APOGEE & 3855 \\
    mwm\_snc\_100pc & Volume limited census of the solar neighbourhood & BOTH & 322222 \\
    mwm\_snc\_250pc & Volume limited census of the solar neighbourhood, extension to earlier types & BOTH & 417431 \\
    mwm\_tessrgb\_core & TESS observed red giant asteroseismology follow-up & APOGEE & 1004714 \\
    mwm\_wd\_core & White dwarfs selected from Gaia + SDSS-V \citep{GentileFusillo2019} & BOSS & 192534 \\
    mwm\_yso\_cluster & Candidates identified through phase space clustering, targeted with APOGEE & BOTH & 45461 \\
    mwm\_yso\_cmz & Candidates located in Central Molecular Zone+inner Galactic disk & APOGEE & 1282 \\
    mwm\_yso\_ob & Selection of the hottest and youngest OB stars & BOTH & 8670 \\
    mwm\_yso\_s1 & Optically bright candidates identified through infrared excess & BOTH & 28832 \\
    mwm\_yso\_s2 & Optically faint candidates identified through infrared excess & APOGEE & 11086 \\
    mwm\_yso\_s2-5 & Candidates found towards areas of high nebulosity & APOGEE & 1112 \\
    mwm\_yso\_s3 & Low mass candidates identified through Gaia variability & BOTH & 52691 \\
\enddata
\tablecomments{\tablenotetext{1}{See online documentation for complete selection details: \url{https://www.sdss.org/dr19/mwm/programs/cartons/}}
\tablenotetext{2}{\textbf{Available targets} is the number of targets that satisfy the carton selection function in the targeting database. The number of targets ultimately observed for each carton will be smaller than this value.}}
\end{deluxetable*}

\include{sections/mwm_cartons/mwm_cartons_appendix}

\subsection{BHM Plate Targets}
\label{sec:bhm_cartons_appendix}
Here we provide details of the BHM targeting cartons that were used during
the SDSS-V plate program. The five SPIDERS/eFEDS cartons used during SDSS-V plate operations (\texttt{bhm\_spiders\_agn-efeds},
\texttt{bhm\_spiders\_clusters-efeds-ls-redmapper},
\texttt{bhm\_spiders\_clusters-efeds-sdss-redmapper},
\texttt{bhm\_spiders\_clusters-efeds-hsc-redmapper} and \texttt{bhm\_spiders\_clusters-efeds-erosita})
were previously described by \citet{almeida2023} (their appendix B), and so to avoid duplication we do not list them here.

For each listed target carton, we provide the following information:
Software versioning information, a short summary of the content of the target carton,
a description of the carton selection criteria, target meta-data (``priority'' and ``cadence''),
links to target selection code, and finally, the number of targets which pass all of the carton selection criteria.

\include{sections/bhm_cartons/bhm_target_cartons_v0}

\subsection{Plate Calibration Targets  }
\label{sec:plate_calib}
During the plate operations phase we used early versions of cartons containing spectrophotometric
standard stars and sky locations suitable for the placement of sky fibers.

\noindent
\textit{BOSS and APOGEE Skies}:
During the plate operations phase we used two cartons to describe sky locations:
\texttt{ops\_sky\_boss} and \texttt{ops\_sky\_apogee} (both plan=0.1.0, tag=0.1.0).
These sky locations were selected in an algorithmically similar way to the later FPS sky location cartons, but relied on object
avoidance rules which were derived from the set of external catalogs available in the SDSS-V database at that time
(specifically the Gaia DR1, LegacySurvey DR8, Tycho2, and 2MASS catalogs).

\noindent
\textit{BOSS Spectrophotometric standards:}
During the plate operations phase we defined four (overlapping) cartons of BOSS spectrophotometric standard stars.
Three of these cartons were intended to be as broad as possible, including standards suited for dark or bright time observations:
\texttt{ops\_std\_boss}, \texttt{ops\_std\_boss-red} (both plan=0.1.0, tag=0.1.0), and \texttt{ops\_std\_boss\_tic} (plan=0.1.3, tag=0.1.4).
An additional carton provided fainter standards only suitable for use in dark time: \texttt{ops\_std\_eboss} (both plan=0.1.0, tag=0.1.0).
These standard star cartons are selected in an identical way to their later FPS era namesakes (described by \citealt{almeida2023}),
except that they rely on the plate version of the catalog crossmatch table, as well as minor changes to meta-data.

\noindent
\textit{APOGEE Telluric standards:}
In order to remove telluric lines from the APOGEE spectra, broad-lined stellar spectra are observed. In the earliest days of the SDSS-V plate program, legacy scripts following \citep{Zasowski_2017_apogee2targeting} were used. 
From platerun 2021.01.b.mwm-bhh (MWM\_04) onward we selected APOGEE standards from the \texttt{ops\_apogee\_stds carton}.
These stars were selected from the 2MASS-PSC catalogue, then they were 
filtered selecting magnitudes 7$<H<$11, colors -0.25$<(J-K)<$0.5, magnitude uncertainties $J_{\rm {err}}<0.1$, $H_{\rm {err}}<0.1$,
$K_{\rm {err}}<0.1$, read flag equal to ``1'' or ``2'' for each JHK
band, photometric quality flag equal to ``A'' or ``B'' for each JHK
band, extended source contamination flag equal to ``0'', confusion
flag equal to ``000'', extended key ID equal to $Null$, and farther than
6 arcseconds from the nearest neighbor star. After this pool of
candidates is created, the sky is divided into an nside = 128 HEALPIX 
skymap, and finally for each healpix the 5 sources with the smallest ($J-K$) color values are selected and saved for the output carton.

\subsection{Additional FPS-era BHM target cartons}
For completeness, we describe here a single BHM target carton (\texttt{bhm\_csc\_apogee}, plan 0.5.0)
that predates the ``v0.5.3'' targeting generation that was released in SDSS DR18, but that was used briefly for initial
FPS commissioning observations (as part of targeting generation ``v0.5.epsilon-7-core-0'') before being
superseded by an updated version.


\hypertarget{bhm_csc_apogee_plan0.5.0}{%
\subsubsection{bhm\_csc\_apogee}\label{bhm_csc_apogee_plan0.5.0}}

\noindent\textbf{target\_selection plan:} 0.5.0

\noindent\textbf{target\_selection tag:}
\href{https://github.com/sdss/target_selection/tree/0.3.0/}{0.3.0}

\noindent\textbf{Summary:} X-ray sources from the CSC2.0 source catalog with
NIR counterparts in 2MASS PSC

\noindent\textbf{Simplified description of selection criteria:} Starting from the
parent catalog of CSC2.0 sources associated with optical/IR counterparts
(\texttt{bhm\_csc}). Select entries satisfying the following criteria: i) NIR
counterpart is from the 2MASS catalog, ii) 2MASS H-band magnitude
measurement is in the estimated range for SDSS-V:
$10.0 < H < 15.0$.

\noindent\textbf{Target priority options:} 4000

\noindent\textbf{Cadence options:} bhm\_csc\_apogee\_3x1

\noindent\textbf{Implementation:}
\href{https://github.com/sdss/target_selection/blob/0.3.0/python/target_selection/cartons/bhm_csc.py}{bhm\_csc.py}

\noindent\textbf{Number of targets:} 13437

\subsection{Additional FPS-era MWM target cartons}
There were several FPS cartons that were not part of DR 18  \citep{almeida2023}. The newly released cartons are in  Table \ref{tab:mwm_fps_cartons} which are in the same format as described in Section \ref{sec:mwm_cartons_appendix}. Some of these cartons were not in previous releases because they were used for technical work. For instance, the \texttt{manual\_offset*} cartons where created as part of the learning how to do offset observations with BOSS. Other cartons like \texttt{mwm\_tess\_ob} just had placeholder stars, so that we could run robostrategy keeping fibers open for when the actual targets were delivered in the carton \texttt{manual\_mwm\_tess\_ob}. These cartons like the ones in Section \ref{sec:mwm_cartons_appendix} have more information available in the upcoming target selection paper and on the DR19 website.

\begin{deluxetable*}{llll}
\label{tab:mwm_fps_cartons}
\tablecaption{Milky Way Mapper FPS Cartons not in DR18}
\tablehead{\colhead{\multirow{2}{*}{Carton Name}} & \colhead{\multirow{2}{*}{Selection Summary\tablenotemark{1}}} &
\colhead{\multirow{2}{*}{Instrument}} & \colhead{Available} \\ [-0.2cm]
\colhead{} & \colhead{} & \colhead{} & \colhead{Targets\tablenotemark{2}}}

\startdata
	manual\_offset\_mwmhalo\_off00 & Offset Test Carton & BOSS & 1330 \\
    manual\_offset\_mwmhalo\_off05 & Offset Test Carton & BOSS & 1330 \\
    manual\_offset\_mwmhalo\_off10 & Offset Test Carton & BOSS & 1330 \\
    manual\_offset\_mwmhalo\_off20 & Offset Test Carton & BOSS & 1330 \\
    manual\_offset\_mwmhalo\_off30 & Offset Test Carton & BOSS & 1330 \\
    manual\_offset\_mwmhalo\_offa & Offset Test Carton & BOSS & 1330 \\
    manual\_offset\_mwmhalo\_offb & Offset Test Carton & BOSS & 1330 \\
    manual\_validation\_apogee & Calibration targets selected from a variety of validation catalogs &APOGEE& 24518 \\
    manual\_validation\_boss & Calibration targets selected from a variety of validation catalogs &BOSS& 130218 \\
    mwm\_halo\_bb\_boss & Halo metal-poor star selection using \citet{Schlaufman2014} & BOSS & 640058 \\
    mwm\_halo\_sm\_boss & Halo metal-poor star selection using Skymapper \citep{Chiti2021}  & BOSS & 138298 \\
    mwm\_tess\_ob & Placeholder targets for manual\_mwm\_tess\_ob & APOGEE & 767 \\
    mwm\_yso\_ob\_apogee & Selection of the hottest and youngest OB stars & APOGEE & 8670 \\
    mwm\_yso\_ob\_boss & Selection of the hottest and youngest OB stars & BOSS & 8670 \\
\enddata
\tablecomments{\tablenotetext{1}{See online documentation for complete selection details: \url{https://www.sdss.org/dr19/mwm/programs/cartons/}}
\tablenotetext{2}{\textbf{Available targets} is the number of targets that satisfy the carton selection function in the targeting database. The number of targets ultimately observed for each carton will be smaller than this value.}}
\end{deluxetable*}

\subsection{Additional FPS-era Calibration target cartons} 
A small number of FPS sky fiber location cartons
predate the ``v0.5.3'' targeting generation previously released in SDSS DR18, but were used briefly for initial
FPS commissioning observations (as part of targeting generation ``v0.5.epsilon-7-core-0'').
Those cartons are as follows: \texttt{ops\_sky\_boss\_best}, \texttt{ops\_sky\_boss\_good},
\texttt{ops\_sky\_apogee\_best}, and \texttt{ops\_sky\_apogee\_good} (all have plan=0.5.8, tag=0.3.8).
These cartons are identical to the sky cartons described in \citet{almeida2023}, except for very minor differences in meta-data.

In addition, a single APOGEE calibration carton (\texttt{ops\_std\_apogee}, plan 0.5.22) was introduced after targeting generation ``v0.5.3''. This carton is described in section \ref{sec:plate_calib}. The only significant change from the plate program to the FPS-era is how the APOGEE standard stars are distributed across the FPS focal plane. This distribution is described in more detail in \citet{Blanton2025}.

\bibliography{finalbib}
\bibliographystyle{aasjournal}

\end{document}

%% file: sections/MOS_Targeting/semaphore.tex
In prior data releases, targeting information was provided to end users via an ever-increasing number of bitmask columns in the various data products. However, with the large number (251 as of DR19) of target cartons in SDSS-V, this method proved untenable. Therefore, the decision was made to replace the SDSS-V targeting bitmasks (like those used in DR18) with a byte array, where a bit corresponds to a single version of a carton. The mapping and the code to encode and decode these targeting flags are contained in the SDSS Sempahore product\footnote{\url{https://github.com/sdss/semaphore/tree/0.2.4}}, with a SDSS carton-to-bit version (\acronym{SDSSC2BV}) $= 1$ in DR19.

Each bit also has other metadata associated with it that can be used for select targets of interest:
\begin{itemize}
    \item label: carton name + version
    \item carton\_pk: carton table primary key value in the SDSS-V Targeting database
    \item program: targeting program of the carton
    \item version: targeting version used to build the carton
    \item name: the name of the carton as given in the SDSS-V Targeting database
    \item mapper: the SDSS-V Mapper associated with the carton: ops, bhm, mwm, open
    \item alt\_program: a more human readable and uniform version of the carton program
    \item alt\_name: a more human readable and uniform version of the carton name
\end{itemize}
Any of these (except version) can be used to easily filter the spectra in the Astra data analysis (see also section~\ref{astra}) or BOSS data reduction pipeline (\texttt{idlspec2d}, see also section~\ref{boss_drp}) summary files as the SDSS5\_target\_flags column. The DR19 APOGEE summary files do not include these flags, but can be filtered by joining to the Astra summary files. Guidance and use examples can be found in the GitHub repository or in the DR19 tutorial: \url{https://www.sdss.org/dr19/tutorials/semaphore/} (see also section~\ref{sec:demos}).

%% file: sections/MOS_Targeting/sdss_id.tex
The SDSS-V survey has spanned a long range in time and has been in progress during several major data releases of external astronomical surveys. Namely, \textit{Gaia} eDR3 was released on December 3rd, 2020 (and its successor DR3, released June 13, 2022) with a record number of parallaxes and magnitudes. With the release of these data and a desire to target sources based on the most up-to-date information, the internal SDSS crossmatch was recomputed incorporating these new catalogs (described in detail in \S\ref{sec:mos_target_product}). Each crossmatch had different ranked priorities for surveys, unique identifiers (called \textsc{catalogid}s), and matched objects differently in small ways. SDSS-V has targeted observations based on three different crossmatches (with version names: \textsc{0.1.0}, \textsc{0.5.0}, and \textsc{1.0.0}) and has matched in its legacy data (before SDSS-V) within this schema. In DR19, we only make observations public which were targeted based on versions \textsc{0.1.0} and \textsc{0.5.0} or are pre-SDSS-V legacy spectra, with observations based on \textsc{1.0.0} to come in the next data release.

These new crossmatches were necessary, but complicate the idea of an ``object" observed in SDSS-V. For example, a star may be primarily sourced in the \textsc{0.1.0} crossmatch based on its membership in the TESS Input Catalog, but then is observed later in the survey as a \textit{Gaia} eDR3 source in the \textsc{1.0.0} crossmatch. In order to tie these crossmatches together, the \textsc{sdss\_id} was created, which matches the unique crossmatch \textsc{catalogid}s together between versions. 

The meta-crossmatch table ({\tt mos\_sdss\_id\_stacked}) was completed incrementally via a series of left joins. Beginning with the \textsc{catalogid}s in \textsc{0.1.0}, objects are matched to those in \textsc{0.5.0} which can both be attributed to the prior's source catalog. \textsc{catalogid}s which do not match between versions are appended to the table. Then the same process is completed matching \textsc{catalogid}s from \textsc{0.5.0} to \textsc{1.0.0}, with mismatches appended to the table. The final result is a table with three columns related to each crossmatch \textsc{catalogid21}, \textsc{catalogid25}, and \textsc{catalogid31} (relating to \textsc{0.1.0}, \textsc{0.5.0}, and \textsc{1.0.0}, respectively). Each row is thus a unique combination of \textsc{catalogid}s and is given an incremental \textsc{sdss\_id}. This allows more objects to be added to {\tt mos\_sdss\_id\_stacked} as new catalogs are added into the crossmatch, such as the legacy SDSS data and targets-of-opportunity.

Each row of \textsc{catalogid}s is a unique triplet in {\tt mos\_sdss\_id\_stacked} but, importantly, the relation between \textsc{catalogid} and \textsc{sdss\_id} is not necessarily unique. Issues within a crossmatch or between versions can lead to \textsc{catalogid}s matching to multiple new sources (e.g., a TESS Input Catalog source becoming a resolved binary in \textit{Gaia} DR3). In general, these issues are rare but can occur. To account for this, we provide {\tt mos\_sdss\_id\_flat} which is a pivoted version of the {\tt mos\_sdss\_id\_stacked} where each row contains a unique \textsc{sdss\_id}/\textsc{catalogid} pair. When a \textsc{catalogid} matches to multiple \textsc{sdss\_id}s, we prioritize the lowest numerical value as the chosen \textsc{sdss\_id}. This information is stored in the \textsc{rank} column of {\tt mos\_sdss\_id\_flat} where the \textsc{rank} = 1 association is preferred.

%% file: sections/pipelines/Overview.tex
Each generation of SDSS has provided spectroscopic pipelines for the instruments that were used to observe the targets selected for each program, starting with SDSS-I and the original SDSS spectrographs, with optical plates containing 640 fibers~\citep{York.2000AJ....120.1579Y}. The pipelines for each mapper are usually split into a data reduction pipeline (DRP) and a data analysis pipeline (DAP). 

The original BOSS spectroscopic pipeline was designed in the \texttt{IDL} programming language and reduced an entire plate of 15 minute exposures observed by each modified julian date (MJD). The \texttt{idlspec2d} code had stages to extract, combine, and calibrate the SDSS spectra of individual objects from the two-dimensional raw CCD data, followed by the \texttt{idlspec1d} analysis pipeline which performed best fits using principal component analysis to derive the model spectra including object classifications and precise redshifts.  The SDSS spectrograph's reduced spectra, identified by versions with \texttt{run2d} (26, 103, 104) have not been re-processed since their finalization in DR8, and form a legacy dataset from which the full cosmological sample is derived. At the beginning of SDSS-III, with the introduction of the 1000 fiber BOSS spectrograph, the \texttt{idlspec2d/1d} pipelines were continuously adapted and improved to yield a better quality reduction, including an effort to incorporate manual inspections performed by members of the collaborations~\citep{Bolton2012,Dawson_2013_boss}.  This practice continued into SDSS-IV with the eBOSS survey moving to higher redshifts and lower signal-to noise~\citep{Ahumada2020_sdss_dr16}.  The final SDSS-IV plates are now also legacy reductions, identified by the version with \texttt{run2d} v5\_13\_2, since they also have not been re-reduced by the BHM data reduction pipeline, which is optimized for SDSS-V plates, during the limited plate program; and FPS fields as discussed in Section \ref{sec.fps}.  The new observations for DR19, which are identified by the version with \texttt{run2d} v6\_1\_3, are discussed in \ref{boss_drp}; and the complete optical cosmological data set consists of the combined DR8 and DR17 legacy reductions alongside the SDSS-V observations, including \texttt{run2d} in the set (26, 103, 104, v5\_13\_2, v6\_1\_3).

New instruments were introduced in SDSS-III, including the APOGEE infrared spectrograph; and in SDSS-IV, the MaNGA integral field unit (IFU) which took data from the BOSS spectrograph and required a new MaNGA data reduction pipeline~\citep[MaNGA DRP;][]{Law.2016AJ....152...83L,Law.2021AJ....161...52L} which repurposed many of the core \texttt{idlspec2d} algorithms, followed by a distinct MaNGA data analysis pipeline, which was the first python programming language pipeline developed for SDSS-IV~\citep[MaNGA DAP,][]{Westfall.2019AJ....158..231W}.  Similarly to the optical pipeline which is a superset of legacy and SDSS-V reductions, the APOGEE data reduction pipeline \citep[DRP,][]{2015AJ....150..173N} has not re-reduced any of the final SDSS-IV plates, which are identified by the \texttt{apred\_vers} `dr17'.  The new observations for DR19, which are identified by the version with \texttt{apred\_vers} 1.3, are discussed in \ref{apogee_drp}; and the complete infrared spectroscopic data set consists of the DR17 legacy reductions alongside the SDSS-V observations, including \texttt{apred\_vers} in the set (dr17, 1.3).

The original APOGEE Stellar Parameters and Abundances Pipeline~\citep[ASPCAP;][]{ASPCAP.2016AJ....151..144G} developed during SDSS-III and improved during SDSS-IV has been factored into the SDSS-V astra data analysis pipeline which includes numerous other pipelines, as discussed in \ref{astra}.

%% file: sections/pipelines/BOSS_DRP.tex
\label{boss_drp}
The optical BOSS data released in DR19 are processed with version \texttt{v6\_1\_3} of the BOSS pipeline software \texttt{idlspec2d} \citep[][Morrison et al.\,in prep]{Bolton2012,Dawson_2013_boss}. The implementation of the FPS, and its associated science requirements, necessitated significant changes to the pipeline from the DR18 version \texttt{v6\_0\_4}. All SDSS-V versions of \texttt{idlspec2d} are available for download from the SDSS GitHub, with the version described here available at \url{https://github.com/sdss/idlspec2d/releases/tag/v6_1_3}. 

One of the most significant changes involved the modification of the exposure coadding schema and code. This update is centered on two separate changes: the coadding of the exposures within an individual coadd, and the selection of exposures to coadd. Within an individual coadd, the schema has changed to a 2 step coadd, where initially the matching red and blue frames are combined with an outlier rejection. This is then followed by a catalogid based coadding of the exposures. This differs from the DR18 and earlier versions, where all exposures (red and blue) were combined in a single outlier rejection step on a fiberid baseline. The exposures selected to be included in a given coadd  has also changed in DR19, where the baseline products only combine exposures taken on a single MJD, rather than over multiple nights. This is supplemented by field-epoch coadds (called `epoch' in the pipeline outputs), which are similar to the DR18 and earlier versions, with RM plates being combined with a max epoch of 3 days, general (non-RM) plates being coadded within a single plugging, and FPS fields being combined using their FPS field epoch cadence requirements. For some BHM targets, as selected by a FirstCarton entry matching `*spiders*' or `*bhm\_gua*' or `*bhm\_csc*' or `*mwm\_erosita*' or `*bhm\_colr\_galaxies*', an additional target level coadd is produced for all exposures of these targets, regardless of the MJD. These are called `allepoch' by the pipeline, and are analogous to the DR18 `eFeds' custom coadds, but only include the selected targets and not all targets on the designs. These were built using the red+blue exposure level coadds from the daily pipeline as a basis and processed through the remaining pipeline as pseudo-fields with MJDs of the last exposure being used as the coadd MJD. 

Spectral fluxes in BOSS spectra are affected by dust and reddening. As such, the addition of MWM targeted data in the Galactic Plane necessitated the replacement of the SFD dust extinction maps \citep{Schlegel1998} with more sophisticated maps of the dust in the Milky Way. In DR19, \citet{Green2015} maps were added using \texttt{Dustmaps}\footnote{\url{https://dustmaps.readthedocs.io/en/latest/}} \citep{Green2018} for targets with galactic latitude $<15$\degree.

The DR19 version of the pipeline had a number of other small changes, including additional metadata, monitoring of spectrophotometric accuracy, the addition of the new SDSS-V targeting flags (\autoref{sec:semaphore}), and SDSS-IDs (\autoref{sec:sdssid}). Additionally, since the \texttt{pyXCSAO} \citep{marina_kounkel_2022_6998993} fits are now included (since DR18), the Elodie model \citep{Prugniel2001} fit parameters have been depreciated from the final spAll summary files, but are still available in the intermediate produces and used in the classification of the spectra. 

%% file: sections/data/access.tex
There are a number of simple and advanced methods to access the SDSS DR19 data products, ranging from single file and bulk download methods of data products from the Science Archive Server (SAS), as described in Section \ref{sec:data.sas}, to the catalog level search interface provided by the Catalog Archive Server (CAS), described in Section \ref{sec:data.cas}; and SciServer Compute, which supports a programmatic interface with Jupyter notebooks, described in Section \ref{sec:data.sciserver}. All data products available on both the SAS and the SciServer Compute environment contain a hierarchical file structure with individual file species documented by their respective data models, as described in Section \ref{sec:data.datamodel}.  New for SDSS-V is the introduction of the Zora webapp, which replaces the previous Science Archive Webapp for DR19 BHM and MWM spectra, as described in Section \ref{sec:data.dataviz}. Each of these data access options are summarized on the SDSS website\footnote{\url{https://www.sdss.org/dr19/data_access/}}.

%% file: sections/data/sas.tex
For those users interested in the reduced images, reduced spectra, and the full output of the analysis pipelines, we recommend that they access these data products through the SDSS Science Archive Server (SAS, \url{https://data.sdss.org/sas/}). All of these data products were derived through the official SDSS data reduction pipelines (see section~\ref{sec:pipeline}, which are also publicly available through the SDSS GitHub Organization\footnote{\url{https://github.com/sdss/}}, with individual software documentation\footnote{\url{https://www.sdss.org/dr19/software/}}. The SAS and SciServer Compute also contain the VACs that science team members have contributed to the data releases (see Section \ref{sec:VAC}), as well as the raw and intermediate data products. All files available through the SAS have a data model that explains their content (\url{https://data.sdss.org/datamodel/}). Data products can be downloaded from the SAS either directly through browsing, or by using methods such as wget, rsync and Globus Online\footnote{\url{https://www.sdss.org/dr19/data_access/bulk}}.

%% file: sections/data/cas.tex
The Catalog Archive Server (CAS) for DR19 retains the same collaborative science platform, SciServer \citep{sciserver}, as well as the stalwart services SciServer Compute (Jupyter based), SkyServer and CasJobs for synchronous and asynchronous data access, but adds several new enhancements and functionalities.

Two new apps have been added to work with the SkyServer tools.
The first is a new and more simple Cross-Match tool \footnote{\url{https://skyserver.sdss.org/dr19/CrossMatchTools/CrossMatch}} 
to replace the client web interface within SkyServer for the deprecated SkyQuery cross-match application \citep{Budavari2013}. This new tool allows users to interactively run 2-way, 2-dimensional spatial cross-matches across more than 50 astronomical catalogs stored as public tables in the CasJobs \textit{xmatch} database context. This tool embeds the standalone SQLxMatch cross-match interface \footnote{\url{https://skyserver.sdss.org/xmatch}} \citep{sciserver_xmatch}, which exposes a schema browser to explore the metadata associated with the catalogs, and helps users to dynamically build a SQL query that executes a cross-match between 2 catalog tables in CasJobs.

This SQL query can be run either synchronously (with a cross-match table returned to the user), or as an asynchronous batch job (with output table stored in MyDB). The query results in the execution of a dedicated stored procedure that implements the \textit{zones} cross-matching algorithm \citep{gray2007zones}, which leverages relational database algebra and B-Trees, and takes advantage of co-located catalog tables on fast RAID 6 NVMe storage for an efficient in-database cross-match.

For more advanced use cases, SciServer-Compute allows users to run python Jupyter Notebooks with the capability of running these cross-match queries on CasJobs in a programmatic way, including for example the cross-match against MyDB tables. An example of such Notebooks is provided in a Github repository for tutorials, discussed in Section \ref{sec:demos}.

\begin{figure*}
\centering
\includegraphics[width=1\textwidth]{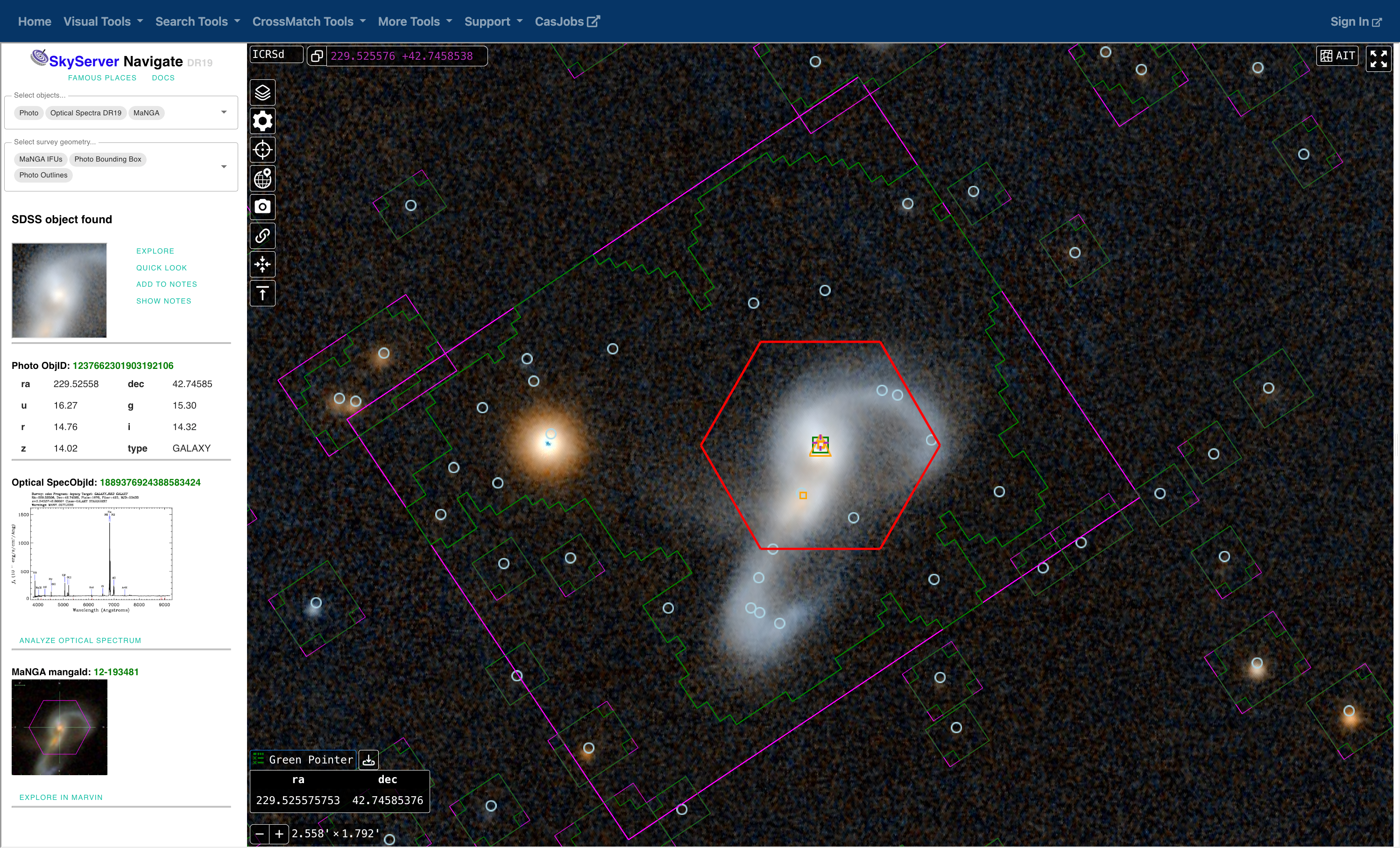}
\caption{SkyServer-Navigate application, fully redesigned.}
\label{fig:SkyServer-Navigate}
\end{figure*}

A major new feature for DR19 is the fully redesigned SkyServer Navigate App shown in Figure \ref{fig:SkyServer-Navigate}, which is now a new tool with an interactive sky-maps web interface 
\footnote{\url{https://skyserver.sdss.org/dr19/VisualTools/navi}}. The new Navigate incorporates more features and overcomes previous performance constraints. Earlier versions of Navigate, although very useful, were only able to display medium-to-high zoom levels of the sky regions within the SDSS legacy photometric footprint, thus being unable to display objects outside this footprint that were observed in subsequent SDSS spectroscopic surveys.
The new SkyServer-Navigate \citep{skyserver-navigate}, based on the Vue.js \footnote{\url{https://vuejs.org}} javascript framework, leverages the Aladin Lite viewer \footnote{\url{https://aladin.cds.unistra.fr/AladinLite/doc}} \citep{baumann2022} to greatly expand its functionality, being able to display maps from many other astronomical surveys at multiple wavelengths and on selectable spherical projections.

Navigate now uses the HiPS \footnote{\url{https://aladin.cds.unistra.fr/hips}} tiling mechanism \citep{Fernique2015} to store SDSS object catalogs across zoom levels in a hierarchical directory structure, so that the right amount of objects can be quickly overlaid on demand over the map at any zoom level without needlessly overcrowding the map. These objects can be then freely clicked by the users for further inspection of their spectra, close-up picture, and/or properties as observed in several SDSS surveys.

Moreover, Navigate also allows for selecting geometrical objects related to the different SDSS surveys, such as spectral plate and IFU outlines, survey masks, object bounding boxes, or footprint boundaries, which are displayed as colored polygons or circles overlaid on the maps.
For reproducibility and shareability, Navigate implements a full REST API \footnote{\url{https://restfulapi.net}}, allowing users to copy a URL link with all parameters needed to replicate the current map view, and share it with their collaborators or students for further discussion or dissemination.


In addition, CasJobs, which is the batch mode query submission service, hosts the public ``DR19'' context containing the catalog data for this data release, along with all the previous SDSS data sets. As described above, CasJobs will facilitate the cross-matching of DR19 and older SDSS data with other astronomical catalogs either directly in the app or via Jupyter notebooks in SciServer Compute. SciServer Compute updates for DR19 are described in a separate subsection below. Several new tutorials are available on SciServer that describe how to access the DR19 data.

\subsection{Sciserver Compute }
\label{sec:data.sciserver}

SciServer Compute - the Jupyter notebook based analysis environment available within SciServer - will facilitate the use of DR19 data in important ways. The DR19 SAS (file-based) data will be available locally for Jupyter notebooks to access, along with the catalog data available via CasJobs and SkyServer. The instructions for setting up your DR19 environment within SciServer Compute are listed below, and available on-line\footnote{\url{https://www.sdss.org/dr19/data_access/sciserver_compute/}}.

An addition to SciServer Compute for DR19 is the availability of a new public Docker image - named simply ``SDSS" - for creating containerized Jupyter Notebook sessions with the software and configurations needed to help users support their science use cases involving SDSS V data. This software environment includes dedicated Python libraries created specifically for SDSS, such as \textit{sdss-access} and \textit{sdss-semaphore}, as well as others related to astronomy in general, such as \textit{astropy}, \textit{astroquey}, and \textit{specutils}.

When creating a container, users need to attach the {\tt sdss\_sas} data volume, 
so that the DR19 data would become available as a folder under the 
\texttt{ \textasciitilde/workspace/sdss\_sas/dr19} path in their Jupyter Notebook session for 
fast analysis and/or exploration of the co-located SDSS data. This folder also includes a local copy of SDSS Jupyter notebook tutorials with demos and documentation that will help users to better understand and use the DR19 data, as explained in Section \ref{sec:demos}.

%% file: sections/data/datamodel.tex
With each data release, there are a variety of file species that are archived on the SAS that have a common data model.  The majority of theses file species are FITS files, which are organized on the SAS by the environmental variable of their respective data products\footnote{\url{https://dr19.sdss.org/sas/dr19/env}}.  This includes the spectroscopic pipeline output, with environmental variables including {\tt APOGEE\_REDUX}, {\tt BOSS\_SPECTRO\_REDUX}, {\tt LVM\_SPECTRO\_REDUX} and {\tt MWM\_ASTRA}; the raw observatory data, with environmental variables including {\tt APOGEE\_DATA\_N}, {\tt BOSS\_SPECTRO\_DATA\_N}, {\tt FCAM\_DATA\_N} and {\tt GCAM\_DATA\_N}; and also includes environmental variables for targetting and value-added catalogs (see section~\ref{sec:VAC}), with the data volume for each environmental variable tabulated at \url{https://www.sdss.org/dr19/data_access/volume/}.

In addition to the environmental variable, each file species with a data model has a file species name and an abstract path, enabling general access to a file species by keywords that are substituted into the abstract path to resolve into a specific file on the SAS, which is supported by the {\tt sdss\_access} software package for efficient downloading and a complete listing of available file species abstract paths\footnote{\url{https://sdss-access.readthedocs.io}}.

The combination of an environmental variable and a file species name is sufficient to determine a unique data model, which are listed at \url{https://data.sdss.org/datamodel}.  Within each data model, there is a section on the basic information, and HDU list (for FITS files) including HDU extension column names, descriptions and units.

%% file: sections/data/dataviz.tex
For DR19, the original Science Archive Webapp (SAW) used for previous SDSS data releases is complemented by a new two-component web framework consisting of: \texttt{zora}, a modern front-end user interface, and \texttt{valis}, a backend API for programmatic data access.  This new system is built atop the SDSS Software Framework, a set of Python tools that understand the SDSS environment, its databases, products, and data models.  The use of the SDSS Software Framework as the core backbone allows for the development of new tools and services in a flexible manner, agnostic to data-specific details. 

In addition to improving the accessibility of SDSS-V data within the SDSS Software Framework, work has also been done to make SDSS-V data accessible via community open-source tools like \texttt{specutils}~\citep{specutils2024} and \texttt{Jdaviz}~\citep{Jdaviz2022}, an open-source, data visualization and analysis package for astronomy data within both Jupyter and web-based environments.  This helps make SDSS data more accessible to the community at large, outside of SDSS-specific tooling.   

\subsubsection{Valis API Backend}

\texttt{valis} is a new back-end programmatic API built in the python \texttt{fastapi} framework, which provides distinctive endpoints for accessing SDSS information and data of interest.  With access to the SDSS databases, and Science Archive Server (SAS), \texttt{valis} provides convenient public access for users, via simple, HTTP GET/POST requests, to SDSS data.  It is meant to be a general, easily extensible, framework to  support both legacy and future SDSS surveys long-term. 

\texttt{valis} is structured as groups of related endpoints around common tasks, for example: \textbf{query:} Querying for SDSS-V targets by various mechanisms and criteria, \textbf{target:} Retrieving target-specific metadata, reduction parameters, or spectra, 
\textbf{info:} Retrieving general information about SDSS data products and releases,
\textbf{maskbits:} Working with traditional SDSS maskbits.  For a complete list, see the official \texttt{valis} documentation\footnote{\label{valisdocs}https://api.sdss.org/valis/docs}. 

As an independent API, \texttt{valis} powers the \texttt{zora} front-end, but is also useable on its own within python scripts and Jupyter notebooks.  By cleanly separating the back-end and front-end, users can access SDSS data in their preferred manner, or even build their own web interfaces on top of the API.     

\subsubsection{Zora Frontend}

\texttt{zora}, is a modern, reactive, component-based User Interface (UI) framework, built in \texttt{Vue}.  As the main user entry point, it provides quick data discovery, exploration, and visualization via a simple web interface to SDSS-V data.  For DR19, \texttt{zora} focuses on new data available from the BHM and MWM surveys, and is available to the public at \url{https://dr19.sdss.org/zora}. 

\begin{figure*}
\centering
\includegraphics[width=0.9\textwidth]{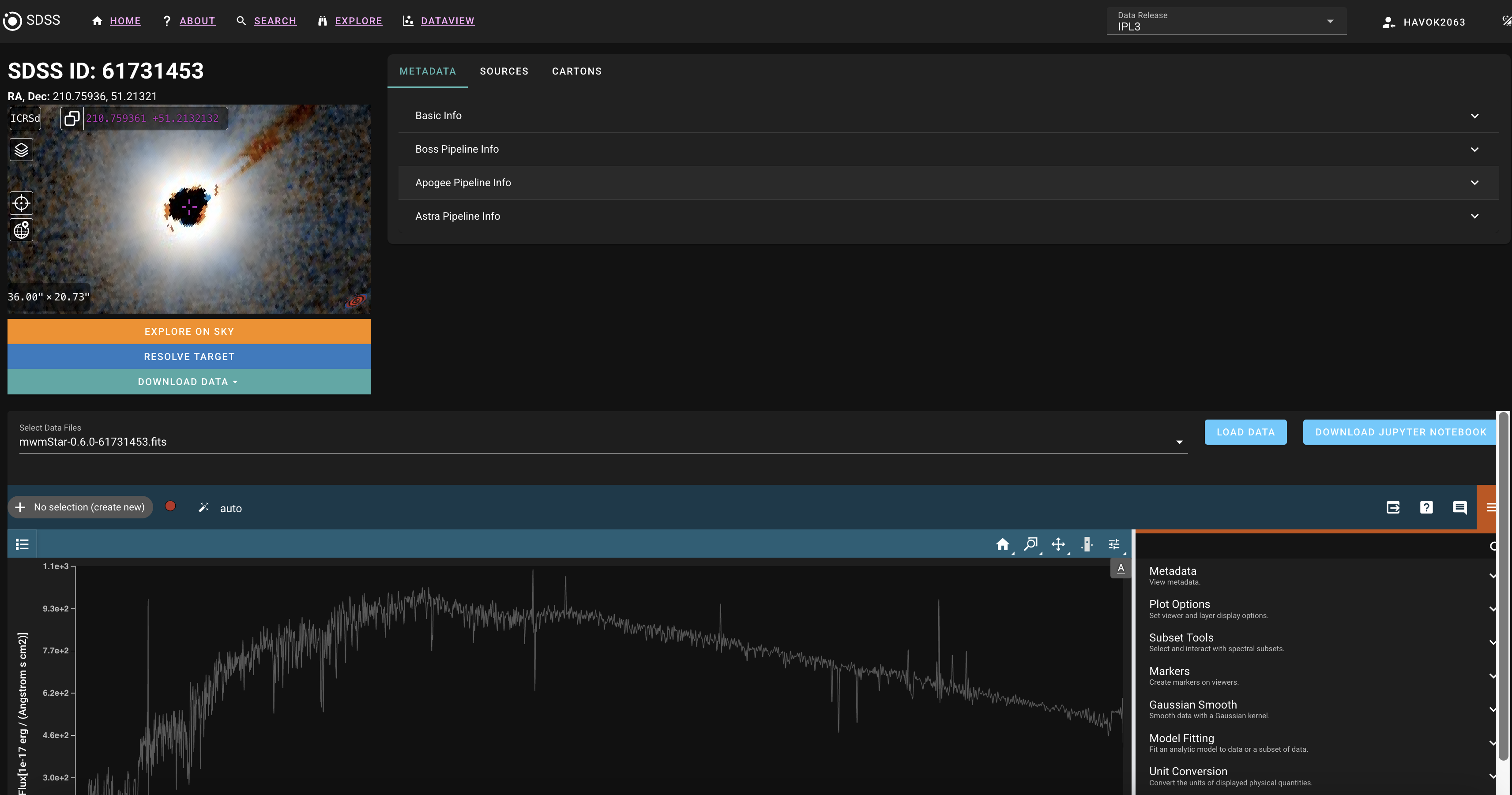}
\caption{Zora target page, with interactive spectral display, and pipeline and metadata parameters.}
\label{fig:zora-target}
\end{figure*}

\begin{figure*}
\centering
\includegraphics[width=0.9\textwidth]{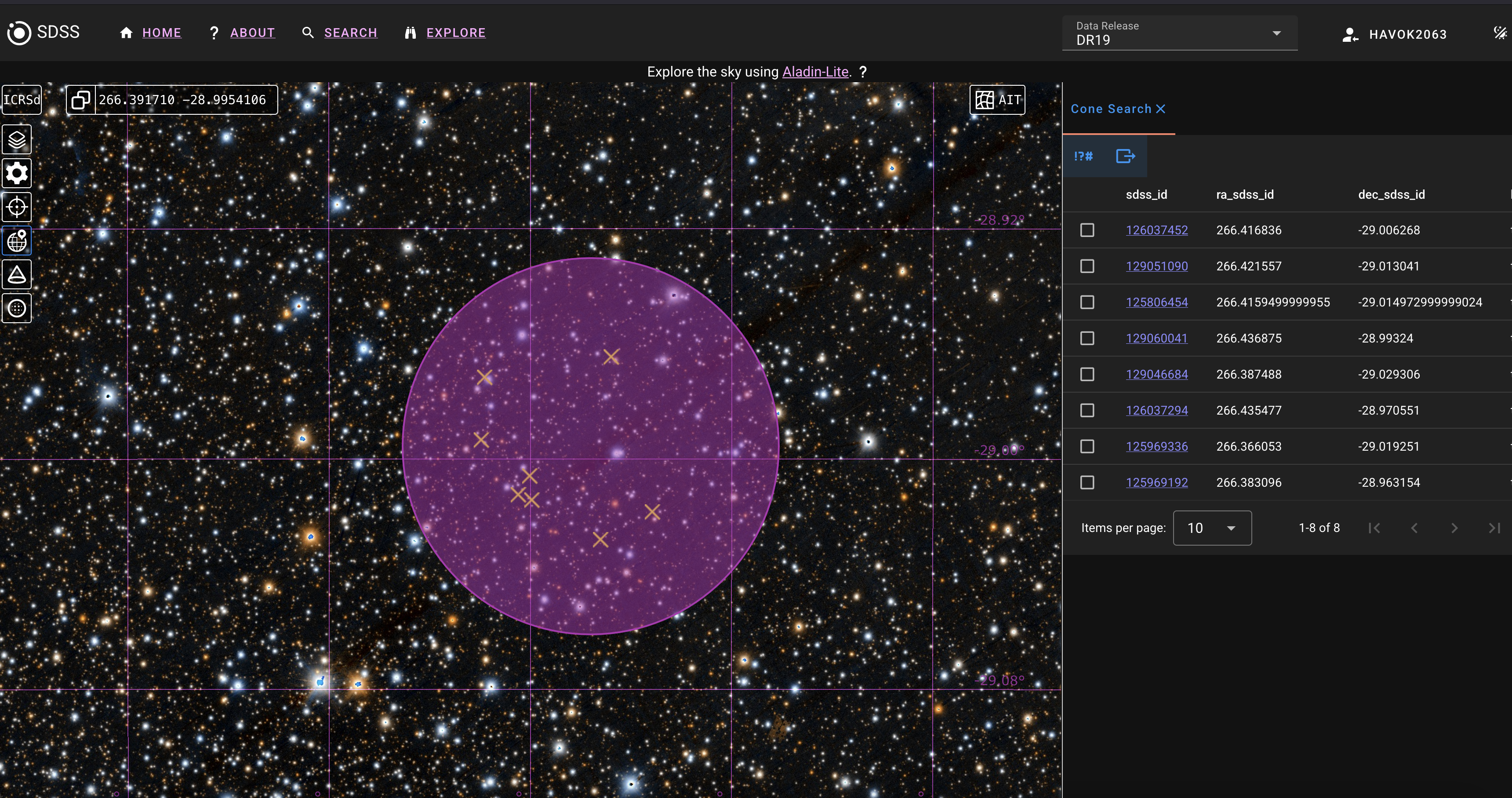}
\caption{Zora on-sky explorer, with cone search overlay and results.}
\label{fig:zora-explore}
\end{figure*}

\begin{figure*}
\centering
\includegraphics[width=0.9\textwidth]{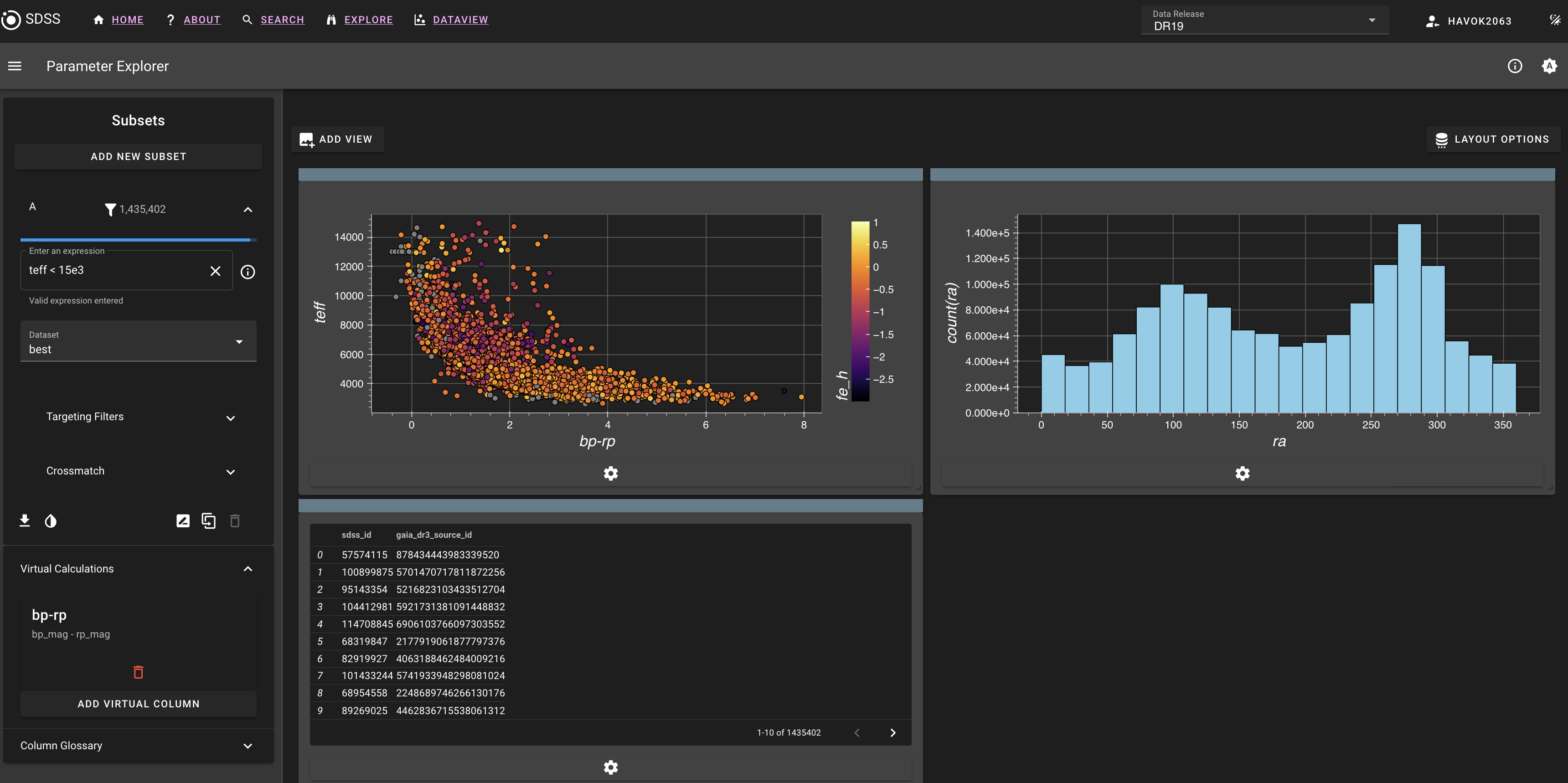}
\caption{Zora dataview dashboard, with interactive catalog filtering and chart plotting.}
\label{fig:zora-dataview}
\end{figure*}

Currently \texttt{zora} provides the following web functionality: \textbf{search:} search for targets by name, id, or coordinate cone search; \textbf{target:} visualize a specific target with detailed metadata, output pipeline parameters, and an interactive spectral display (Figure~\ref{fig:zora-target}); \textbf{explore:} explore targets on sky via an interactive sky viewer overlaying SDSS-V data alongside other astronomical datasets and surveys using \texttt{AladinLite}~\citep{aladinlite2024} (Figure~\ref{fig:zora-explore}); and \textbf{dataview:} a dynamic data dashboard rendering millions of SDSS-V catalog parameters using \texttt{solara} and \texttt{vaex} (Figure~\ref{fig:zora-dataview}). 

%% file: sections/mwm_cartons/mwm_cartons_appendix.tex
Below, we give detailed descriptions for three of the MWM plate target cartons.
For each target carton, we provide the following information: 
The {\bf Description of selection criteria} provides a short summary of the carton selection in  human-readable terms.  It also includes a list of the limits in color, magnitude, parallax, and other quantities applied to the carton's target candidates.
{\bf Data Sources} gives the catalogs from which these quantities are drawn. In {\bf Target priority options}, we indicate which priority is given to targets in this carton for observing; smaller priorities are more likely to be assigned fibers. The {\bf Cadence options} describes which exposure time requirement(s) are assigned to sources in the carton. 
The {\bf DR18 Counterpart} refers the reader to the corresponding science carton that was released in \citet{almeida2023}.  In some cases (e.g., \ref{mwm_galactic_core_plan} and \ref{mwm_yso_disk_apogee_plan}), the selection functions are identical, while in others (e.g. \ref{mwm_rv_long_fps_plan} ) slight differences exist due to differences in the exposure time or sky area available between plate and FPS observations.

\begin{center}\rule{0.5\linewidth}{0.5pt}\end{center}

\subsubsection{mwm\_gg\_core}\label{mwm_galactic_core_plan}
\begin{description}
    \item[Description of selection criteria] A simple color-magnitude cut effectively targets luminous cool giant stars (${\rm median}(\log g) \sim 1.0-1.5$) with little ($<$6\%) contamination from dwarf stars.
    \begin{itemize}
        \item $H < 11$
        \item$ G - H > 3.5$ or Gaia non-detection
         \item gal\_contam == 0
         \item cc\_flg == 0
        \item 0 $<$ rd\_flag $<= 3$
         \item ph\_qual flag is A or B
    \end{itemize}
\item[Data Sources] Gaia DR2 (G), 2MASS PSC (H, gal\_contam, cc\_flg, rd\_flag, ph\_qual)
\item[Target priority options] 2710
\item[Cadence options] 1 x 33~min
\item[DR18 Counterpart] \textit{mwm\_galactic\_core} - Section A.19 of \citet{almeida2023}.
\end{description}

\begin{center}\rule{0.5\linewidth}{0.5pt}\end{center}

\subsubsection{mwm\_rv\_long-bplates}\label{mwm_rv_long_fps_plan}
\begin{description}
    \item[Description of selection criteria] This carton selects stars previously observed at least 3 times with the APOGEE instrument if it was targeted as part of the main APOGEE sample or the APOGEE-2 binary program.
    \begin{itemize}
\item Presence in \texttt{sdss\_apogeeallstarmerge\_r13} file (previously observed with APOGEE 1 and/or 2
\item APOGEE number of visits $\geq6$  
\item $H < 12.2 $  
\item APOGEE TARGFLAGS includes one of the following: APOGEE\_SHORT,
APOGEE\_INTERMEDIATE, APOGEE\_LONG, APOGEE2\_BIN 
\item Gaia-based distance
\item Targets fall within a list of 20-34 hand selected field centers that are viewable for 6-months during the plate program 
    \end{itemize}
\item[Data Sources] APOGEE, 2MASS PSC (H), Gaia DR2
\item[Target priority options] 2526
\item[Cadence options] 6 x 67~min, 12 x 67~min 
\item[DR18 Counterpart] \textit{mwm\_rv\_long\_fps} - Section A.23 of \citet{almeida2023}.
\end{description}

\begin{center}\rule{0.5\linewidth}{0.5pt}\end{center}

\subsubsection{mwm\_yso\_s1}\label{mwm_yso_disk_apogee_plan}
\begin{description}
    \item[Description of selection criteria] YSO's expected to have protoplanetary disks, as identified by larger infrared excesses.
    \begin{itemize}
        \item $H < 13$
        \item $W1-W2 > 0.25$
        \item $W2-W3 > 0.50$
        \item $W3-W4> 1.50$
        \item $\varpi > 0.3$ mas
    \end{itemize}
\item[Data Sources]  Gaia DR2 ($\varpi$), 2MASS PSC (H), AllWise (W1, W2, W3, W4)
\item[Target priority options] 2700
\item[Cadence options] 3 x 67~min 
\item[DR18 Counterpart] \textit{mwm\_yso\_disk\_apogee} and \textit{mwm\_yso\_disk\_boss} -- Sections A.35 and A.36 of \citet{almeida2023}, respectively.
\end{description}

%% file: sections/bhm_cartons/bhm_target_cartons_v0.tex
\begin{center}\rule{0.5\linewidth}{0.5pt}\end{center}

\hypertarget{bhm_aqmes_med_plan0.1.0}{%
\subsubsection{bhm\_aqmes\_med}\label{bhm_aqmes_med_plan0.1.0}}

\noindent\textbf{target\_selection plan:} 0.1.0

\noindent\textbf{target\_selection tag:}
\href{https://github.com/sdss/target_selection/tree/0.1.0/}{0.1.0}

\noindent\textbf{Summary:} Spectroscopically confirmed optically bright SDSS
QSOs, selected from the SDSS QSO catalogue (DR16Q,
\citealt{Lyke2020}). Located in 36 mostly disjoint fields within the SDSS QSO
footprint that were pre-selected to contain higher than average numbers
of bright QSOs and CSC targets. The list of field centres can be found
within
\href{https://github.com/sdss/target_selection/blob/0.1.0/python/target_selection/masks/candidate_target_fields_bhm_aqmes_med_v0.2.1.fits}{the
target\_selection repository}.

\noindent\textbf{Simplified description of selection criteria:} Select all
objects from SDSS DR16 QSO catalogue that have
16.0\textless sdss\_psfmag\_i\textless19.1~AB, that lie within
1.49~degrees of at least one AQMES-medium field location.

\noindent\textbf{Target priority options:} 1100

\noindent\textbf{Cadence options:} bhm\_aqmes\_medium\_10x4

\noindent\textbf{Implementation:}
\href{https://github.com/sdss/target_selection/blob/0.1.0/python/target_selection/cartons/bhm_aqmes.py}{bhm\_aqmes.py}

\noindent\textbf{Number of targets:} 2663

\begin{center}\rule{0.5\linewidth}{0.5pt}\end{center}

\hypertarget{bhm_aqmes_med-faint_plan0.1.0}{%
\subsubsection{bhm\_aqmes\_med-faint}\label{bhm_aqmes_med-faint_plan0.1.0}}

\noindent\textbf{target\_selection plan:} 0.1.0

\noindent\textbf{target\_selection tag:}
\href{https://github.com/sdss/target_selection/tree/0.1.0/}{0.1.0}

\noindent\textbf{Summary:} Spectroscopically confirmed optically faint SDSS QSOs,
selected from the SDSS QSO catalogue (DR16Q,
\citealt{Lyke2020}). Located in 36 mostly disjoint fields within the SDSS QSO
footprint that were pre-selected to contain higher than average numbers
of bright QSOs and CSC targets. The list of field centres can be found
within
\href{https://github.com/sdss/target_selection/blob/0.1.0/python/target_selection/masks/candidate_target_fields_bhm_aqmes_med_v0.2.1.fits}{the
target\_selection repository}.

\noindent\textbf{Simplified description of selection criteria:} Select all
objects from SDSS DR16 QSO catalogue that have
19.1\textless sdss\_psfmag\_i\textless21.0~AB, that lie within
1.49~degrees of at least one AQMES-medium field location.

\noindent\textbf{Target priority options:} 3100

\noindent\textbf{Cadence options:} bhm\_aqmes\_medium\_10x4

\noindent\textbf{Implementation:}
\href{https://github.com/sdss/target_selection/blob/0.1.0/python/target_selection/cartons/bhm_aqmes.py}{bhm\_aqmes.py}

\noindent\textbf{Number of targets:} 16853

\begin{center}\rule{0.5\linewidth}{0.5pt}\end{center}

\hypertarget{bhm_aqmes_wide2_plan0.1.0}{%
\subsubsection{bhm\_aqmes\_wide2}\label{bhm_aqmes_wide2_plan0.1.0}}

\noindent\textbf{target\_selection plan:} 0.1.0

\noindent\textbf{target\_selection tag:}
\href{https://github.com/sdss/target_selection/tree/0.1.0/}{0.1.0}

\noindent\textbf{Summary:} Spectroscopically confirmed optically bright SDSS
QSOs, selected from the SDSS QSO catalogue (DR16Q,
\citealt{Lyke2020}). Located in 330 fields within the SDSS QSO footprint,
where the choice of survey area prioritized fields that overlapped with
the SPIDERS footprint (approx 180\textless b\textless360~deg), and/or
had higher than average numbers of bright QSOs and CSC targets. The list
of field centres can be found
\href{https://github.com/sdss/target_selection/blob/0.1.0/python/target_selection/masks/candidate_target_fields_bhm_aqmes_wide_v0.2.1.fits}{within
the target\_selection repository}.

\noindent\textbf{Simplified description of selection criteria:} Select all
objects from SDSS DR16 QSO catalogue that have
16.0\textless sdss\_psfmag\_i\textless19.1~AB, and that lie within
1.49~degrees of at least one AQMES-wide2 field location.

\noindent\textbf{Target priority options:} 1210

\noindent\textbf{Cadence options:} bhm\_aqmes\_wide\_2x4

\noindent\textbf{Implementation:}
\href{https://github.com/sdss/target_selection/blob/0.1.0/python/target_selection/cartons/bhm_aqmes.py}{bhm\_aqmes.py}

\noindent\textbf{Number of targets:} 18376

\begin{center}\rule{0.5\linewidth}{0.5pt}\end{center}

\hypertarget{bhm_aqmes_wide2-faint_plan0.1.0}{%
\subsubsection{bhm\_aqmes\_wide2-faint}\label{bhm_aqmes_wide2-faint_plan0.1.0}}

\noindent\textbf{target\_selection plan:} 0.1.0

\noindent\textbf{target\_selection tag:}
\href{https://github.com/sdss/target_selection/tree/0.1.0/}{0.1.0}

\noindent\textbf{Summary:} Spectroscopically confirmed optically faint SDSS QSOs,
selected from the SDSS QSO catalogue (DR16Q,
\citealt{Lyke2020}). Located in 330 fields within the SDSS QSO footprint,
where the choice of survey area prioritized field that overlapped with
the SPIDERS footprint (approx 180\textless b\textless360~deg), and/or
had higher than average numbers of bright QSOs and CSC targets. The list
of field centres can be found
\href{https://github.com/sdss/target_selection/blob/0.1.0/python/target_selection/masks/candidate_target_fields_bhm_aqmes_wide_v0.2.1.fits}{within
the target\_selection repository}.

\noindent\textbf{Simplified description of selection criteria:} Select all
objects from SDSS DR16 QSO catalogue that have
19.1\textless sdss\_psfmag\_i\textless21.0~AB, and that lie within
1.49~degrees of at least one AQMES-wide2 field location.

\noindent\textbf{Target priority options:} 3210

\noindent\textbf{Cadence options:} bhm\_aqmes\_wide\_2x4

\noindent\textbf{Implementation:}
\href{https://github.com/sdss/target_selection/blob/0.1.0/python/target_selection/cartons/bhm_aqmes.py}{bhm\_aqmes.py}

\noindent\textbf{Number of targets:} 63816

\begin{center}\rule{0.5\linewidth}{0.5pt}\end{center}

\hypertarget{bhm_aqmes_wide3_plan0.1.0}{%
\subsubsection{bhm\_aqmes\_wide3}\label{bhm_aqmes_wide3_plan0.1.0}}

\noindent\textbf{target\_selection plan:} 0.1.0

\noindent\textbf{target\_selection tag:}
\href{https://github.com/sdss/target_selection/tree/0.1.0/}{0.1.0}

\noindent\textbf{Summary:} Spectroscopically confirmed optically bright SDSS
QSOs, selected from the SDSS QSO catalogue (DR16Q,
\citealt{Lyke2020}). Located in 95 fields within the SDSS QSO footprint,
where the choice of survey area prioritized fields having higher than
average numbers of bright QSOs and CSC targets. The list of field
centres can be found
\href{https://github.com/sdss/target_selection/blob/0.1.0/python/target_selection/masks/candidate_target_fields_bhm_aqmes_wide_v0.2.1.fits}{within
the target\_selection repository}.

\noindent\textbf{Simplified description of selection criteria:} Select all
objects from SDSS DR16 QSO catalogue that have
16.0\textless sdss\_psfmag\_i\textless19.1~AB, and that lie within
1.49~degrees of at least one AQMES-wide3 field location.

\noindent\textbf{Target priority options:} 1200

\noindent\textbf{Cadence options:} bhm\_aqmes\_wide\_3x4

\noindent\textbf{Implementation:}
\href{https://github.com/sdss/target_selection/blob/0.1.0/python/target_selection/cartons/bhm_aqmes.py}{bhm\_aqmes.py}

\noindent\textbf{Number of targets:} 5785

\begin{center}\rule{0.5\linewidth}{0.5pt}\end{center}

\hypertarget{bhm_aqmes_wide3-faint_plan0.1.0}{%
\subsubsection{bhm\_aqmes\_wide3-faint}\label{bhm_aqmes_wide3-faint_plan0.1.0}}

\noindent\textbf{target\_selection plan:} 0.1.0

\noindent\textbf{target\_selection tag:}
\href{https://github.com/sdss/target_selection/tree/0.1.0/}{0.1.0}

\noindent\textbf{Summary:} Spectroscopically confirmed optically faint SDSS QSOs,
selected from the SDSS QSO catalogue (DR16Q,
\citealt{Lyke2020}). Located in 95 fields within the SDSS QSO footprint,
where the choice of survey area prioritized fields having higher than
average numbers of bright QSOs and CSC targets. The list of field
centres can be found
\href{https://github.com/sdss/target_selection/blob/0.1.0/python/target_selection/masks/candidate_target_fields_bhm_aqmes_wide_v0.2.1.fits}{within
the target\_selection repository}.

\noindent\textbf{Simplified description of selection criteria:} Select all
objects from SDSS DR16 QSO catalogue that have
19.1\textless sdss\_psfmag\_i\textless21.0~AB, and that lie within
1.49~degrees of at least one AQMES-wide2 field location.

\noindent\textbf{Target priority options:} 3200

\noindent\textbf{Cadence options:} dark\_2x4

\noindent\textbf{Implementation:}
\href{https://github.com/sdss/target_selection/blob/0.1.0/python/target_selection/cartons/bhm_aqmes.py}{bhm\_aqmes.py}

\noindent\textbf{Number of targets:} 35803

\begin{center}\rule{0.5\linewidth}{0.5pt}\end{center}

\hypertarget{bhm_aqmes_bonus-dark_plan0.1.0}{%
\subsubsection{bhm\_aqmes\_bonus-dark}\label{bhm_aqmes_bonus-dark_plan0.1.0}}

\noindent\textbf{target\_selection plan:} 0.1.0

\noindent\textbf{target\_selection tag:}
\href{https://github.com/sdss/target_selection/tree/0.1.0/}{0.1.0}

\noindent\textbf{Summary:} Spectroscopically confirmed SDSS QSOs, with magnitudes
in the range appropriate for observations in dark time, selected from
the SDSS QSO catalogue (DR16Q,
\citealt{Lyke2020}). Located anywhere within the SDSS DR16Q footprint.

\noindent\textbf{Simplified description of selection criteria:} Select all
objects from SDSS DR16 QSO catalogue that have
16.5\textless sdss\_psfmag\_i\textless21.5~AB

\noindent\textbf{Target priority options:} 3300

\noindent\textbf{Cadence options:} dark\_spiders\_1x4

\noindent\textbf{Implementation:}
\href{https://github.com/sdss/target_selection/blob/0.1.0/python/target_selection/cartons/bhm_aqmes.py}{bhm\_aqmes.py}

\noindent\textbf{Number of targets:} 579590

\begin{center}\rule{0.5\linewidth}{0.5pt}\end{center}

\hypertarget{bhm_aqmes_bonus-bright_plan0.1.0}{%
\subsubsection{bhm\_aqmes\_bonus-bright}\label{bhm_aqmes_bonus-bright_plan0.1.0}}

\noindent\textbf{target\_selection plan:} 0.1.0

\noindent\textbf{target\_selection tag:}
\href{https://github.com/sdss/target_selection/tree/0.1.0/}{0.1.0}

\noindent\textbf{Summary:} Spectroscopically confirmed SDSS QSOs, with magnitudes
in the range appropriate for observations in bright time, selected from
the SDSS QSO catalogue (DR16Q,
\citealt{Lyke2020}). Located anywhere within the SDSS DR16Q footprint.

\noindent\textbf{Simplified description of selection criteria:} Select all
objects from SDSS DR16 QSO catalogue that have
14.0\textless sdss\_psfmag\_i\textless18.0~AB

\noindent\textbf{Target priority options:} 4040

\noindent\textbf{Cadence options:} bhm\_boss\_bright\_3x1

\noindent\textbf{Implementation:}
\href{https://github.com/sdss/target_selection/blob/0.1.0/python/target_selection/cartons/bhm_aqmes.py}{bhm\_aqmes.py}

\noindent\textbf{Number of targets:} 10848

\begin{center}\rule{0.5\linewidth}{0.5pt}\end{center}

\hypertarget{bhm_rm_ancillary_plan0.1.0}{%
\subsubsection{bhm\_rm\_ancillary}\label{bhm_rm_ancillary_plan0.1.0}}

\noindent\textbf{target\_selection plan:} 0.1.0

\noindent\textbf{target\_selection tag:}
\href{https://github.com/sdss/target_selection/tree/0.1.0/}{0.1.0}

\noindent\textbf{Summary:} A supporting sample of candidate QSOs which have been
selected by the Gaia-unWISE AGN catalog
(\citealt{Shu2019}) and/or the SDSS XDQSO catalog
(\citealt{Bovy2011}). These targets are located within five (+1 backup) well
known survey fields (SDSS-RM, COSMOS, XMM-LSS, ECDFS, CVZ-S/SEP, and
ELIAS-S1).

\noindent\textbf{Simplified description of selection criteria:} Starting from a
parent catalogue of optically selected objects in the RM fields (as
presented by
\citealt{Yang2022}), select candidate QSOs that satisfy all of the
following: i) are identified via external ancillary methods
(photo\_bitmask \& 3 != 0); ii) have
15\textless psfmag\_i\textless21.5~AB; iii) do not have significant
detections (\textgreater3$\sigma$) of parallax and/or proper motion in Gaia
DR2; iv) are not classified as a STAR in SDSS DR16 spectroscopy; and v)
do not lie in the SDSS-RM field

\noindent\textbf{Target priority options:} 1004-1050

\noindent\textbf{Cadence options:} bhm\_rm\_174x8

\noindent\textbf{Implementation:}
\href{https://github.com/sdss/target_selection/blob/0.1.0/python/target_selection/cartons/bhm_rm.py}{bhm\_rm.py}

\noindent\textbf{Number of targets:} 948

\begin{center}\rule{0.5\linewidth}{0.5pt}\end{center}

\hypertarget{bhm_rm_core_plan0.1.0}{%
\subsubsection{bhm\_rm\_core}\label{bhm_rm_core_plan0.1.0}}

\noindent\textbf{target\_selection plan:} 0.1.0

\noindent\textbf{target\_selection tag:}
\href{https://github.com/sdss/target_selection/tree/0.1.0/}{0.1.0}

\noindent\textbf{Summary:} A sample of candidate QSOs selected via the methods
presented by
\citet{Yang2022}. These targets are located within five (+1 backup) well
known survey fields (SDSS-RM, COSMOS, XMM-LSS, ECDFS, CVZ-S/SEP, and
ELIAS-S1).

\noindent\textbf{Simplified description of selection criteria:} Starting from a
parent catalogue of optically selected objects in the RM fields (as
presented by
\citealt{Yang2022}), select candidate QSOs that satisfy all of the
following: i) are identified via the Skew-T algorithm (skewt\_qso == 1,
and skewt\_qso\_prior == 1 for targets in the CVZ-S/SEP field); ii) have
17\textless psfmag\_i\textless21.5~AB; iii) do not have significant
detections (\textgreater3$\sigma$) of parallax and/or proper motion in Gaia
DR2; iv) are not classified as a STAR in SDSS DR16 spectroscopy; vi)
have detections in all of the gri bands (a Gaia detection is sufficient
in the CVZ-S/SEP field); and vii) do not lie in the SDSS-RM field

\noindent\textbf{Target priority options:} 1002-1050

\noindent\textbf{Cadence options:} bhm\_rm\_174x8

\noindent\textbf{Implementation:}
\href{https://github.com/sdss/target_selection/blob/0.1.0/python/target_selection/cartons/bhm_rm.py}{bhm\_rm.py}

\noindent\textbf{Number of targets:} 3811

\begin{center}\rule{0.5\linewidth}{0.5pt}\end{center}

\hypertarget{bhm_rm_var_plan0.1.0}{%
\subsubsection{bhm\_rm\_var}\label{bhm_rm_var_plan0.1.0}}

\noindent\textbf{target\_selection plan:} 0.1.0

\noindent\textbf{target\_selection tag:}
\href{https://github.com/sdss/target_selection/tree/0.1.0/}{0.1.0}

\noindent\textbf{Summary:} A sample of candidate QSOs selected via their optical
variability properties, as presented by
\citet{Yang2022}. These targets are located within five (+1 backup) well
known survey fields (SDSS-RM, COSMOS, XMM-LSS, ECDFS, CVZ-S/SEP, and
ELIAS-S1).

\noindent\textbf{Simplified description of selection criteria:} Starting from a
parent catalogue of optically selected objects in the RM fields (as
presented by
\citealt{Yang2022}), select candidate QSOs that satisfy all of the
following: i) have significant variability in the DES or PanSTARRS1
multi-epoch photometry (var\_sn{[}g{]}\textgreater3 and
var\_rms{[}g{]}\textgreater0.05); ii) have
17\textless psfmag\_i\textless20.5~AB; iii) do not have significant
detections (\textgreater3$\sigma$) of parallax and/or proper motion in Gaia
DR2; iv) are not classified as a STAR in SDSS DR16 spectroscopy; and vi)
do not lie in the SDSS-RM field

\noindent\textbf{Target priority options:} 1003-1050

\noindent\textbf{Cadence options:} bhm\_rm\_174x8

\noindent\textbf{Implementation:}
\href{https://github.com/sdss/target_selection/blob/0.1.0/python/target_selection/cartons/bhm_rm.py}{bhm\_rm.py}

\noindent\textbf{Number of targets:} 992

\begin{center}\rule{0.5\linewidth}{0.5pt}\end{center}

\hypertarget{bhm_rm_known_spec_plan0.1.0}{%
\subsubsection{bhm\_rm\_known\_spec}\label{bhm_rm_known_spec_plan0.1.0}}

\noindent\textbf{target\_selection plan:} 0.1.0

\noindent\textbf{target\_selection tag:}
\href{https://github.com/sdss/target_selection/tree/0.1.0/}{0.1.0}

\noindent\textbf{Summary:} A sample of known QSOs identified through optical
spectroscopy from various projects, as collated by
\citet{Yang2022}. These targets are located within five (+1 backup) well
known survey fields (SDSS-RM, COSMOS, XMM-LSS, ECDFS, CVZ-S/SEP, and
ELIAS-S1).

\noindent\textbf{Simplified description of selection criteria:} Starting from a
parent catalogue of optically selected objects in the RM fields (as
presented by
\citealt{Yang2022}), select targets which satisfy all of the following: i)
are flagged as having a spectroscopic identification in the parent
catalogue; ii) have 15\textless psfmag\_i\textless21.7~AB (SDSS-RM,
CDFS, ELIAS-S1, CVZ-S/SEP fields), or
15\textless psfmag\_i\textless21.5~AB (COSMOS and XMM-LSS fields); iii)
have a spectroscopic redshift in the range 0.005\textless z\textless7;
iv) are not classified as a STAR in SDSS DR16 spectroscopy;

\noindent\textbf{Target priority options:} 1001-1050

\noindent\textbf{Cadence options:} bhm\_rm\_174x8

\noindent\textbf{Implementation:}
\href{https://github.com/sdss/target_selection/blob/0.1.0/python/target_selection/cartons/bhm_rm.py}{bhm\_rm.py}

\noindent\textbf{Number of targets:} 2992

\begin{center}\rule{0.5\linewidth}{0.5pt}\end{center}

\hypertarget{bhm_csc_apogee_plan0.1.0}{%
\subsubsection{bhm\_csc\_apogee}\label{bhm_csc_apogee_plan0.1.0}}

\noindent\textbf{target\_selection plan:} 0.1.0

\noindent\textbf{target\_selection tag:}
\href{https://github.com/sdss/target_selection/tree/0.1.0/}{0.1.0}

\noindent\textbf{Summary:} X-ray sources from the CSC2.0 source catalogue with
NIR counterparts in 2MASS PSC

\noindent\textbf{Simplified description of selection criteria:} Starting from the
parent catalogue of CSC2.0 sources with optical/IR counterparts
(bhm\_csc). Select entries satisfying the following criteria: i)
counterpart is from the 2MASS catalogue, ii) 2MASS H-band magnitude
measurement is in the accepted range for SDSS-V:
10.0\textless H\textless15.0.

\noindent\textbf{Target priority options:} 4000

\noindent\textbf{Cadence options:} bhm\_csc\_apogee\_3x1

\noindent\textbf{Implementation:}
\href{https://github.com/sdss/target_selection/blob/0.1.0/python/target_selection/cartons/bhm_csc.py}{bhm\_csc.py}

\noindent\textbf{Number of targets:} 10633

\begin{center}\rule{0.5\linewidth}{0.5pt}\end{center}

\hypertarget{bhm_csc_boss_dark_plan0.1.0}{%
\subsubsection{bhm\_csc\_boss\_dark}\label{bhm_csc_boss_dark_plan0.1.0}}

\noindent\textbf{target\_selection plan:} 0.1.0

\noindent\textbf{target\_selection tag:}
\href{https://github.com/sdss/target_selection/tree/0.1.0/}{0.1.0}

\noindent\textbf{Summary:} X-ray sources from the CSC2.0 source catalogue with
counterparts in Panstarrs1-DR1

\noindent\textbf{Simplified description of selection criteria:} Starting from the
parent catalogue of CSC sources with optical/IR counterparts (bhm\_csc).
Select entries satisfying the following criteria: i) counterpart is from
the PanSTARRS1 catalogue, ii) optical flux/magnitude is in a range
suited to SDSS-V dark time observations:
16.0\textgreater psfmag\_i\textgreater22.0~AB.

\noindent\textbf{Target priority options:} 3000

\noindent\textbf{Cadence options:} bhm\_csc\_boss\_1x4

\noindent\textbf{Implementation:}
\href{https://github.com/sdss/target_selection/blob/0.1.0/python/target_selection/cartons/bhm_csc.py}{bhm\_csc.py}

\noindent\textbf{Number of targets:} 65350

\begin{center}\rule{0.5\linewidth}{0.5pt}\end{center}

\hypertarget{bhm_csc_boss-bright_plan0.1.0}{%
\subsubsection{bhm\_csc\_boss-bright}\label{bhm_csc_boss-bright_plan0.1.0}}

\noindent\textbf{target\_selection plan:} 0.1.0

\noindent\textbf{target\_selection tag:}
\href{https://github.com/sdss/target_selection/tree/0.1.0/}{0.1.0}

\noindent\textbf{Summary:} X-ray sources from the CSC2.0 source catalogue with
counterparts in Panstarrs1-DR1

\noindent\textbf{Simplified description of selection criteria:} Starting from the
parent catalogue of CSC sources with optical/IR counterparts (bhm\_csc).
Select entries satisfying the following criteria: i) counterpart is from
the PanSTARRS1 catalogue, ii) optical flux/magnitude is in a range
suited to SDSS-V bright time observations:
13.0\textgreater psfmag\_i\textgreater18.0~AB.

\noindent\textbf{Target priority options:} 4000

\noindent\textbf{Cadence options:} bhm\_csc\_boss\_1x1

\noindent\textbf{Implementation:}
\href{https://github.com/sdss/target_selection/blob/0.1.0/python/target_selection/cartons/bhm_csc.py}{bhm\_csc.py}

\noindent\textbf{Number of targets:} 22173

\begin{center}\rule{0.5\linewidth}{0.5pt}\end{center}

\hypertarget{bhm_gua_dark_plan0.1.0}{%
\subsubsection{bhm\_gua\_dark}\label{bhm_gua_dark_plan0.1.0}}

\noindent\textbf{target\_selection plan:} 0.1.0

\noindent\textbf{target\_selection tag:}
\href{https://github.com/sdss/target_selection/tree/0.1.0/}{0.1.0}

\noindent\textbf{Summary:} A sample of optically faint candidate AGN lacking
spectroscopic confirmations, derived from the parent sample presented by
\citet{Shu2019}, who applied a machine-learning approach to select QSO
candidates from a combination of the Gaia DR2 and unWISE catalogues.

\noindent\textbf{Simplified description of selection criteria:} Starting with the
\citet{Shu2019} catalogue, select targets which satisfy the following
criteria: i) have a Random Forest probability of being a QSO
of\textgreater0.8, ii) are in the (dereddened) magnitude range suitable
for BOSS spectroscopy in dark time
(G\textsubscript{dered}\textgreater16.5 and
RP\textsubscript{dered}\textgreater16.5, as well as
G\textsubscript{dered}\textless21.2 or
RP\textsubscript{dered}\textless21.0,~Vega mag), iii) do not have good
quality optical spectroscopic measurements in SDSS DR16

\noindent\textbf{Target priority options:} 3400

\noindent\textbf{Cadence options:} bhm\_spiders\_1x4

\noindent\textbf{Implementation:}
\href{https://github.com/sdss/target_selection/blob/0.1.0/python/target_selection/cartons/bhm_gua.py}{bhm\_gua.py}

\noindent\textbf{Number of targets:} 2085729

\begin{center}\rule{0.5\linewidth}{0.5pt}\end{center}

\hypertarget{bhm_gua_bright_plan0.1.0}{%
\subsubsection{bhm\_gua\_bright}\label{bhm_gua_bright_plan0.1.0}}

\noindent\textbf{target\_selection plan:} 0.1.0

\noindent\textbf{target\_selection tag:}
\href{https://github.com/sdss/target_selection/tree/0.1.0/}{0.1.0}

\noindent\textbf{Summary:} A sample of optically bright candidate AGN lacking
spectroscopic confirmations, derived from the parent sample presented by
\citet{Shu2019}, who applied a machine-learning approach to select QSO
candidates from a combination of the Gaia DR2 and unWISE catalogues.

\noindent\textbf{Simplified description of selection criteria:} Starting with the
\citet{Shu2019} catalogue, select targets which satisfy the following
criteria: i) have a Random Forest probability of being a QSO
of\textgreater0.8, ii) are in the (dereddened) magnitude range suitable
for BOSS spectroscopy in bright time
(G\textsubscript{dered}\textgreater13.0 and
RP\textsubscript{dered}\textgreater13.5, as well as
G\textsubscript{dered}\textless18.5 or
RP\textsubscript{dered}\textless18.5,~Vega mags), iii) do not have good
quality optical spectroscopic measurements in SDSS DR16.

\noindent\textbf{Target priority options:} 4040

\noindent\textbf{Cadence options:} bhm\_boss\_bright\_3x1

\noindent\textbf{Implementation:}
\href{https://github.com/sdss/target_selection/blob/0.1.0/python/target_selection/cartons/bhm_gua.py}{bhm\_gua.py}

\noindent\textbf{Number of targets:} 237236